\documentclass[journal]{IEEEtran}
\usepackage[OT1]{fontenc}
\usepackage{amsmath,amsopn,amssymb}
\usepackage{graphicx,xspace,color,soul,enumerate}
\usepackage{epsfig,subfigure}
\usepackage{booktabs,longtable,multirow}
\usepackage{array,float,enumitem}
\usepackage{cite,stfloats}
\usepackage[ruled,vlined]{algorithm2e}
\usepackage{bm,epstopdf,url}

\allowdisplaybreaks[4]
\def\ie{{\textit{i.e.}}}

\def\etc{{\textit{etc}}}
\def\etal{{\textit{et al.~}}}

\def\e{{\mathbf e}}

\def\u{{\mathbf u}}

\def\x{{\mathbf x}}

\def\A{{\mathbf A}}

\def\D{\mathbf{D}}

\def\I{{\mathbf I}}

\def\U{{\mathbf U}}
\def\W{{\mathbf W}}

\def\bfLambda{{\boldsymbol \Lambda}}

\def\cL{{\mathcal L}}

\begin{document}

\title{Generic Reversible Visible Watermarking Via Regularized Graph Fourier Transform Coding}

\author{
    Wenfa~Qi{\small},~\IEEEmembership{}
    Sirui~Guo{\small},~\IEEEmembership{}
    Wei~Hu{\small},~\IEEEmembership{Senior~Member,~IEEE}
    \thanks{W. Qi and W. Hu are with Wangxuan Institute of Computer Technology, Peking University, No. 128, Zhongguancun North Street, Beijing, China (e-mail:qiwenfa@pku.edu.cn; forhuwei@pku.edu.cn).

S. Guo is with State Key Laboratory of Integrated Services Networks, Xidian University, No. 2 South Taibai Road, Xi'an, China (e-mail: srguo@stu.xidian.edu.cn).
}
\thanks{Corresponding author: Wei Hu (forhuwei@pku.edu.cn).}
 }

\maketitle
\begin{abstract}
Reversible visible watermarking (RVW) is an active copyright protection mechanism. It not only transparently superimposes copyright patterns on specific positions of digital images or video frames to declare the copyright ownership information, but also completely erases the visible watermark image and thus enables restoring the original host image without any distortion. However, existing RVW algorithms mostly construct the reversible mapping mechanism for a specific visible watermarking scheme, which is not versatile. Hence, we propose a generic RVW framework to accommodate various visible watermarking schemes.
In particular, we obtain a reconstruction data packet---the compressed difference image between the watermarked image and the original host image, which is embedded into the watermarked image via any conventional reversible data hiding method to facilitate the blind recovery of the host image.
The key is to achieve compact compression of the difference image for efficient embedding of the reconstruction data packet.
To this end, we propose regularized Graph Fourier Transform (GFT) coding, where the difference image is smoothed via the graph Laplacian regularizer for more efficient compression and then encoded by multi-resolution GFTs in an approximately optimal manner.
Experimental results show that the proposed framework has much better versatility than state-of-the-art methods. Due to the small amount of auxiliary information to be embedded, the visual quality of the watermarked image is also higher.
\end{abstract}


\IEEEpeerreviewmaketitle

\section{Introduction}
\label{sec:intro}
Reversible data hiding (RDH) has attracted increasing attention in the field of information hiding \cite{b1,b2,b3,b4,b5,b6,b7,AE1,AE2,NEW_PVO,NEW_DE}.
It enables not only accurate extraction of the secret data from the watermarked image, but also lossless restoration of the original host image, thus avoiding permanent distortion on the host image induced by conventional information hiding. 
RDH has been widely applied in many sensitive scenes including image authentication \cite{b8,AE0}, medical image processing \cite{b9,b10}, and video error-concealment coding \cite{b11}.
Based on the visibility of hidden information, RDH can be divided into two categories: invisible and visible RDH. 
Among them, the reversible visible watermarking (RVW) is considered to be an active copyright protection mechanism. 
RVW superimposes the owner's copyright patterns to a specific location of a digital image or a video frame in a translucent manner, so that the ownership can be identified directly, which makes illegal uses difficult. 
Further, the visible watermark image can be completely erased and the host image can be losslessly restored, which promotes the applications of RDH in many fields, such as content protection \cite{b17}, copyright identification \cite{b18} and advertising \cite{b19}. 

 \begin{figure}[t]
  \centering
   \includegraphics[width=0.48\textwidth]{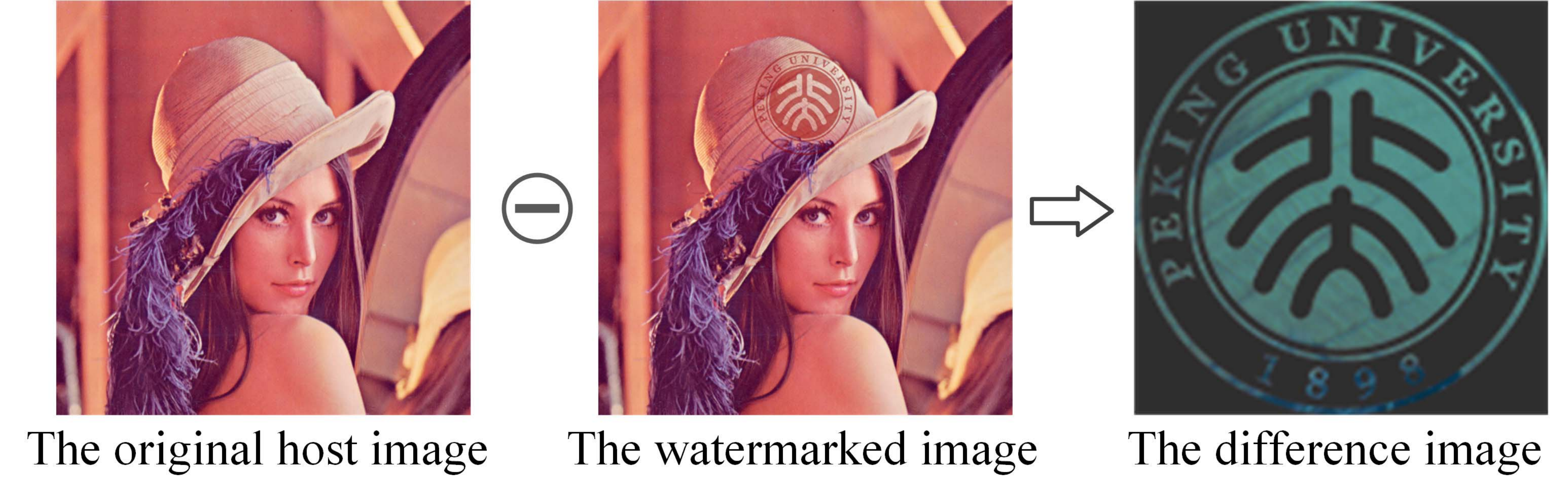}
   \caption{An example of the difference image between the original host image and the watermarked image. The difference image is magnified for clear demonstration.}
  \label{fig:data_packet}
 \end{figure}
 
Most previous RVW works \cite{b21,b22,b23,b24,b25,b26,b27,b28,b29,b30,b31,b32,b33,b34,b35} focus on how to improve the visual fusion effect of the watermarked image. Few attempts study the generic construction of reversible mapping relationship (RMR), through which the visible watermark image can be embedded into the host image appropriately and removed in a reversible manner. 
In particular, there exist two major challenges in conventional RVW methods:
1) The RMR is constructed for a {\it specific} visible watermarking algorithm, which cannot be generalized to other visible watermarking approaches. Thus, the {\it versatility} of existing RVW methods is limited \cite{b22,b23,b25}.
2) The balance between the reversibility and the visual quality of the watermarked image is hardly achieved \cite{b21,b28,b31}. The construction of RMR often limits the flexibility of visible watermarking, which further affects the visual quality of the watermarked image.

{While attempts \cite{AE0, b27} have been made to use any visible data hiding technique, there still exist some limitations.  
In \cite{AE0}, the authors proposed a block-based RVW method that never requires any location map. In the watermark embedding, any arbitrary visible watermarking algorithm is able to be used for serving image authentication. However, the reversing process corresponding to visible watermark embedding is needed to make the visibly watermarked bitplane unmarked. 
Qin \etal \cite{b27} consider the scenario of visible-watermark removal for the image generated by any visible watermarking algorithm, which can remove the embedded visible watermark and also exactly recover the original image with the assist of an image inpainting technique. However, this method has the following two disadvantages: 1) The image restoration process needs the access of the original visible watermark image, which is limited in practical applications. 2) The lossless compression algorithm used in this method is inefficient and the visual quality of the final watermarked image is poor. In all, the versatility of the existing methods is greatly limited.  }


To address these challenges, we propose a reversible visible watermarking framework that is versatile to construct reversible mapping relationship and meanwhile leads to satisfactory visual quality of the watermarked image.
Firstly, in order to achieve the versatility of the proposed framework, \ie, the construction of the RMR is independent of specific visible watermarking schemes, we introduce a {\it reconstruction data packet} to facilitate the removal of the visible watermark image. 
In particular, after the visible watermark image is embedded, a reconstruction data packet is acquired from the compressed {\it difference image} between the watermarked image and host image, and is embedded into the watermarked image as the auxiliary information.
During the restoration of the host image, we extract the auxiliary information from the watermarked image first, and reconstruct the data packet to facilitate the removal of the visible watermark image, which leads to the construction of a general RMR. 
Also, the proposed method performs the watermarking extraction in a blind manner, which does not resort to the original visible watermark image.

Secondly, due to the large amount of data in the difference image for embedding as the auxiliary information, it is in demand to compress the difference image efficiently. 
A compact representation of the difference image would allow for flexible visible watermarking, thus leading to the balance between the reversibility and the visual quality of the watermarked image.
Our proposed compression method is based on two basic assumptions: 1) The texture complexity of an appropriate visible watermark image is moderate, which is beneficial to data compression. 
2) Appropriate modification of the visible-watermarked region is acceptable, since visible watermarking has already caused obvious visual impact on the host image, which means lossy compression can be introduced.
As shown in Fig.~\ref{fig:data_packet}, a typical difference image often exhibits sharp contours and smooth interior surfaces with certain amounts of gradual transitions, which is referred to as generalized piecewise smoothness (GPWS) \cite{b40}. 
In order to efficiently encode the difference image, we propose to smooth it via the graph Laplacian regularizer (GLR) \cite{pang2017graph,Hu20GSP} first, which enforces the difference image to be {\it piecewise smooth} (PWS) with few transitions. 
Then, we represent the smoothed difference image on graphs and leverage the multi-resolution GFT (MR-GFT) coding method in \cite{b40}.    
As we pre-process the difference image by GLR-regularized smoothing, we reformulate the rate-distortion objective in \cite{b40} to account for the distortion caused by the smoothing process for approximately optimal decorrelation, which is referred to as regularized GFT (RGFT) coding. We further develop an efficient algorithm for the RGFT coding, which ensures a good trade-off between the reversibility and the visual quality of the watermarked image. 

To summarize, our main contributions include:
\begin{itemize}
    \item We propose a reversible visible watermarking framework that is {\it versatile} to construct reversible mapping relationship and {\it achieve reversibility for any visible watermarking algorithm}, by introducing a reconstruction data packet to facilitate the removal of the visible watermark image. 
    In addition, the proposed method performs the watermarking extraction blindly, which does not resort to the original visible watermark image. 
    
    \item We propose Regularized Graph Fourier Transform (RGFT) coding for approximately optimal compression tailored to the difference image between the original host image and the watermarked image, leading to a compact reconstruction data packet. This compact representation allows for flexible embedding of the reconstruction data packet, thus leading to the balance between the reversibility and the visual quality of the watermarked image.
    
    
    \item Experimental results show that the proposed framework has much better versatility and higher compression efficiency than state-of-the-art methods. Due to the small amount of auxiliary information to be embedded, the visual quality of the watermarked image is also higher. 
\end{itemize}

The rest of this paper is organized as follows. We review previous works on RVW in Section~\ref{sec:related}. 
We then overview the proposed versatile framework in Section~\ref{sec:method}, and elaborate on our regularized GFT coding in Section~\ref{sec:graph}. 
Finally, experimental results and conclusions are presented in Section~\ref{sec:results} and Section~\ref{sec:conclude}, respectively.

\section{Related Works}
\label{sec:related}
 Previous attempts on RVW mainly involve data compression \cite{b21,b22,b24}, spatial-domain modification \cite{AE0,b25,b26,b27,b28,b29}, or frequency-domain modification \cite{b23,b30,b31,b32,b33,b34,b35}. 
 Please note that, the key property of RVW is the perfect recovery of both the host image and the embedded secret data. While any distortion / processing for removal (e.g., in-painting) of the visible watermark will lead to inaccurate recovery of the host image, this is {\it beyond the scope of this work}.

\subsection{Data-compression-based RVW}
Hu \etal \cite{b21} first proposed a RVW algorithm based on bit-plane compression. In particular, this method substitutes a prominent bit plane of the watermarked region with watermark embedding, and the original values of the specified bit plane are compressed and embedded into other non-watermarked regions.
Qin \etal \cite{b27} proposed two schemes for visible watermark removal and reversible image recovery. In the first scheme, a run-length coding based method is utilized to compress the difference between the preliminary recovered image and original image. The second scheme can perfectly remove the embedded visible watermark and exactly recover the original image with the assistance of an image inpainting technique {, which can also collaborate with any visible data hiding technique. For any visibly watermarked image $\I^{w}$ by embedding visible watermark image $\W$ into the host image $\I$, classic image inpainting methods can be used to fill in the image regions covered by the visible watermark image. The pre-processed image $\I^{a}$ with the visible watermark image roughly removed by inpainting is very similar to the original host image $\I$. Then the difference image $\D$ between $\I$ and $\I^{a}$ is calculated and compressed to acquire $D_c$ by using lossless compression methods.  The final watermarked image $\I^{w\prime}$ is obtained by embedding $D_c$ into $\I^{w}$ using the conventional RDH. On the receiver side, the compressed data $D_c$ is extracted from the watermark image $\I^{w\prime}$, and then the difference image $\D$ is obtained after decompression, leading to the recovered watermark image $\I^{w}$. Next, $\I^{w}$ is processed by the same image inpainting method to generate image $\I^{a}$. Finally, the final restored image $\I^{\prime}$ is obtained by adding $\I^{w}$ and $\D$, that is, $\I^{\prime}=\I$. Therefore, this method can be applied to any visible watermarking algorithm, no matter whether the sender and the receiver know the algorithms or not.} Yang \etal \cite{b39} proposed a RVW algorithm by establishing a packet according to the difference image between the host image and its approximate version, which can improve the efficiency of data compression. However, the generation of the approximation depends on the specific visible watermarking algorithm, which means that this framework can only be applied to specific algorithms with no versatility. Also, the data volume of the embedded packet is still large. 

\subsection{Spatial-modification-based RVW}
RVW algorithms based on spatial modification are relatively more popular. This class of methods mainly realize the embedding of the visible watermark image by modifying pixels in the spatial domain, and then construct the RMR to recover the original host image. 
 Han \etal \cite{AE0} proposed a block-based RVW algorithm with authentication ability. Firstly, the host image is divided into sub-blocks, and the feature of each sub-block is generated and hidden by RDH. Then, blocks are chosen to be visibly marked according to the watermark pattern using any visible watermarking algorithm. The hidden image features play the role of positioning visual watermarked areas during the image recovery and are used for image authentication. Chen \etal \cite{b25} proposed a RVW algorithm based on an improved conventional Difference Expansion (DE) method. Firstly, the host image is divided into sub-blocks. Then, pixels in each block are divided into two groups, and the visible watermark image is embedded by using the difference expansion between the sum of two groups of pixel values. Qi \etal \cite{b26} and Qin \etal \cite{b27} proposed the corresponding improved methods to address the problems of more overflow image pixels and monotonous visual effect of the visible watermark image in \cite{b25}, respectively. 
Liu \etal \cite{b28} proposed a RVW scheme based on hybrid mapping, which can map a pixel value to another close value by the composite mapping to embed the visible watermark image. In addition, the mapping is reversible, but only by using the original visible watermark image can the original carrier image be restored. In \cite{b29}, a RVW scheme based on pixel-value-interval-mapping is proposed, which takes into account the characteristics of visibility, transparency and robustness of the visible watermarking.

\subsection{Spectral-modification-based RVW}
This class of methods perform modification in the spectral domain. 
In \cite{b23}, a new reversible visible watermarking scheme based on the Absolute Moment Block Truncation Coding (AMBTC) domain is proposed, which achieves high visual quality of stego images and recovers the original BTC-compressed image losslessly. Moreover, it is robust against common signal processing attacks. 
Yang \etal \cite{b32} proposed a RVW algorithm based on the DCT transform.
Ohura  \etal \cite{b33} proposed  a new RVW method based on the dyadic lifting scheme. This method generates a visible watermarked image by embedding a binary logo into an image using the dyadic wavelet transform (DYWT) and interval arithmetic (IA). 
Yeh \etal \cite{b44-2008} proposed a visible watermarking removal method in the JPEG compression domain which depends on the user's secret key.

\section{The Proposed Versatile Framework}
\label{sec:method}
\begin{figure}[t]
  \centering
  \includegraphics[width=0.48\textwidth]{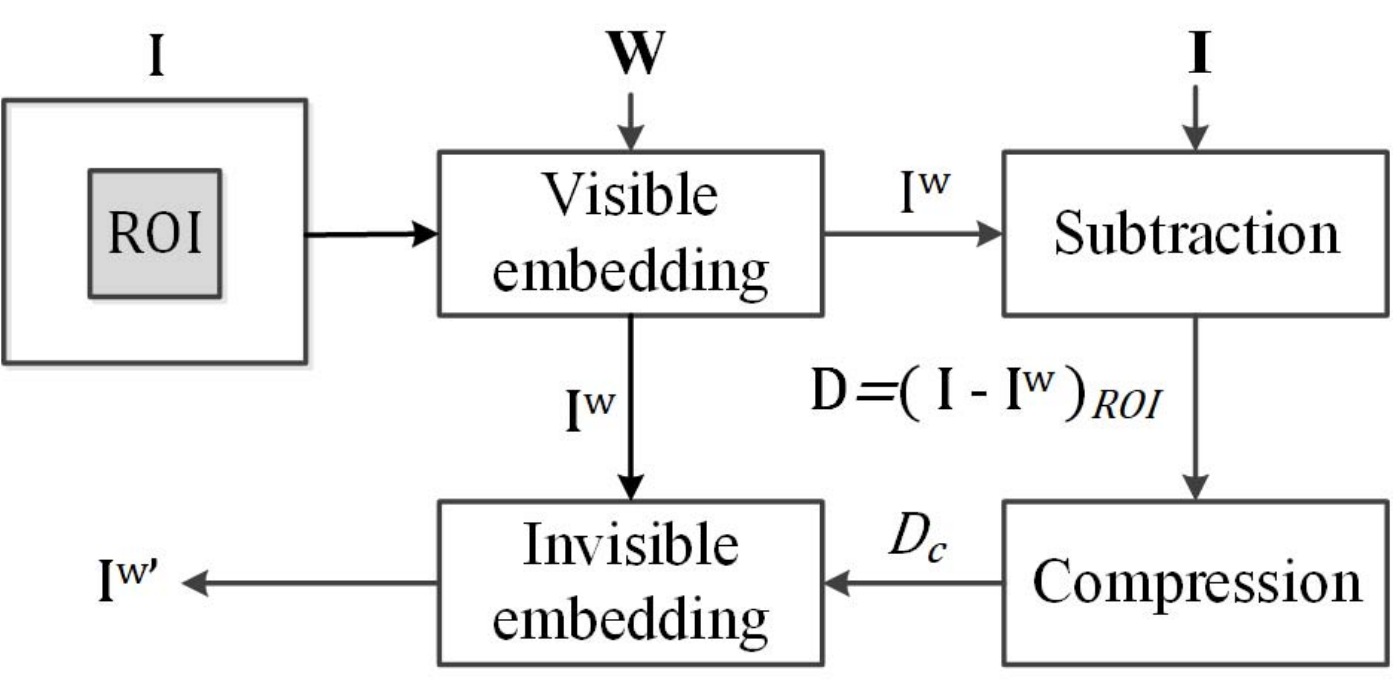}
  \caption{An ideal framework for reversible visible watermarking.}
  \label{fig2.1}
\end{figure}

We first provide an overview of the proposed framework for reversible visible watermarking. 
We start from the motivation, and then present the framework including data embedding and image restoration.

\begin{figure*}[t]
  \centering
  \includegraphics[width=0.8\textwidth]{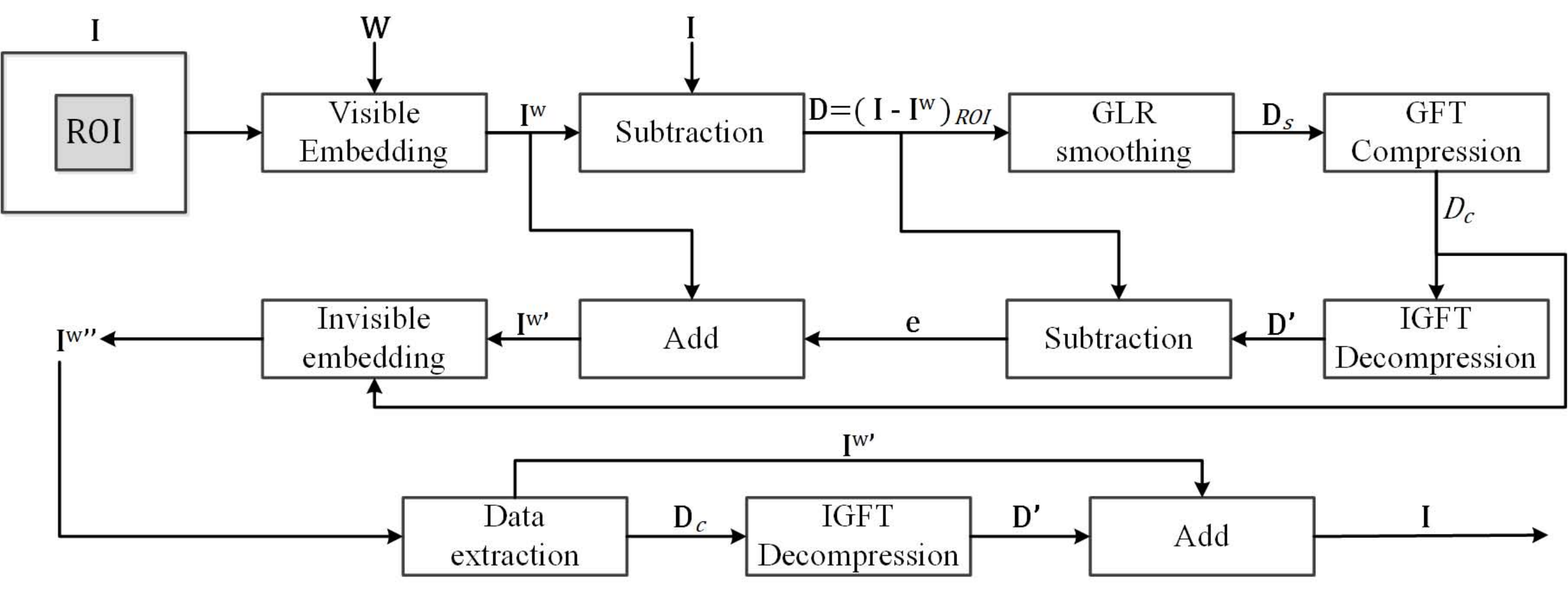}
  \caption{The proposed versatile framework for reversible visible watermarking based on Regularized Graph Fourier Transform coding.}
  \label{fig3.1}
\end{figure*}

\subsection{Motivation}
Given that the visible watermark image $\W$ and host image $\I$ are incorporated to obtain the watermarked image $\I^{w\prime}$ with the region of interest (ROI) embedded, the goal of reversible visible watermarking is to remove the watermark image $\W$ completely from $\I^w$ and restore the original host image $\I$ losslessly.
As demonstrated in Fig. \ref{fig2.1}, an ideal framework is to first compute the difference image $\D$ between $\I$ and $\I^w$ as follows
\begin{equation}
\D=(\I-\I^w)_{ROI}.
\label{eq2}
\end{equation}
Then, we can compress $\D$ to acquire a reconstruction data packet $D_c$. 
Next, the conventional reversible data hiding technology can be used to invisibly embed $D_c$ into a non-ROI region of $\I^w$ to obtain the watermarked image $\I^{w\prime}$. 

In the subsequent restoration process of the host image $\I$, we can first extract the compressed $D_c$ from $\I^{w\prime}$ to acquire the watermarked image $\I^w$. 
Then, $D_c$ is decompressed to reconstruct the difference image $\D$. 
Trough the inverse process of \eqref{eq2}, $\I$ can be losslessly restored by taking the sum of the reconstructed difference image $\D$ and watermarked image $\I^w$, removing the visible watermark image $\W$. 
This framework is ideal with versatility, as the generic RMR can be constructed by requiring the access to only $\I$ and $\I^w$, without the knowledge of specific visible watermarking schemes as in previous works such as \cite{b39}.

However, the data amount of $D_c$ is usually large if the compression is not efficient, which easily leads to excessive distortion in the watermarked image $\I^w$ by directly embedding $D_c$. In extreme cases, the data $D_c$ may not even be completely embedded. 
This motivates our proposed regularized Graph Fourier Transform coding for compact encoding of $\D$. 
The proposed complete framework is shown in Fig.~\ref{fig3.1}, which will be discussed in detail in the following.   

\subsection{Data Embedding}
In a nutshell, the data embedding process embeds the watermark image $\W$ into the ROI region of the original host image $\I$, and infers and embeds the reconstruction data packet $\D_c$ into the non-ROI region of $\I$. 
The key is to acquire a compact $\D_c$ from the proposed RGFT coding.      

Above all, we obtain the watermarked image $\I^w$ by embedding the image $\W$ into $\I$ in a semi-transparent way, which is referred to as the first visible embedding. 
Then, we compute the difference image $\D$ as in \eqref{eq2}.
Due to the large data amount of $\D$, efficient compression of $\D$ is required with a good rate-distortion performance. 
Assuming the difference image $\D$ is generalized piecewise smooth, we perform smoothing on $\D$ which is regularized by the GLR, so that the smoothed difference image $\D_s$ exhibits piecewise smoothness for efficient compression, which will be elaborated in Section~\ref{subsec:smoothing}. 
$\D_s$ is then encoded to acquire $\D_c$ via the proposed optimized GFT coding as will be described in Section~\ref{subsec:coding}. 

\begin{figure}[t]
\centering
\subfigure[]{
\includegraphics[width=0.2\textwidth]{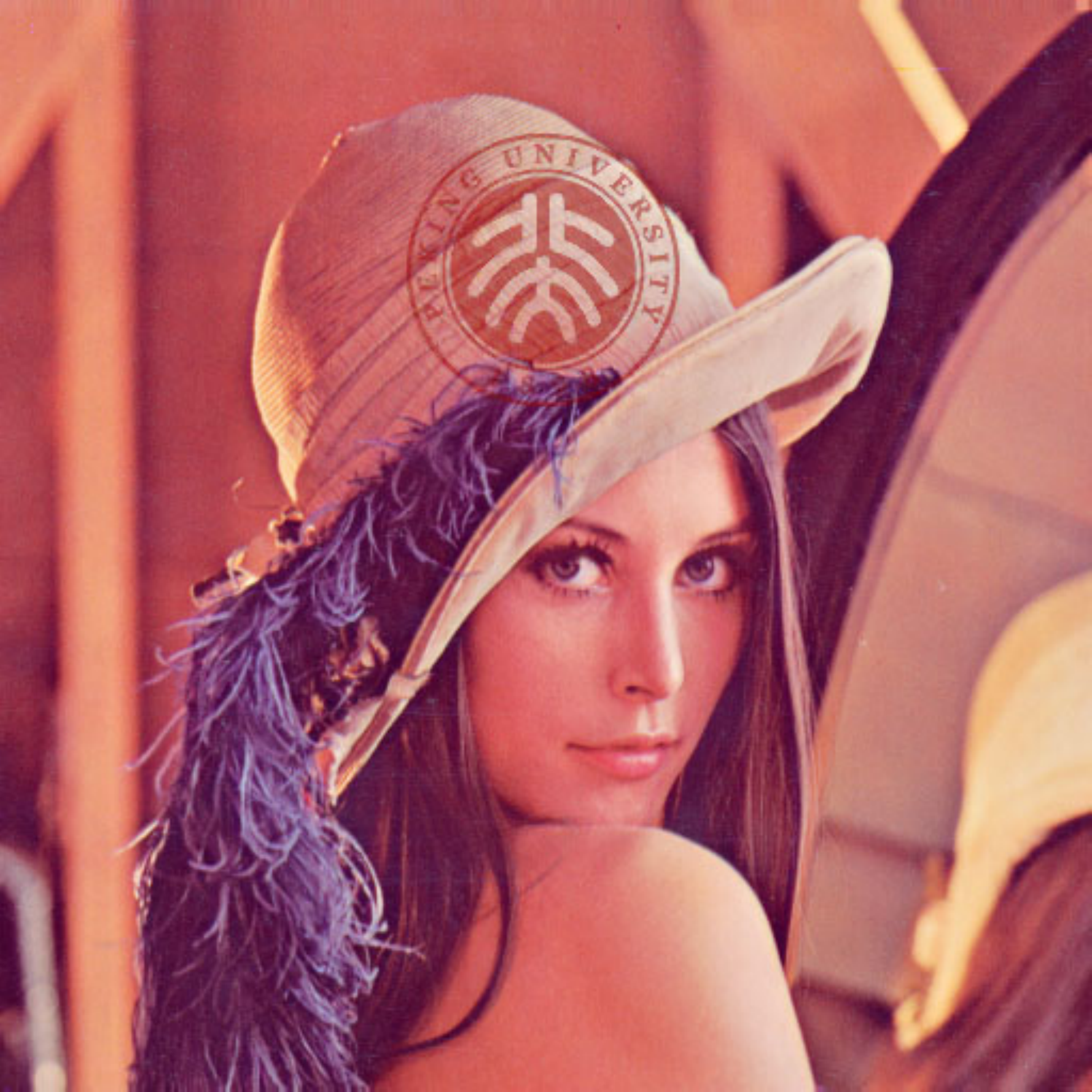}
\label{fig3.4a}
}
\subfigure[]{
\includegraphics[width=0.2\textwidth]{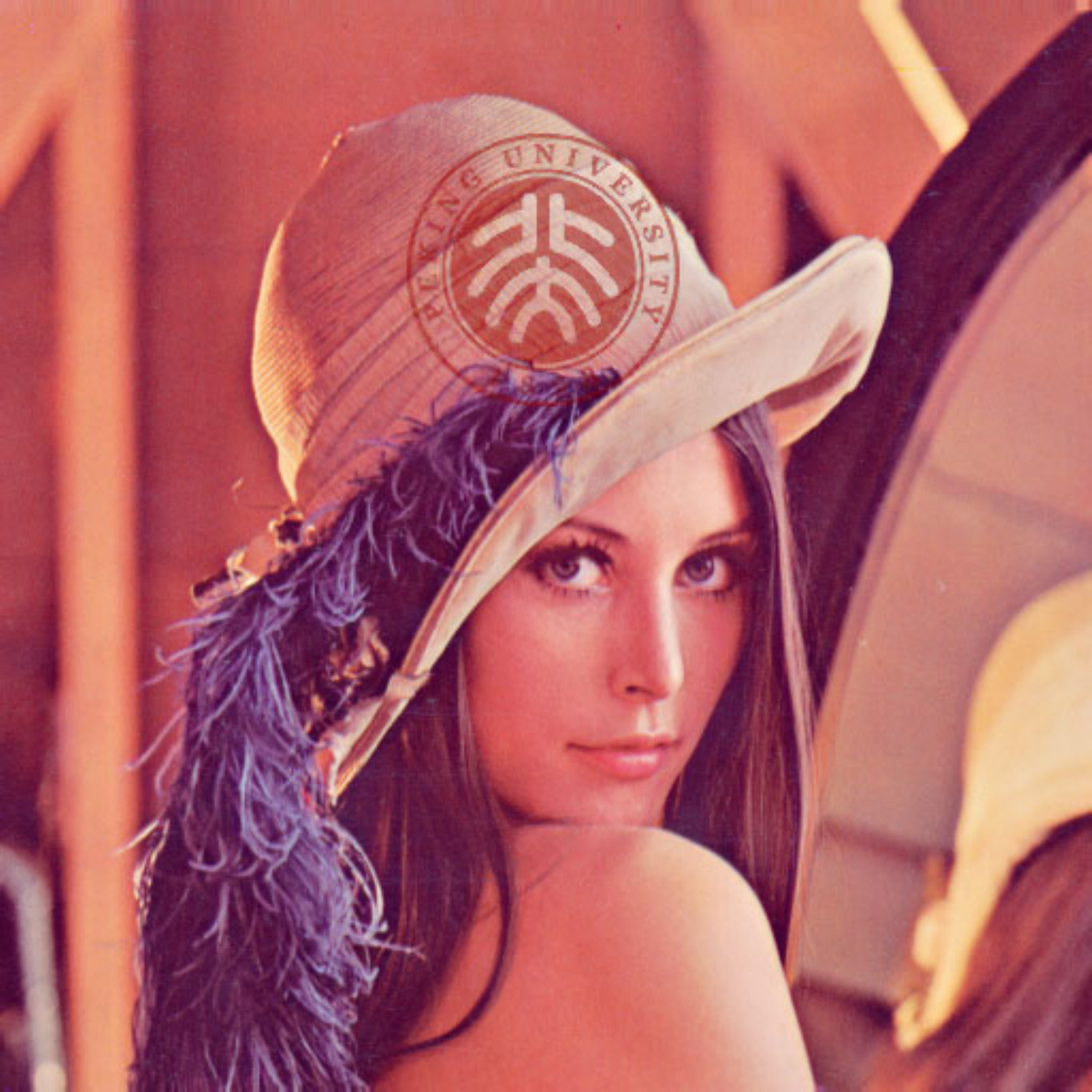}
\label{fig3.4b}
}
\caption{The modified watermarked images: (a) the watermarked image $\I^{w\prime}$ by adding $\e$ to $\I^{w}$; (b) the watermarked image $\I^{w\prime\prime}$ after the second invisible data embedding, which is visually satisfactory thanks to the efficient RGFT coding.}
\label{fig3.4}
\end{figure}

Next, $D_c$ is decoded to reconstruct the difference image  $\D^{\prime}$, which differs from $\D$ to some extent due to the lossy compression.
Hence, we calculate the error matrix $\e=\D-\D^{\prime}$, which is added to the visible-watermarked region of $\I^w$ to acquire $\I^{w\prime}$ as shown in Fig.~\ref{fig3.4}(a), \ie,
\begin{equation}
\I^{w\prime}=\I^{w}+\e=\I^{w}+\D-\D^{\prime}.
\label{eq9}
\end{equation}
Through the conventional RDH technology based on different techniques, such as histogram-modification (HM) \cite{b5}, most-significant-bit (MSB) prediction \cite{AE2}, pixel-value-ordering (PVO) \cite{NEW_PVO} and difference expansion (DE) \cite{NEW_DE,AE1}, $D_c$ is embedded into the non-visible-watermarked area of $\I^{w\prime}$, leading to the final watermarked image $\I^{w\prime\prime}$ as shown in Fig.~\ref{fig3.4}(b), which is referred to as the second invisible data embedding and visually acceptable thanks to the efficient RGFT coding. 

The final output watermarked image $\I^{w\prime\prime}$ is obtained by additionally modifying $\I^{w}$ twice.
However, thanks to the efficient RGFT coding method, the auxiliary information is relatively small and the slight distortion of the watermarked image $\I^{w}$ is acceptable.

\subsection{Image Restoration}
\label{subsec:restoration}

The process of image restoration removes the visible watermark and recover the host image. 
Firstly, we extract $D_c$ from the watermarked image $\I^{w\prime\prime}$ to reconstruct the watermarked image $\I^{w\prime}$. 
Then, we employ the Inverse GFT coding to decode $D_c$ for the restoration of the difference matrix $\D^{\prime}$. 
Combining \eqref{eq9} and \eqref{eq2}, the sum of the watermarked image $\I^{w\prime}$ and $\D^{\prime}$ leads to the original host image $\I$:
\begin{equation}
\I^{w \prime}+\D^{\prime}=\I^{w}+\D-\D^{\prime}+\D^{\prime}=\I^{w}+\D=\I,
\label{eq10}
 \end{equation}
which proves that the proposed framework is reversible.

\section{The Proposed Graph-based Data Coding}
\label{sec:graph}
In this section, we elaborate on the proposed graph-based data coding, including the GLR-based smoothing and regularized GFT-based coding. 
We start from the preliminaries of basic concepts in graph signal processing. Then we introduce the regularized graph Fourier transform coding method. 
Please note that, the proposed compression method is auxiliary to ensure the versatility of the framework. 

\subsection{Preliminaries}
We first provide some preliminaries of basic concepts in graph signal processing (GSP) \cite{Shuman13,Sandryhaila13,Ortega17}, including graph, graph Laplacian, Graph Laplacian Regularizer (GLR) and Graph Fourier Transform (GFT), which will be leveraged in the proposed GLR-based smoothing and regularized GFT-based coding. 

\subsubsection{Graph and Graph Laplacian}
\label{subsec:knn}
We denote an undirected graph $ \mathcal{G}=\{\mathcal{V},\mathcal{E},\A\} $, which is composed of a vertex set $ \mathcal{V} $ of cardinality $|\mathcal{V}|=N$, an edge set $ \mathcal{E} $ connecting vertices, and a weighted \textit{adjacency matrix} $ \mathbf{A} $. 
$ \A \in \mathbb{R}^{N \times N} $ is a real symmetric matrix, where $ a_{i,j} $ is the weight assigned to the edge $ (i,j) $ connecting vertices $ i $ and $ j $. In this paper, we assume non-negative weights to describe the similarity between each pair of pixels, \ie, $ a_{i,j} \geq 0 $. 

Among variation operators in GSP, the commonly used \textit{combinatorial graph Laplacian} \cite{Chung96,b43,b40} is defined as $ \cL:=\mathcal{D}-\A $, where $ \mathcal{D} $ is the \textit{degree matrix}---a diagonal matrix where $ d_{i,i}=\sum_{j=1}^N a_{i,j} $. 
Given real and non-negative edge weights in an undirected graph, $\cL$ is real, symmetric and positive semi-definite.

\subsubsection{Graph Laplacian Regularizer (GLR)}
\label{subsubsec:prior}
A graph signal refers to a scalar or vector residing on each vertex of a graph. For example, we represent each pixel in an image as a vertex and treat the color of each pixel as a graph signal. 

Given a graph signal $ \x $ residing on the vertices of $ \mathcal{G} $ encoded in the graph Laplacian $\cL$, the GLR is expressed as
\begin{equation}
	\x^{\top} \cL \x =\sum_{i \sim j} a_{i,j} \cdot (x_i - x_j)^2,
	\label{eq:prior}
\end{equation}
where $i\sim j$ means vertices $i$ and $j$ are connected. 
The signal $\x$ is smooth with respect to $\mathcal{G}$ if the GLR is small, as connected vertices $ x_i $ and $ x_j $ must be similar for a large edge weight $ a_{i,j} $; for a small $ a_{i,j} $, $x_i$ and $x_j$ can differ significantly. 
Hence, a small GLR enforces $\x$ to adapt to the topology of $ \mathcal{G} $.
This prior also possesses an interpretation in the frequency domain as low-pass filtering, as well as a continuous interpretation as a smoothness functional defined on the underlying Riemannian manifold \cite{Hu20GSP}.
This prior will be employed in our problem formulation of GLR-based smoothing in Section~\ref{subsec:smoothing}.

\subsubsection{Graph Fourier Transform (GFT)}
\label{subsubsec:GFT}

Because $\cL$ is a real symmetric matrix, it admits an eigen-decomposition $\cL = \U \bfLambda \U^{\top}$, where $\U=[\u_1,...,\u_N]$ is an orthonormal matrix containing the eigenvectors $\u_i$, and $\bfLambda = \mathrm{diag}(\lambda_1,...,\lambda_N)$ consists of eigenvalues $\{\lambda_1=0 \leq \lambda_2 \leq ... \leq \lambda_N\}$. 
We refer to the eigenvalue $\lambda_i$ as the {\it graph frequency}, with a smaller eigenvalue corresponding to a lower graph frequency.

For any graph signal $\x \in \mathbb{R}^{N}$ residing on the vertices of $\mathcal G$, its GFT $\hat{\x}  \in \mathbb{R}^{N}$ is defined as \cite{hammond2011wavelets} 
\begin{equation}
    \hat{\x} = \U^{\top} \x. 
\end{equation}
The inverse GFT follows as 
\begin{equation}
    \x = \U \hat{\x}. 
    \label{eq:IGFT}
\end{equation}
With an appropriately constructed graph that captures the signal structure well, the GFT will lead to a compact representation of the graph signal and even optimal decorrelation in the spectral domain \cite{b40}, which is beneficial for image compression and will be deployed in the coding framework in Section~\ref{subsec:coding}.

\subsection{GLR-based Smoothing}
\label{subsec:smoothing}

As the difference image $\D$ is generalized piecewise smooth as shown in Fig.~\ref{fig:data_packet}, instead of compressing $\D$ directly, we propose to perform smoothing on $\D$ prior to coding so that it exhibits piecewise smoothness with fewer high-frequency details for more efficient coding. 

In particular, to achieve piecewise smoothing of the difference image $\D$, we first divide $\D$ into $l \times l$ blocks ($l=8$ in our experiments), which overlap with each other by a step size $t$ ($t=2$ in our setting) so as to keep the smoothed results consistent and avoid blocking artifacts. 
We then construct a fully-connected graph within each block, where the edge weight between nodes $i$ and $j$ is defined as a Gaussian kernel:
\begin{equation}
    a_{i,j} = \exp\{-\frac{|d_i - d_j|^2}{\sigma^2}\},
\end{equation}
where $d_i$ and $d_j$ are the intensity value of node $i$ and $j$, respectively, and $\sigma$ is the standard deviation that is dependent on the intensity difference in the neighborhood to compute proper edge weights. Based on the edge weights, we compute the weighted adjacency matrix $\A$ and then acquire the graph Laplacian $\cL$ according to the definition. 

We then employ the GLR to regularize the smoothing process so as to avoid filtering across image contours. 
In particular, we formulate piecewise smoothing as an optimization problem, where the objective aims to 
1) reconstruct an image $\D_s$ that is similar to $\D$; 
2) perform smoothing in slowly-varying textured regions while keeping sharp image contours. 
The formulation is mathematically written as
\begin{equation}
\label{eq:smoothing}
    \min\limits_{\D_s}\ \left \| \D-\D_s \right \|_2^2 + \mu\cdot\D_s^{\top} \cL \D_s,
\end{equation}
where the first term is the data fidelity term to satisfy the former requirement while the second term is the GLR to meet the latter requirement. $\mu > 0$ is a parameter to strike a balance between the data fidelity term and the GLR. 

Setting the derivative of \eqref{eq:smoothing} to zero yields the closed-form solution: 
\begin{equation}
    \D_s = (\I_N + \mu \cL)^{-1} \D,
    \label{eq:closed_form_GLR}
\end{equation}
where $\I_N \in \mathbb{R}^{N \times N}$ is an identity matrix.   
This is a set of linear equations and can be solved efficiently.

\subsection{Regularized GFT-based Coding}
\label{subsec:coding}
Having acquired the smoothed difference image $\D_s$, we compress $\D_s$ leveraging on the multi-resolution GFT coding method in \cite{b40}. 
Considering the smoothing process, the optimization objective in the GFT coding {\it differs} from \cite{b40} in that the distortion is computed between the decoded difference image $\D'$ and the original difference image $\D$, instead of the classical distortion between the decoded image $\D'$ and the input image $\D_s$ to the encoder. 
Hence, our optimization objective is to minimize the rate-distortion cost, which consists of 1) the overall distortion between the decoded difference image $\D'$ and the original difference image $\D$; and 2) the rate of encoding the smoothed difference image $\D_s$, including the rate of the GFT coefficients as well as the overhead to transmit the GFTs for proper inverse transform at the decoder as they are adaptive to different image blocks. The objective is mathematically written as 
\begin{equation}
    \min\limits_{\mathbf{\A^*},\mu}\ \|\D'(\A^*,\mu) - \D\|_2^2 + \lambda \cdot R(\D_s(\mu), \A^*), 
\end{equation}
where the first term is the Sum of Squared differences (SSD) between the decoded $\D'$ and the original difference image $\D$ as the measure of compression distortion, and the second term is the rate of encoding the smoothed difference image $\D_s$. 
Specifically, $R(\D_s(\mu), \A^*)$ includes the rate of encoding the GFT coefficients dependent on $\D_s(\mu)$ and $\A^*$ as well as the rate of encoding the GFT description $\A^*$, where $\D_s(\mu)$ is a function of $\mu$ as indicated in \eqref{eq:closed_form_GLR}. 
$\lambda$ is a parameter for the rate-distortion trade-off, which is often calculated from the quantization parameter $QP$ as $\lambda = 0.85\times2^{(\frac{QP-6}{3})}$.
$\A^*$ is the weighted adjacency matrix {\it learned} during the compression as in \cite{b40}. 
Further, we consider $\mu$ as an optimization variable that controls the smoothing of the original difference image $\D$ as indicated in \eqref{eq:smoothing}, which affects the smoothed image $\D_s$ and thus the decoded image $\D'$. 

Directly solving $\A^*$ and $\mu$ would be nontrivial.
We propose to optimize the two variables alternately. 
When $\A^*$ is initialized, the objective is non-convex and difficult to optimize. We thus propose a greedy algorithm, where we search a best $\mu$ in a discrete set $ \{10^{0},10^{-1},10^{-2},10^{-3},10^{-4}\} $ empirically that leads to the smallest objective. 
We then fix $\mu$ and update $\A^*$ using the algorithm in \cite{b40}. 
This process is iterated until the convergence of the optimization objective. 

During the encoding of the smoothed difference image $\D_s$, we employ the multi-resolution GFT coding in \cite{b40}.  
The key idea is to downsample each pixel block in the input image to a low-resolution one, so that the piecewise smoothness of the image can be exploited and a low-resolution GFT can be employed for computation efficiency. 
Then at the decoder, we perform upsampling and interpolation adaptively along high-resolution image contours that are losslessly coded, so that sharp object contours can be well preserved. 
Specifically, we describe the multi-resolution GFT coding process as follows.

\subsubsection{Encoder}
We first detect prominent contours in the input smoothed image $\D_s$ adaptively, and encode them losslessly for the adaptive intra prediction and interpolation at the decoder. 
Then, we execute three steps for each pixel block: 
I) contour-aware intra prediction, which efficiently reduces the energy of the prediction error by predicting within the confine of detected contours; 
II) transform coding of the residual block, where we choose between fixed Discrete Cosine Transform (DCT) on the original high-resolution residual block and a pre-computed set of low-resolution GFTs on the down-sampled low-resolution residual block; 
III) quantization and encoding of the resulting transform coefficients for transmission to the decoder, along with the transform index that identifies the chosen transform for proper inverse transform at the decoder. 

\subsubsection{Decoder}
We first perform inverse quantization and inverse transform to reconstruct each residual block, where the transform index is used to identify the transform chosen at the encoder for transform coding.
If the low-resolution GFT is employed, we upsample the reconstructed low-resolution residual block to the original resolution, and interpolate missing pixels adaptively according to the decoded high-resolution image contours. 
Each pixel block is finally reconstructed by adding the intra predictor to the corresponding residual block.

\begin{figure}[t]
\centering
\subfigure[]{
\includegraphics[width=0.12\textwidth]{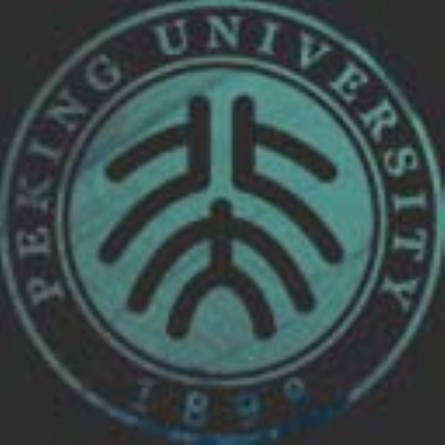}
\label{fig3.2d}
}
\subfigure[]{
\includegraphics[width=0.12\textwidth]{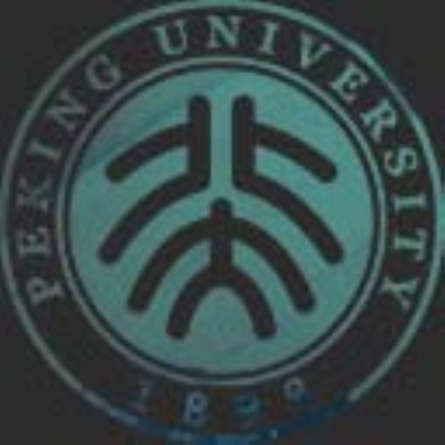}
\label{fig3.2e}
}
\subfigure[]{
\includegraphics[width=0.12\textwidth]{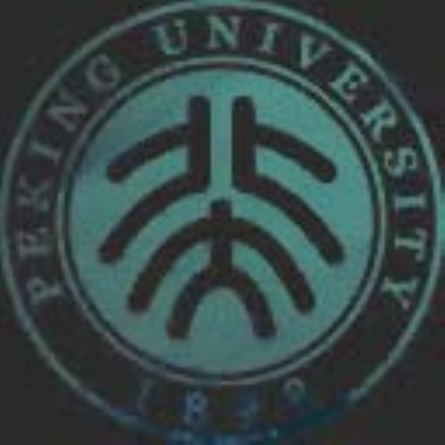}
\label{fig3.2f}
}
\caption {Demonstration of GLR-based smoothing and regularized GFT-based coding for the difference image. (a) the difference image $\D$ as in Fig.~\ref{fig:data_packet}; (b) the GLR-smoothed difference image $\D_s$ ($\mu=0.001$); (c) the decoded difference image $\D^{\prime}$ (the quantization parameter $QP=24$). }
\label{fig3.3}
\end{figure}

Finally, we demonstrate an example in Fig.~\ref{fig3.3}.  
Given an original difference image $\D$ shown in Fig.~\ref{fig3.3}(a), the GLR-smoothed difference image $\D_s$ is presented in Fig.~\ref{fig3.3}(b), which is obviously more piecewise smooth. 
Then we encode and decode $\D_s$ to obtain $\D^{\prime}$, as shown in Fig.~\ref{fig3.3}(c), which preserves sharp contours and prominent details well. 

Along with the image restoration procedure described in Section~\ref{subsec:restoration}, we provide an algorithmic summary in Algorithm 1. 
Please note that we assume the visible watermarking has been conducted to provide one input $\I^w$, and thus the visible watermarking process is not included in Algorithm 1.

\begin{algorithm}[t]
    \caption{The Proposed reversible visible watermarking via regularized GFT coding}
    \label{alg:framework}
    \SetKw{KwEmbedding}{Procedure 1: Data embedding}
    
    \KwIn{The original host image $\I$, the watermarked image $\I^w$}
    
    \KwOut{The final watermarked image $\I^{w\prime\prime}$, the restored original host image $\I$}
    \KwEmbedding\\
    \Begin{

   \lnlset{InResR1}{1}%
    Calculate the difference image $\D$ between the watermarked image $\I^w$ and original host image $\I$ as $\D=(\I-\I^w)_{ROI}$;\\
    
    \lnlset{InResR2}{2}%
    Perform GLR-based smoothing on $\D$ as described in Section \ref{subsec:smoothing}, leading to a smoothed difference image $\D_s$;\\
    
    \lnlset{InResR3}{3}%
    Encode $\D_s$ with the regularized GFT-based coding method in Section \ref{subsec:coding}, leading to a compressed version $D_c$;\\
    
    \lnlset{InResR4}{4}%
    Decode $D_c$ to acquire the reconstructed difference image $\D'$;\\
    
    \lnlset{InResR5}{5}%
    Compute the error matrix $\e=\D-\D'$;\\
    
    \lnlset{InResR6}{6}%
    Add the error matrix $\e$ to $\I^w$ to obtain $\I^{w\prime}=\e+\I^w$;\\
    
    \lnlset{InResR7}{7}%
    Embed data including the reconstruction data packet $D_c$, coordinates of ROI, and auxiliary information of RDH into a non-visible-watermarked region of $\I^{w\prime}$ using the method in \cite{b5}, leading to the final watermarked image $\I^{w\prime\prime}$. \\
    }

    \SetKw{KwRestoration}{Procedure 2: Image restoration}
    \KwRestoration\\
 \Begin{   
    \lnlset{InResR8}{8}%
    Extract $D_c$ from the watermarked image $\I^{w\prime\prime}$ to recover $\I^{w\prime}$;
    
    \lnlset{InResR9}{9}%
    Decode $D_c$ to obtain the difference image $\D'$;\\
    
    \lnlset{InResR0}{10}%
    Restore the original host image $\I$ losslessly by taking the sum of the  watermarked image $\I^{w\prime}$ and the difference image $\D'$ as in \eqref{eq10}.\\
    }
\end{algorithm}


\section{Experimental Results}
\label{sec:results}
We evaluate the versatility and performance of the proposed framework from the following three aspects:

1) \textbf{Compression performance.} We compare and analyze the compression performance of the proposed regularized GFT coding with that of competitive methods including the conventional JPEG 2000 \cite{jpeg_2000} and the MR-GFT coding in \cite{b40}.

2) \textbf{Versatility verification.} In order to verify the compatibility of the proposed framework for various visible watermarking algorithms, we choose different embedding methods for test, including the RVW algorithms in the spatial domain \cite{b25,b26,b28}, in the frequency domain \cite{b32}, and the conventional irreversible alpha fusion method \cite{alpha_fusion}.

3) \textbf{Performance evaluation of the general framework.} We compare the data compression performance and the quality of watermarked images with competitive RVW methods \cite{b27,b39}. 

\subsection{Experimental Setup}
\begin{figure*}[htbp]
\centering
\subfigure[]{
\includegraphics[width=0.1\textwidth]{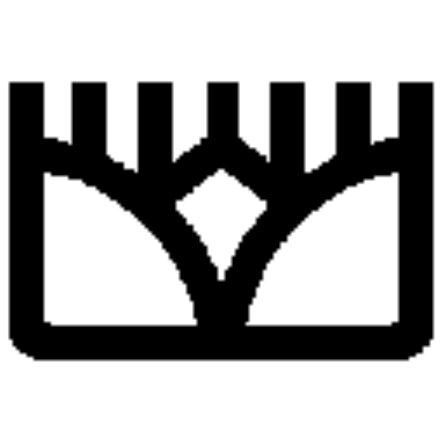}
\label{fig4.1a}
}
\subfigure[]{
\includegraphics[width=0.1\textwidth]{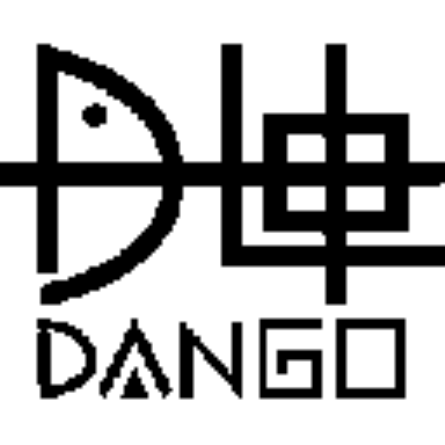}
\label{fig4.1b}
}
\subfigure[]{
\includegraphics[width=0.1\textwidth]{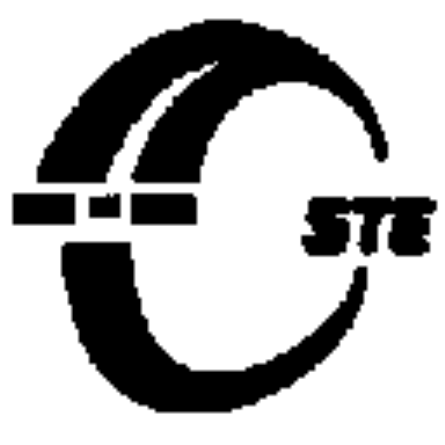}
\label{fig4.1c}
}
\subfigure[]{
\includegraphics[width=0.1\textwidth]{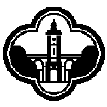}
\label{fig4.1d}
}
\subfigure[]{
\includegraphics[width=0.1\textwidth]{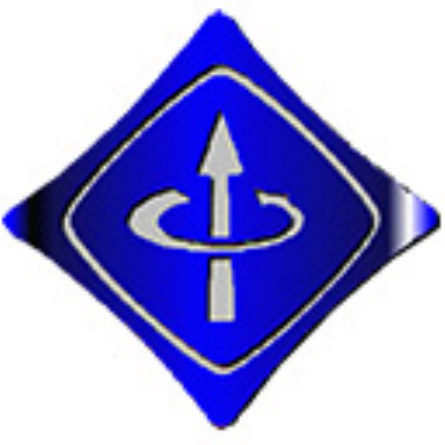}
\label{fig4.1e}
}
\subfigure[]{
\includegraphics[width=0.1\textwidth]{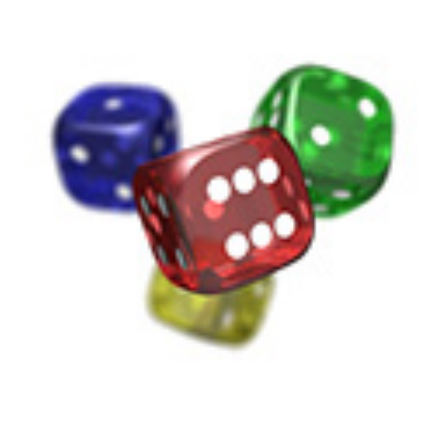}
\label{fig4.1f}
}
\subfigure[]{
\includegraphics[width=0.1\textwidth]{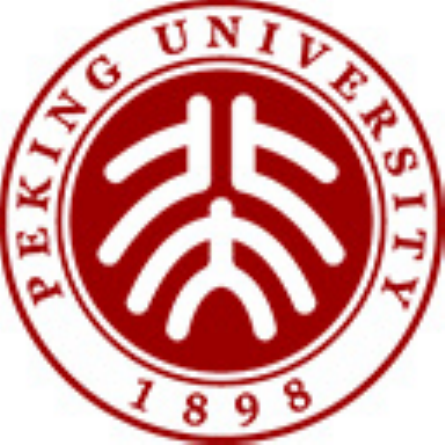}
\label{fig4.1g}
}
\subfigure[]{
\includegraphics[width=0.1\textwidth]{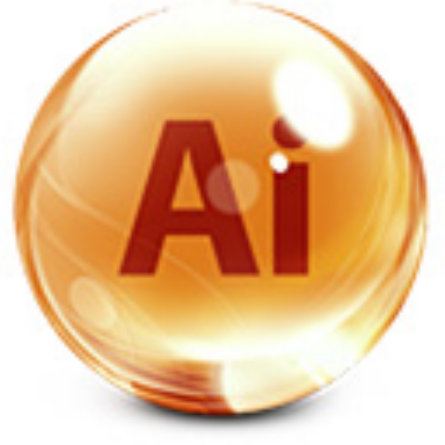}
\label{fig4.1h}
}
\caption{Eight visible watermark images with dimension of $128\times128$ pixels.}
\label{fig4.1}
\end{figure*}

We select eight images with dimension of $128\times128$ as the visible watermark images, including four binary images and four RGB images, as shown in Fig. \ref{fig4.1}. Accordingly, eight gray-scale images and eight RGB images are chosen as the original host images with dimension of $512\times512$, including \textit{F-16}, \textit{Boat} \cite{SIPI}, \textit{Barbara}, \textit{Cameraman}, \textit{Goldhill} \cite{ece533}, \textit{Pepper}, \textit{Sailboat}, \textit{Lena} \cite{Gonzalez}, \textit{Yacht}, \textit{Fruits}, \textit{Pens}, \textit{Flower}, \textit{Couple}, \textit{Cable car}, \textit{Cornfield}, \textit{Girl} \cite{docjava}.
We employ the peak signal-to-noise ratio (PSNR) to measure the image quality. The number of bits to represent an image before compression is denoted as $N_D$ (unit: bit), which is computed as:
\begin{equation}
N_{D}=L \times H \times K_{\text{uint8}},
\label{eq21}
\end{equation}
where $L, H$ are the width and height of the image respectively, and $K_{\text{uint8}}=8$ is the capacity of an 8-bit unsigned integer variable. The number of bits to encode the compressed image is denoted as $N_C$ (unit: bit).
Then the compression ratio is defined as
\begin{equation}
Ratio=\frac{N_{C}}{N_{D}}.
\label{eq22}
\end{equation}

As the proposed framework is versatile to any visible watermark embedding algorithm, how to embed the visible watermark image into the host image (ROI) is not our focus. In the experiments, we select several representative visible watermarking algorithms to embed the visible watermark into the host image, including \cite{b25,b26,b27,b28,b32,b39,alpha_fusion}.

Since the compression efficiency in the proposed method is quite high as shown below, the amount of the compressed data of the difference image $\D$ is small.
Hence, most conventional reversible data hiding methods can be used for the embedding of $\D$ into the non-ROI region.
Specifically, we adopt histogram-modification based RDH\cite{b5} as an example in our experiments for evaluation.

\subsection{Compression Performance}
\label{s4.1}
\begin{figure*}[]
\centering
\subfigure[]{
  \includegraphics[width=0.45\textwidth]{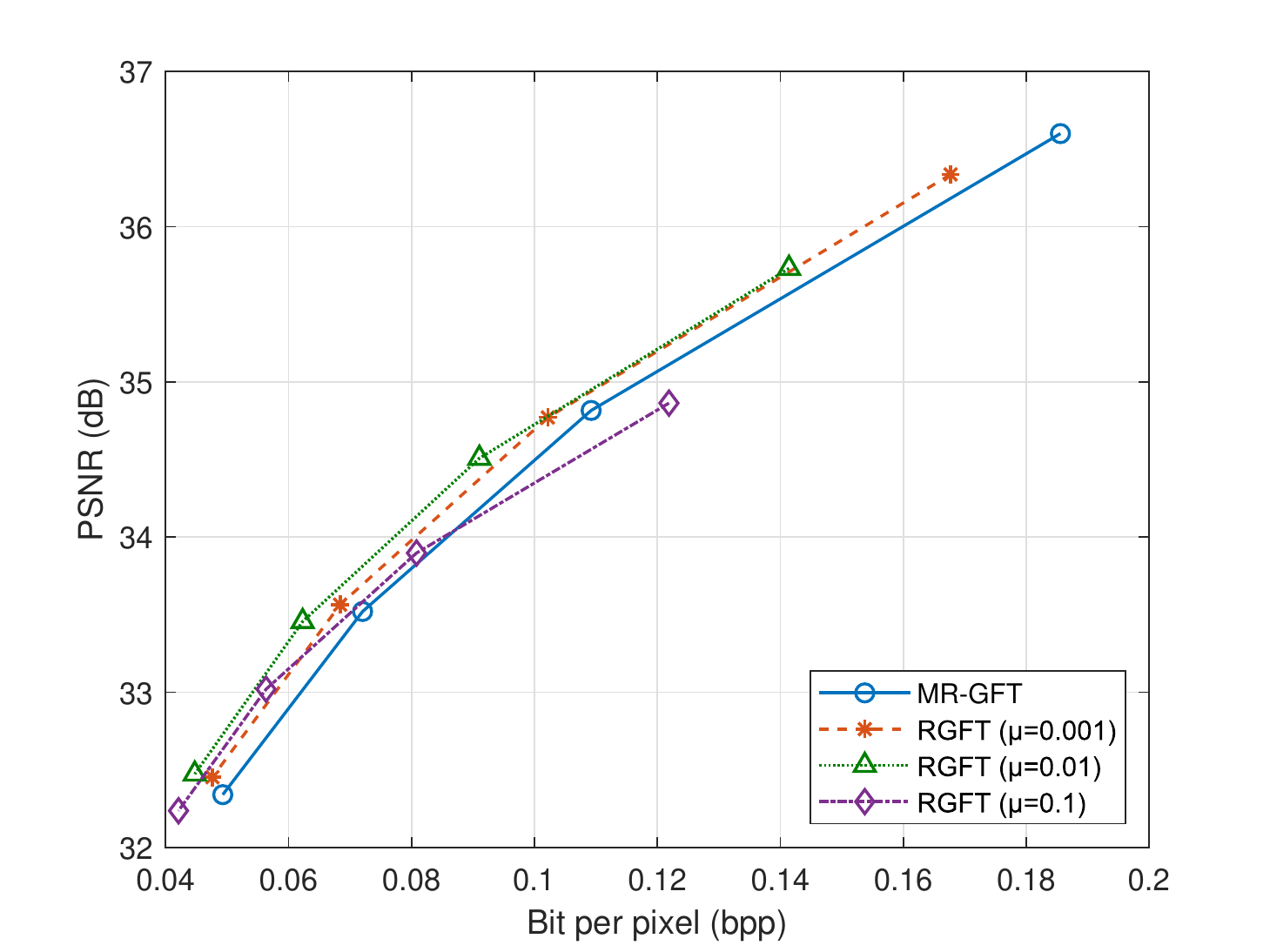}
  \label{fig:rd_mrgft_PSNR}
}
\subfigure[]{
  \includegraphics[width=0.45\textwidth]{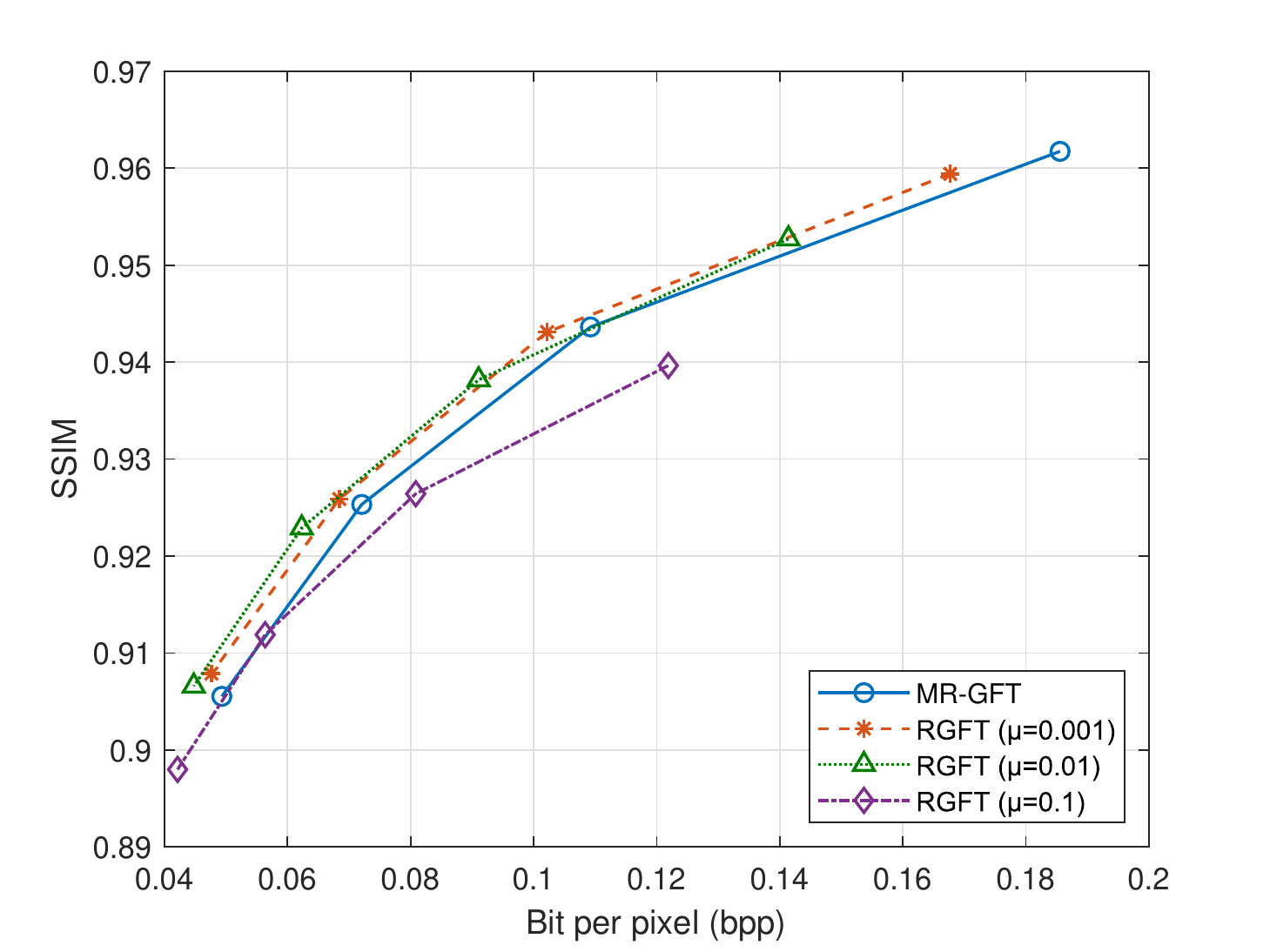}
  \label{fig:rd_mrgft_SSIM}
}
 \caption{The compression performance comparison of our proposed RGFT method with different settings of $\mu$ and the MR-GFT \cite{b40} using different evaluation criteria. (a): PSNR; (b): SSIM. }
   \label{fig:rd_mrgft}
\end{figure*}

\begin{figure*}[]
\centering
\subfigure[]{
  \includegraphics[width=0.45\textwidth]{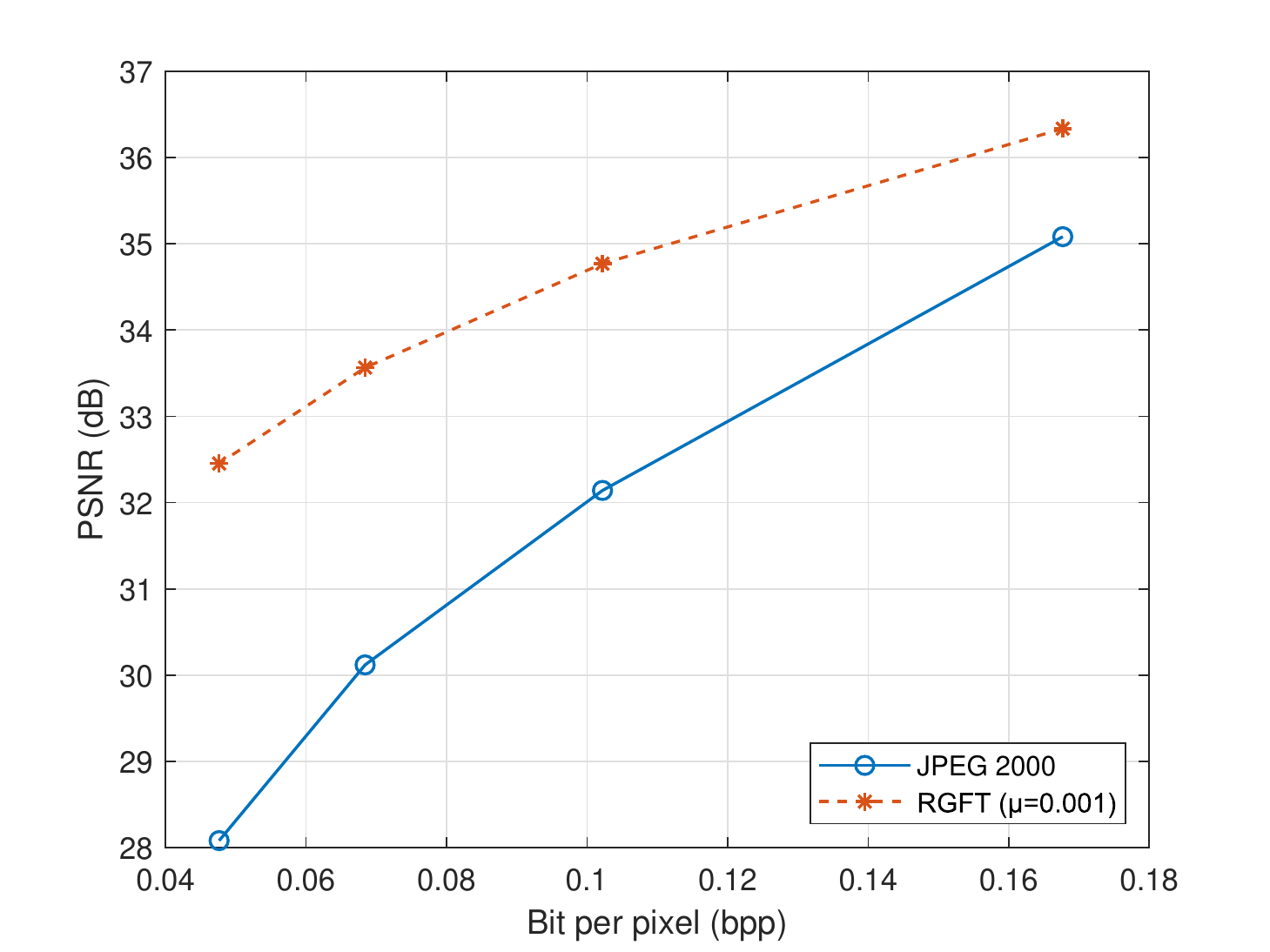}
  \label{fig:rd_jpeg_PSNR}
}
\subfigure[]{
  \includegraphics[width=0.45\textwidth]{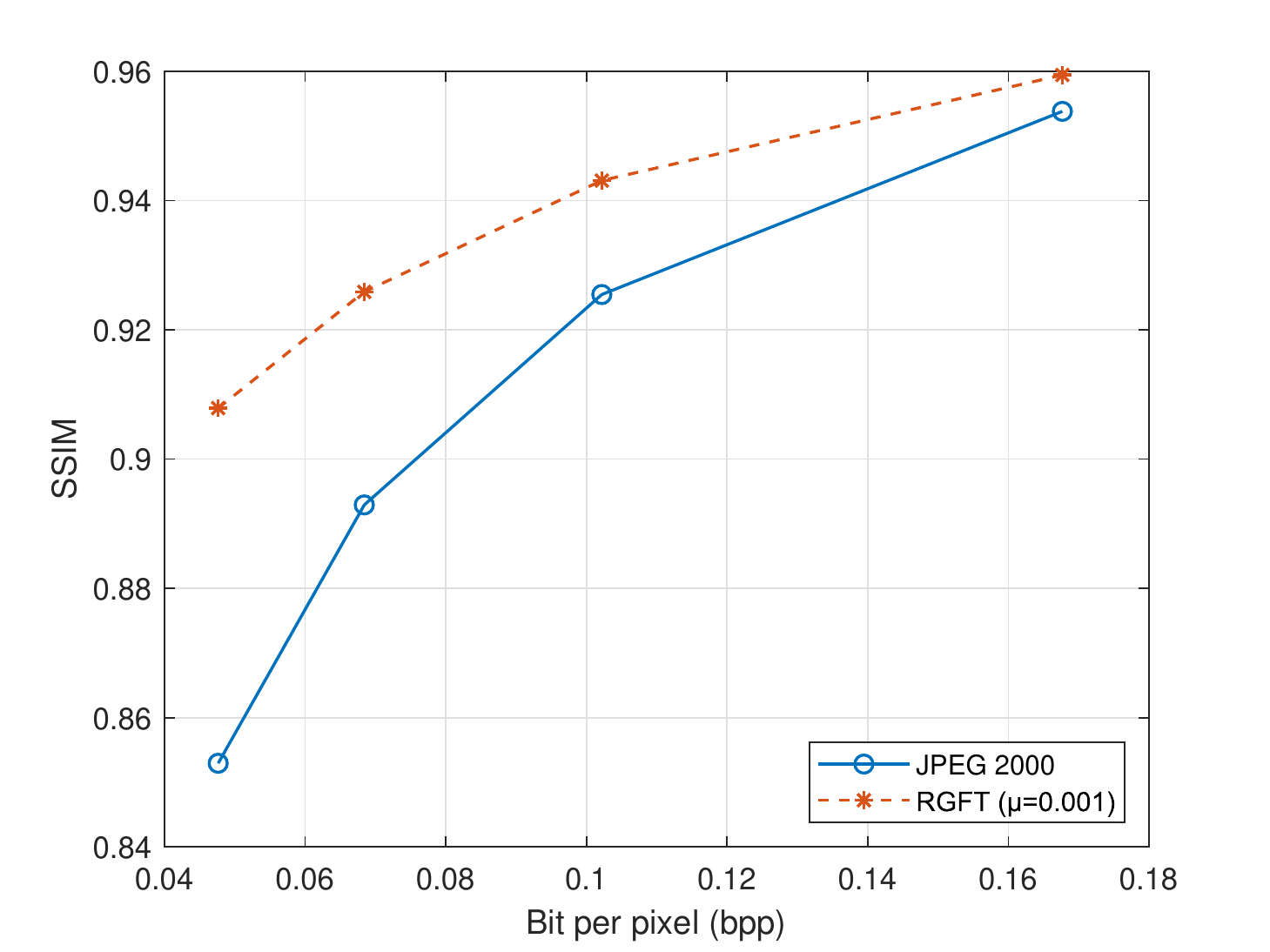}
  \label{fig:rd_jpeg_SSIM}
}
 \caption{The compression performance comparison of our proposed RGFT method and the JPEG 2000 compression method \cite{jpeg_2000} using different evaluation criteria. (a): PSNR; (b): SSIM. }
   \label{fig:rd_jpeg}
\end{figure*}

We first embed the watermark images in Fig. \ref{fig4.1a}-\ref{fig4.1d} into the eight gray-scale host images \textit{F-16}, \textit{Boat}, \textit{Barbara}, \textit{Cameraman}, \textit{Goldhill}, \textit{Pepper}, \textit{Sailboat} and \textit{Lena} using the method in \cite{b26} respectively, leading to 32 gray-scale watermarked images.
Then we embed the watermark images in Fig. \ref{fig4.1e}-\ref{fig4.1h} into the eight RGB host images \textit{Pens}, \textit{Fruits}, \textit{Flower}, \textit{Couple}, \textit{Cornfield}, \textit{Cable car}, \textit{Girl} and \textit{Yacht} via \cite{alpha_fusion}, leading to 32 RGB watermarked images.
Next, we evaluate the compression performance of the difference images with different GLR smoothing parameters $\mu$ and quantization parameters $QP$.
To present the rate-distortion curves, we take the {\it average} of encoded bits for all the difference images, and obtain the average of the PSNR values between all the reconstructed difference images and the original ones.
We compare the proposed RGFT coding approach against the MR-GFT coding method in \cite{b40} and JPEG 2000, where the input $\D$ is presented as a signed integer matrix in all these methods. The results are shown in Fig.~\ref{fig:rd_mrgft} and Fig.~\ref{fig:rd_jpeg}, respectively.
Fig. \ref{fig:rd_mrgft} shows the compression performance of the proposed RGFT with different GLR smoothing coefficients $\mu \in \{0.001,0.01,0.1\}$ and quantization parameters $QP \in \{28,32,36,40\}$, compared with the MR-GFT method measured by both PSNR and SSIM.
Different settings of $\mu$ result in varied rate-distortion performance. When $\mu = 0.01$ or $\mu = 0.001$, the proposed RGFT outperforms the MR-GFT at all the bit rates in both PNSR and SSIM, validating that the RGFT is effective for encoding difference images and preserves the shape well.
When $\mu = 0.1$, while the RGFT outperforms the MR-GFT at low bit rates, the performance becomes worse at high bit rates.
This is reasonable as $\mu = 0.1$ indicates comparatively strong smoothing, which results in more distortions in the difference image and thus leads to worse overall rate-distortion performance at high bit rates.
In general, the compression performance reaches the best when $\mu = 0.01$ or $\mu = 0.001$.

Fig. \ref{fig:rd_jpeg} compares the compression performance of the proposed RGFT and JPEG 2000 measured by both PSNR and SSIM, where $\mu = 0.001, QP \in \{28,32,36,40\}$.
We see that, the RGFT outperforms JPEG 2000 by a large margin at all the bit rates, which validates the effectiveness of the RGFT for encoding difference images.

Based on the comparison, in the subsequent experiments, we set $\mu = 0.001$ and $QP=28$ for the RGFT.

\begin{table*}[t]
  \caption{The compressed file size of $\D$ and $\D^\prime+\e$.}
  \label{newtab1.1}
  \centering
  \resizebox{!}{!}{
  \begin{tabular}{c|c|c|c|c}
  \hline
  \hline
  \textbf{Host Image}&\textbf{Channel}&$\bm{bits(\D)}$&$\bm{bits(\D^\prime)+bits(\bm{e})}$&$\frac{bits(\D^\prime)+bits(\bm{e})}{bits(\D)}$\\
  \hline
 \multirow{3}{*}{Yacht}
  &R&4057&4614&1.137\\
 \cline{2-5}
 &G&3973&4718&1.188\\
 \cline{2-5}
 &B&4023&5437&1.351\\
 \hline

\multirow{3}{*}{Fruits}
 &R&3022&3497&1.157\\
 \cline{2-5}
 &G&3020&3886&1.287\\
 \cline{2-5}
 &B&3048&3914&1.284\\
 \hline
 \multirow{3}{*}{Pens}
 &R&3306&4699&1.421\\
 \cline{2-5}
 &G&2610&3346&1.282\\
 \cline{2-5}
 &B&3071&4034&1.314\\

\hline
 \multirow{3}{*}{Flower}
 &R&1661&2077&1.250\\
 \cline{2-5}
 &G&3347&3600&1.076\\
 \cline{2-5}
 &B&3517&4312&1.226\\

\hline
 \multirow{3}{*}{Couple}
  &R&4001&4721&1.180\\
 \cline{2-5}
 &G&3987&4794&1.202\\
 \cline{2-5}
 &B&3711&3929&1.059\\
\hline
 \multirow{3}{*}{Cable car}
   &R&3759&3853&1.025\\
 \cline{2-5}
 &G&3188&3502&1.098\\
 \cline{2-5}
 &B&3482&3896&1.119\\
\hline
  \multirow{3}{*}{Cornfield}
  &R&2001&2743&1.371\\
 \cline{2-5}
 &G&4557&6326&1.388\\
 \cline{2-5}
 &B&7194&8780&1.220\\
\hline
 \multirow{3}{*}{Girl}
  &R&1695&2145&1.265\\
 \cline{2-5}
 &G&3206&3577&1.116\\
 \cline{2-5}
 &B&3582&4141&1.156\\
 \hline
 AVERAGE&&&&1.216\\

    \hline
    \hline
\end{tabular}
}
\end{table*}

Besides, under the same parameter settings, we conducted experiments to compare the compressed file size of $\D^\prime$ and $\e$ with that of $\D$ to validate the advantage of the proposed compression algorithm. We embed visible watermark images in Fig. \ref{fig4.1e}-\ref{fig4.1h} into the host images \textit{Yacht}, \textit{Fruits}, \textit{Pens}, \textit{Flower}, \textit{Couple}, \textit{Cable car}, \textit{Cornfield} and \textit{Girl} respectively via \cite{alpha_fusion} to obtain watermarked images $\I^w$.
Then the acquired $\D$, $\D^{\prime}$ and $\bm{e}$ are compressed via our proposed RGFT method. The compressed file size of $\D$ and $\D^\prime+\e$ as well as their comparison are presented in Table \ref{newtab1.1}.

The results show that the compressed file size of $\D^\prime$ and $\e$ is close to that of $\D$ with an average ratio of 1.216, which is consistent with that the entropy of $\D$ is equal to that of $\D^\prime$ and $\e$, thus verifying the advantage of the proposed compression algorithm.
\subsection{Versatility Verification}
\label{s4.2}

\begin{figure*}[htbp]
\centering
\begin{minipage}{0.17\textwidth}
  \centerline{$\I^w$}
\end{minipage}
\begin{minipage}{0.17\textwidth}
  \centerline{$\I^{w\prime\prime}$}
\end{minipage}
\begin{minipage}{0.17\textwidth}
  \centerline{$\I^w$}
\end{minipage}
\begin{minipage}{0.17\textwidth}
  \centerline{$\I^{w\prime\prime}$}
\end{minipage}
\\[3pt]
\subfigure[Initial watermarked images $\I^w$ and corresponding final watermarked image $\I^{w\prime\prime}$ obtained by \cite{b26}]{
\includegraphics[width=0.17\textwidth]{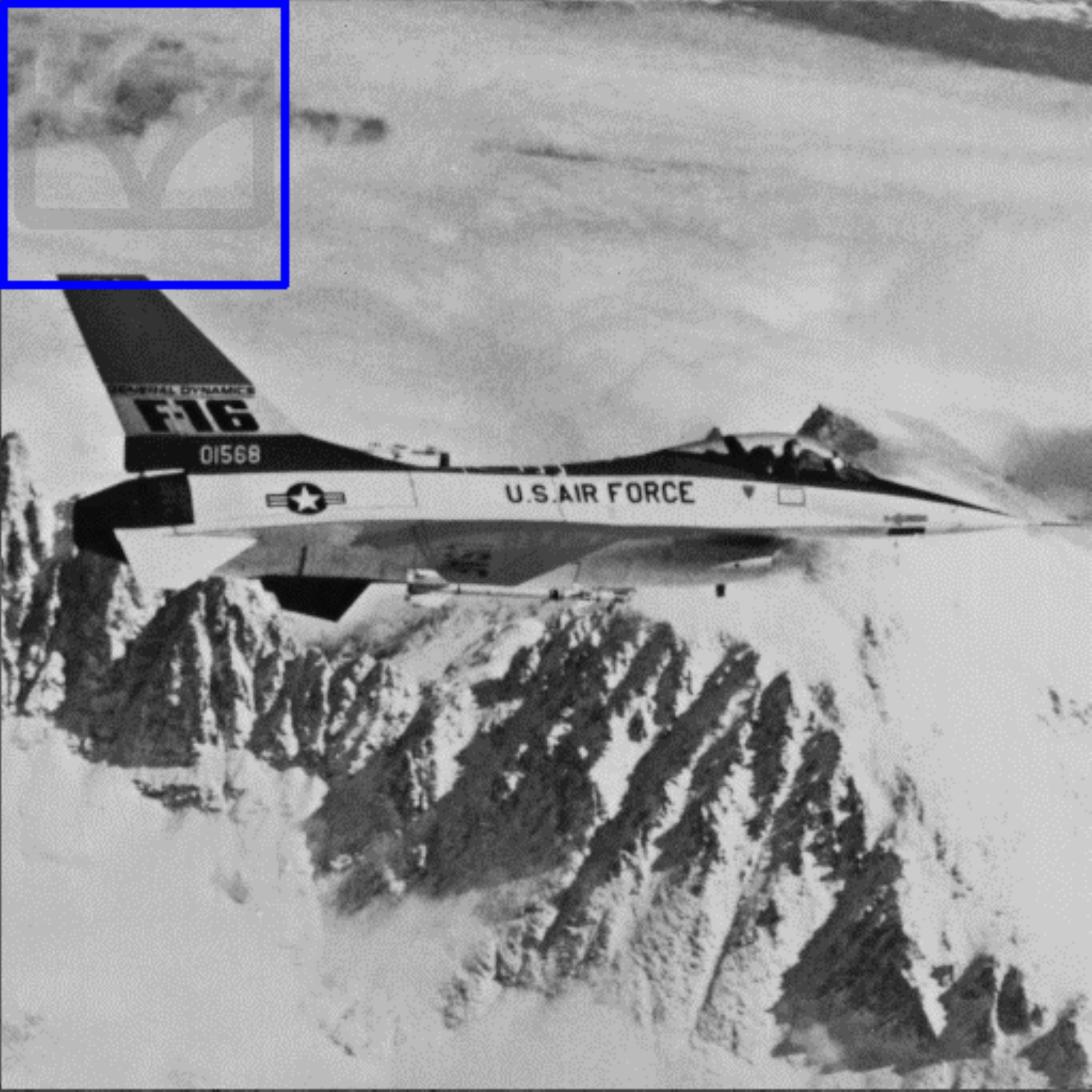}
\includegraphics[width=0.17\textwidth]{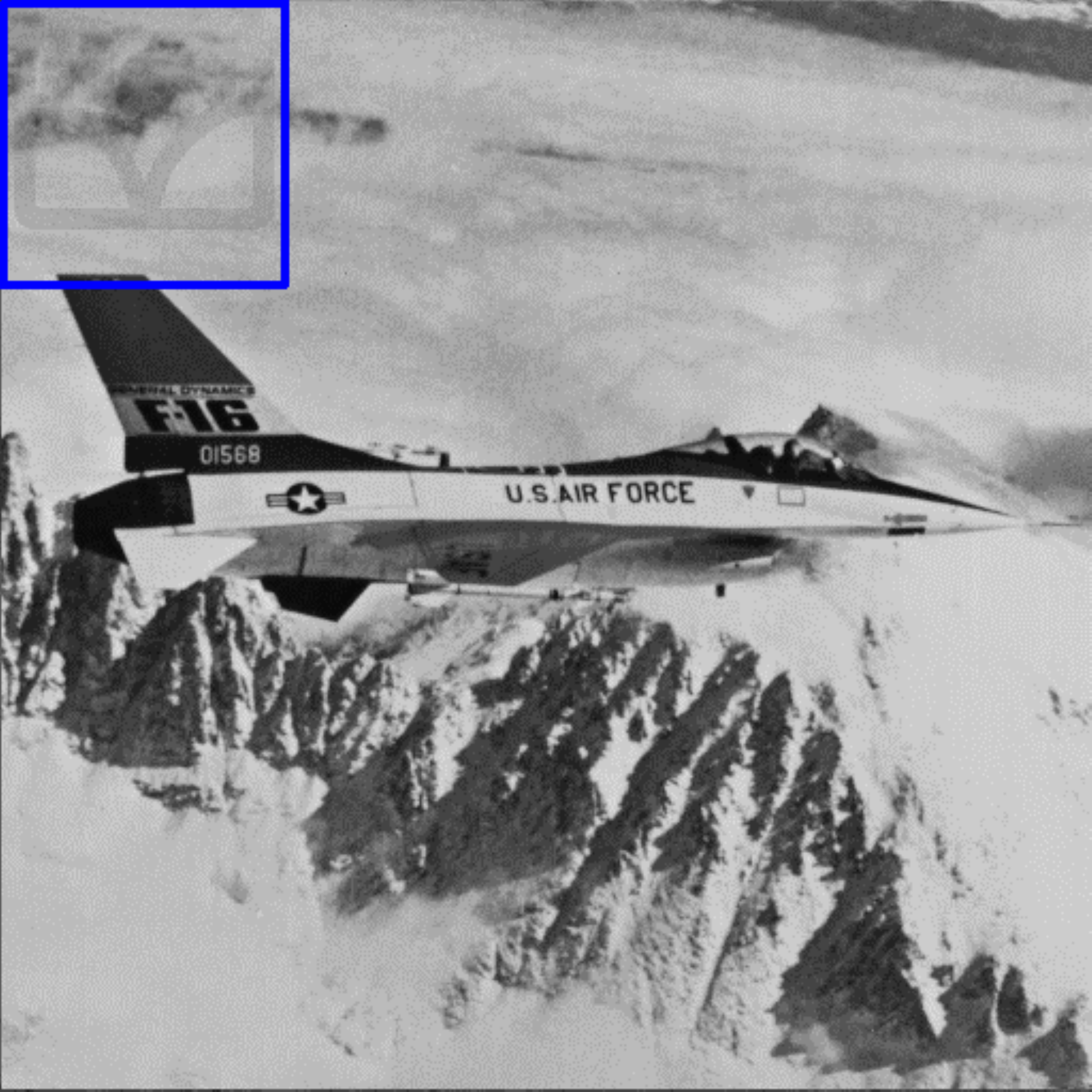}
\includegraphics[width=0.17\textwidth]{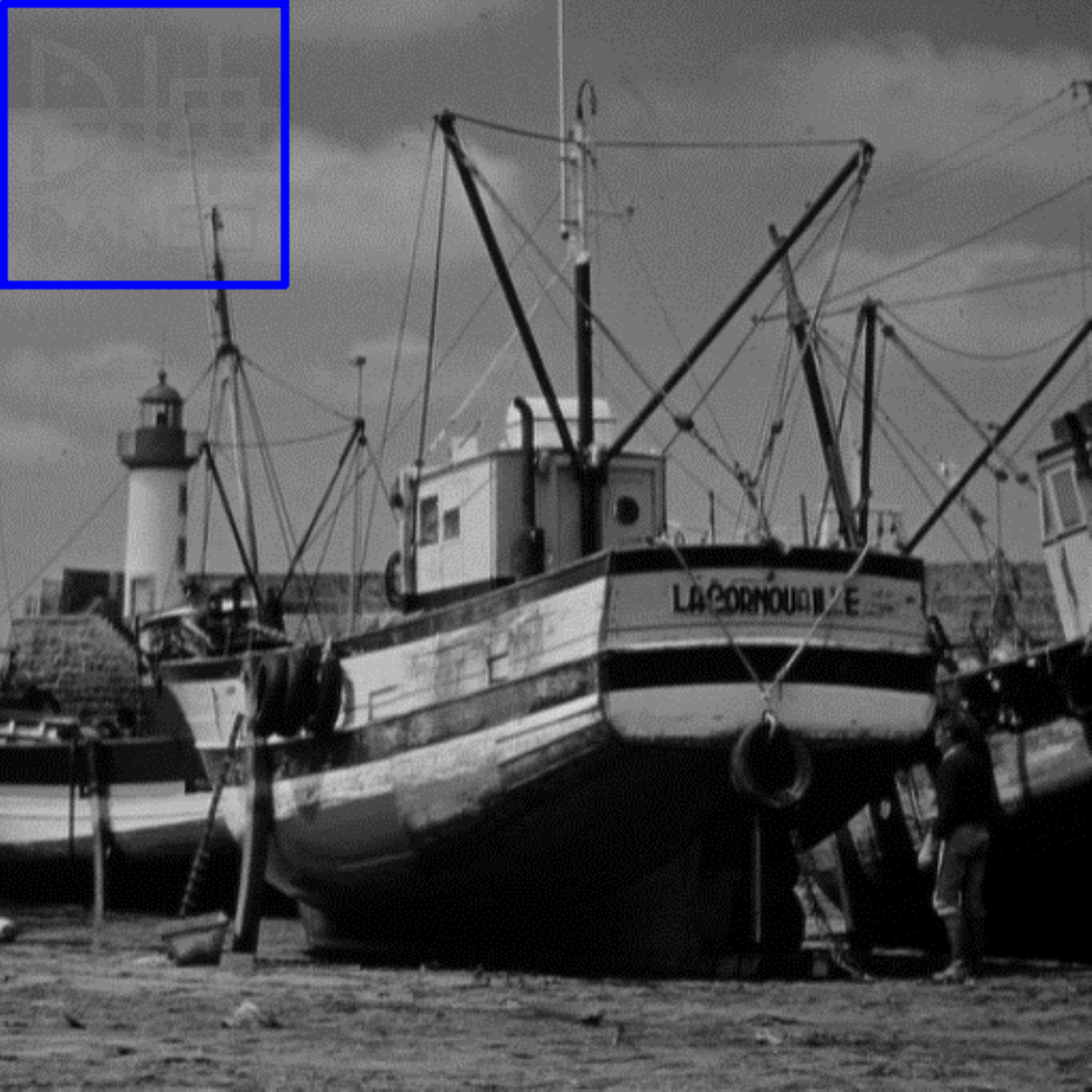}
\includegraphics[width=0.17\textwidth]{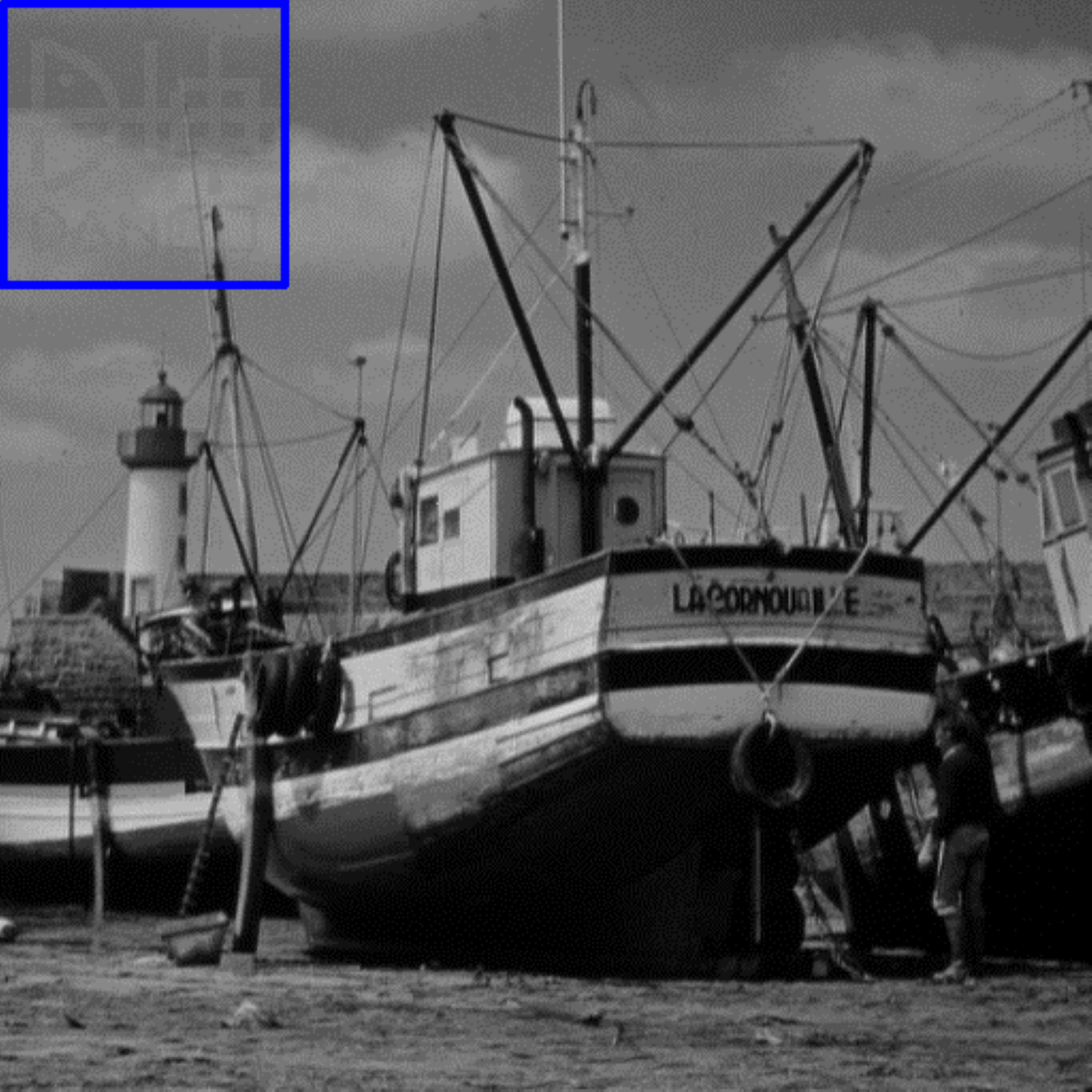}
\label{fig4.8a}
}
\subfigure[Initial watermarked images $\I^w$ and corresponding final watermarked image $\I^{w\prime\prime}$ obtained by \cite{b28}]{
\includegraphics[width=0.17\textwidth]{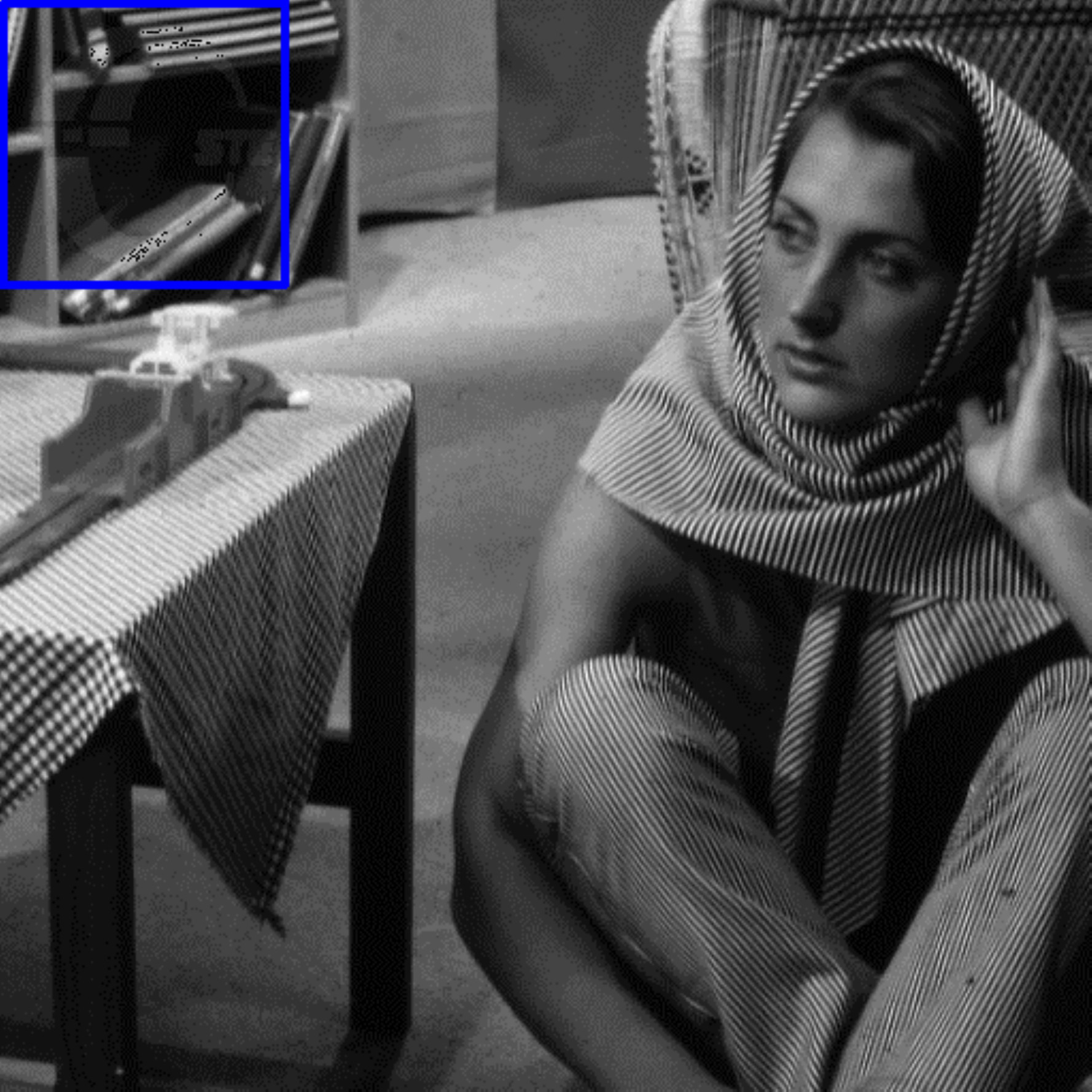}
\includegraphics[width=0.17\textwidth]{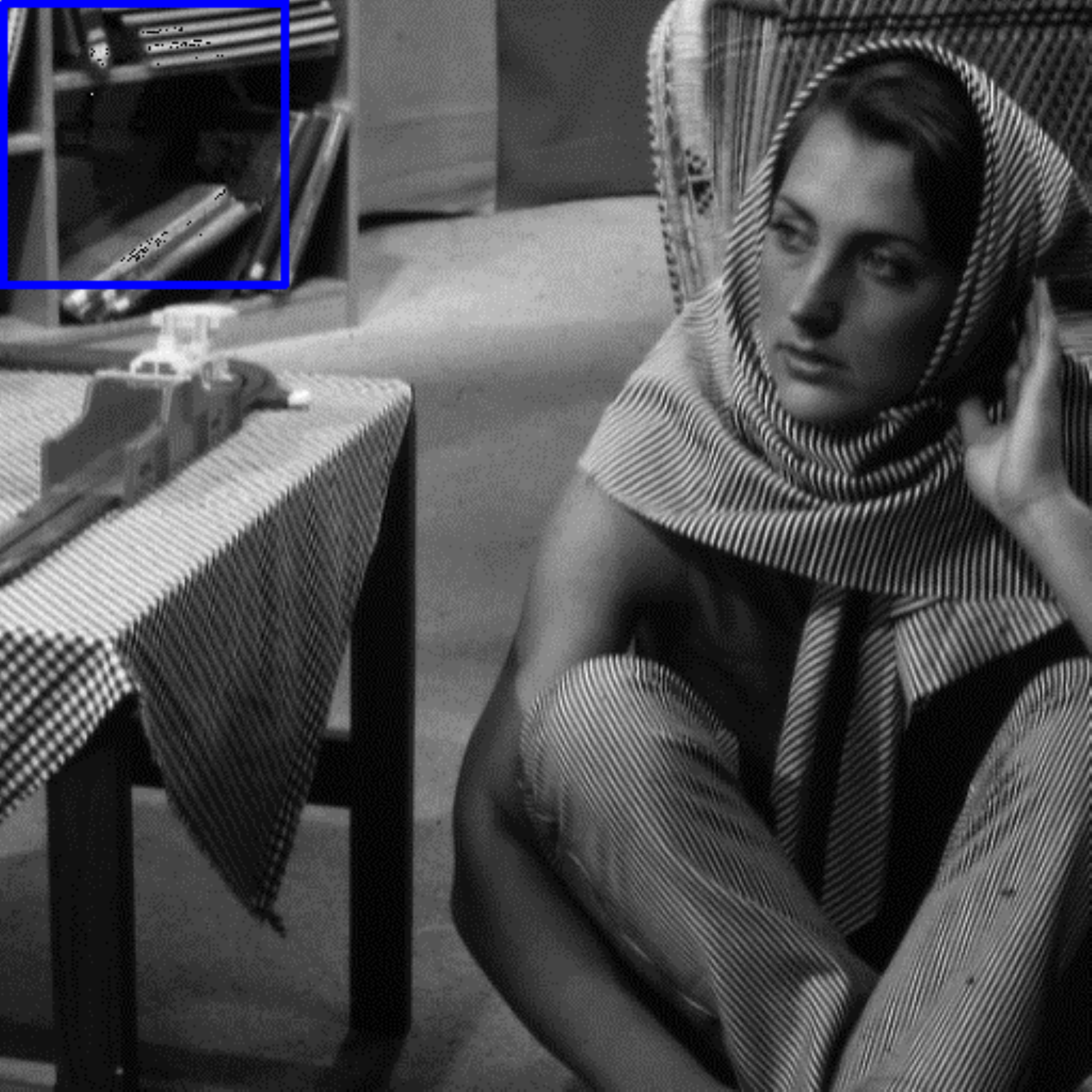}
\includegraphics[width=0.17\textwidth]{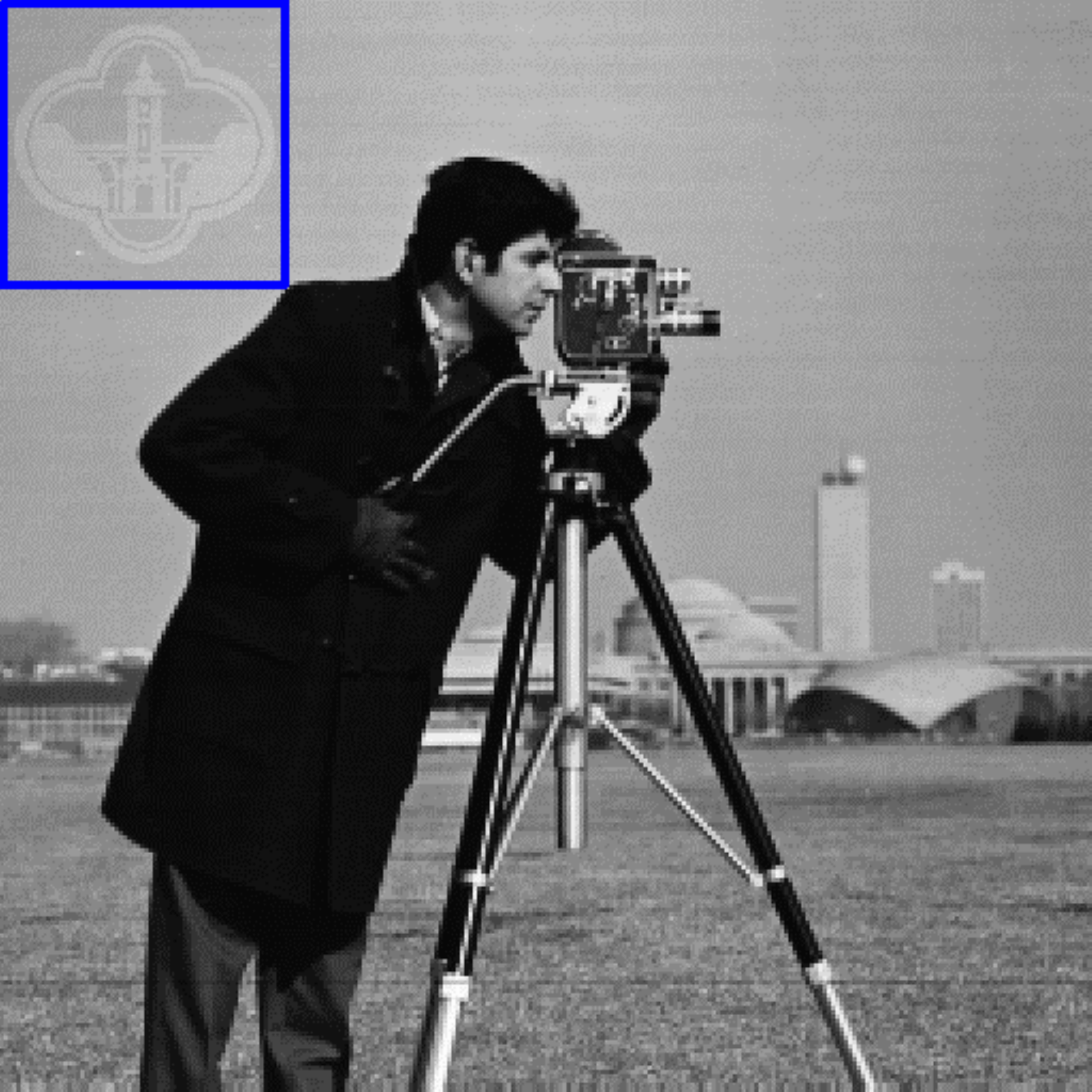}
\includegraphics[width=0.17\textwidth]{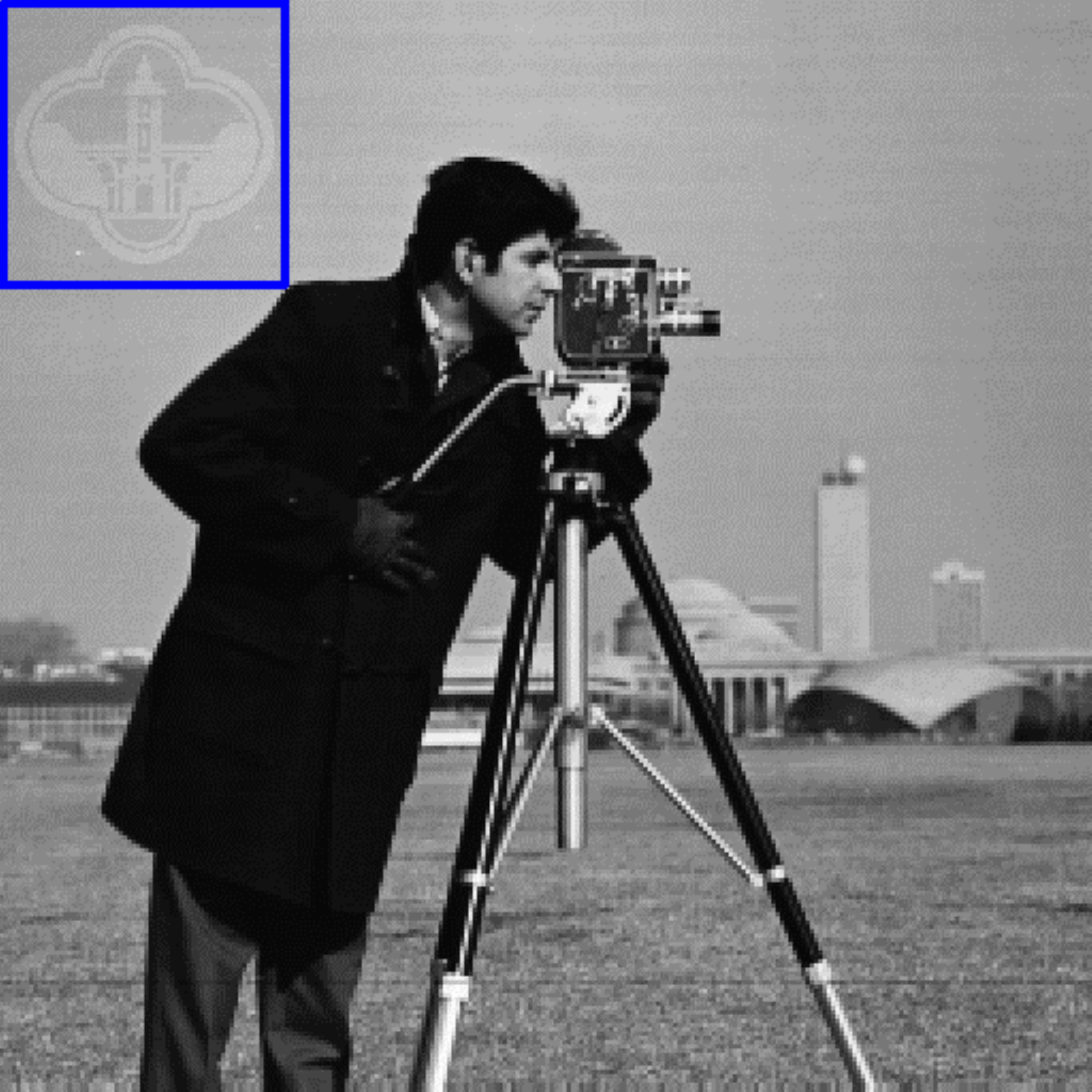}
\label{fig4.8b}
}
\subfigure[Initial watermarked images $\I^w$ \and corresponding final watermarked image $\I^{w\prime\prime}$ obtained by \cite{b25}]{
\includegraphics[width=0.17\textwidth]{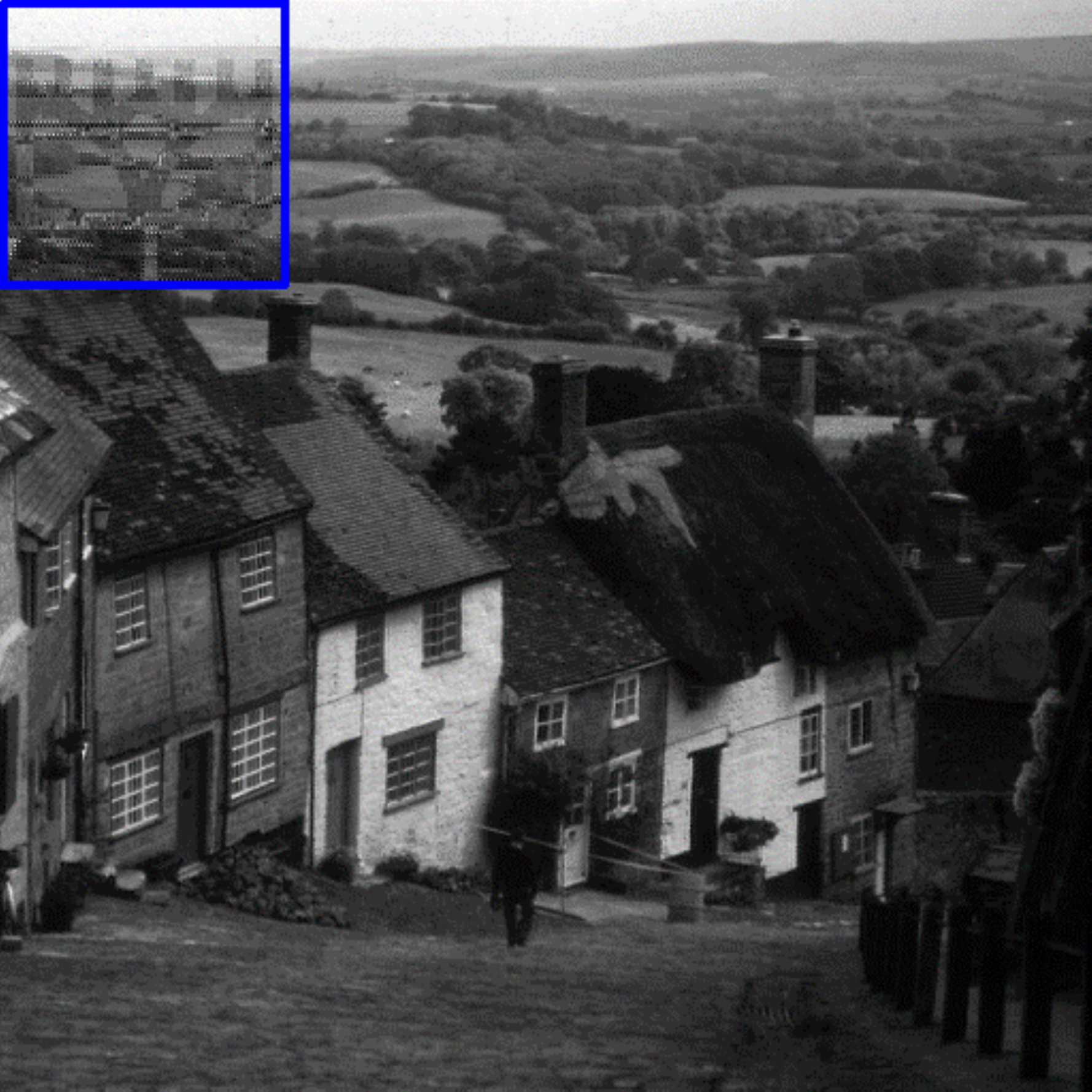}
\includegraphics[width=0.17\textwidth]{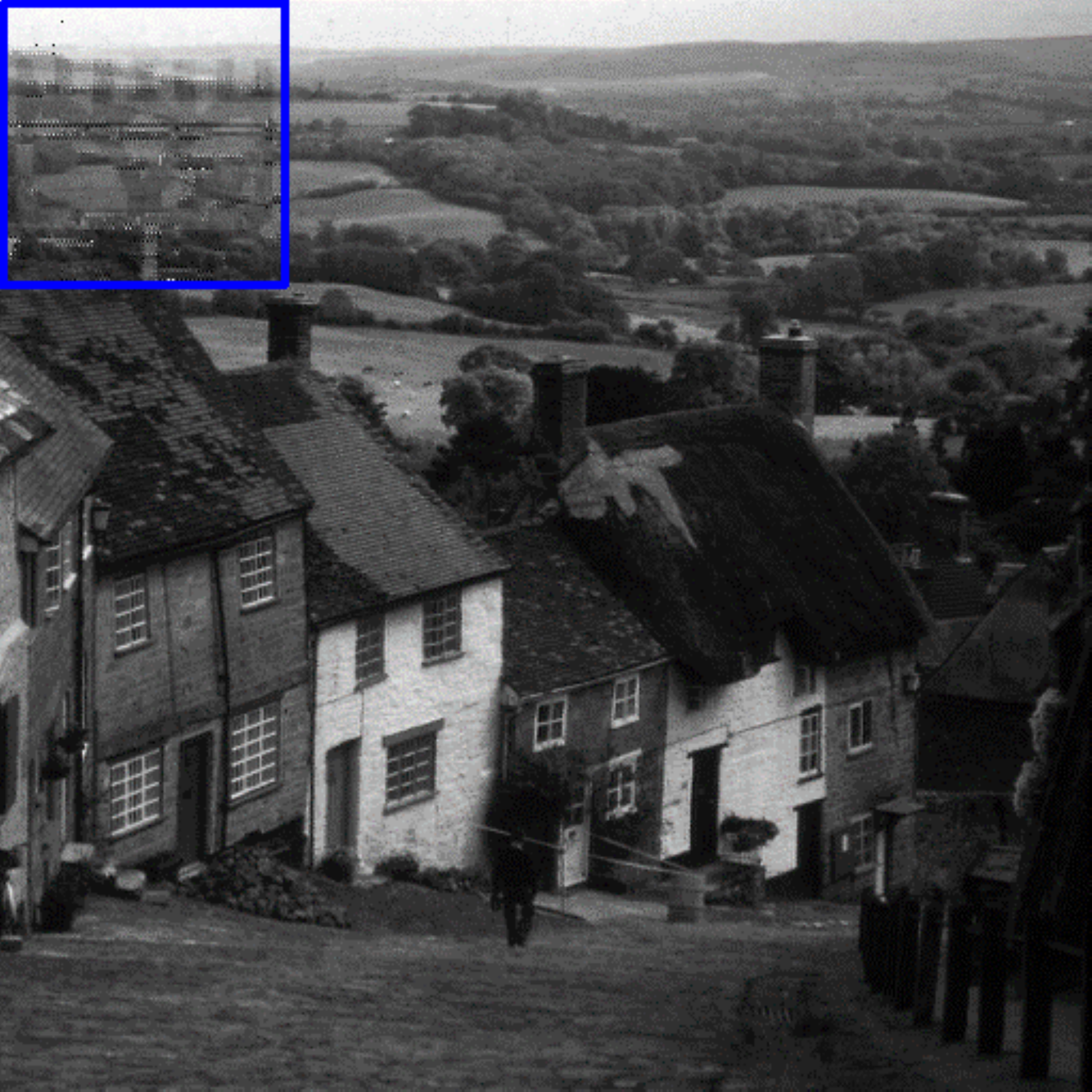}
\includegraphics[width=0.17\textwidth]{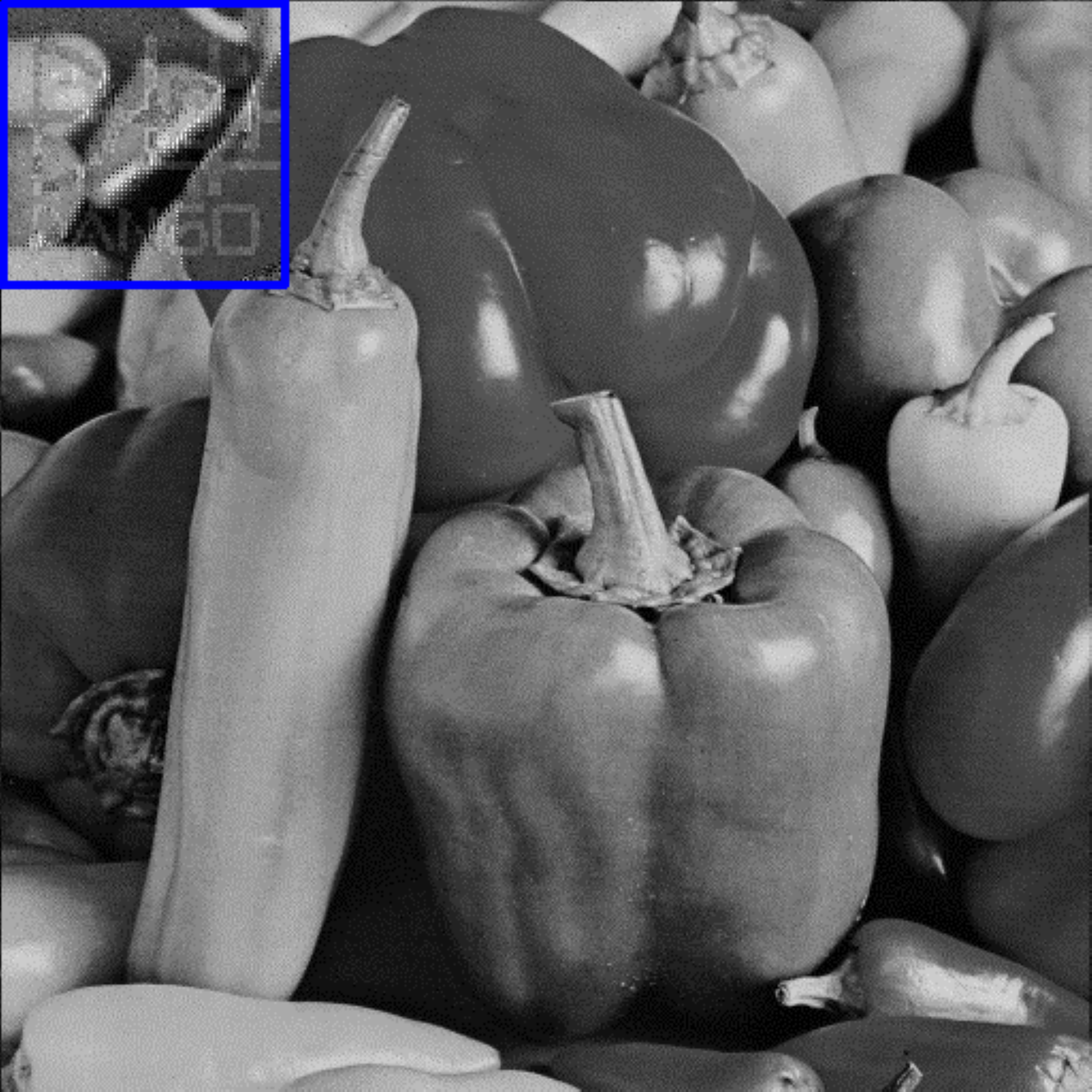}
\includegraphics[width=0.17\textwidth]{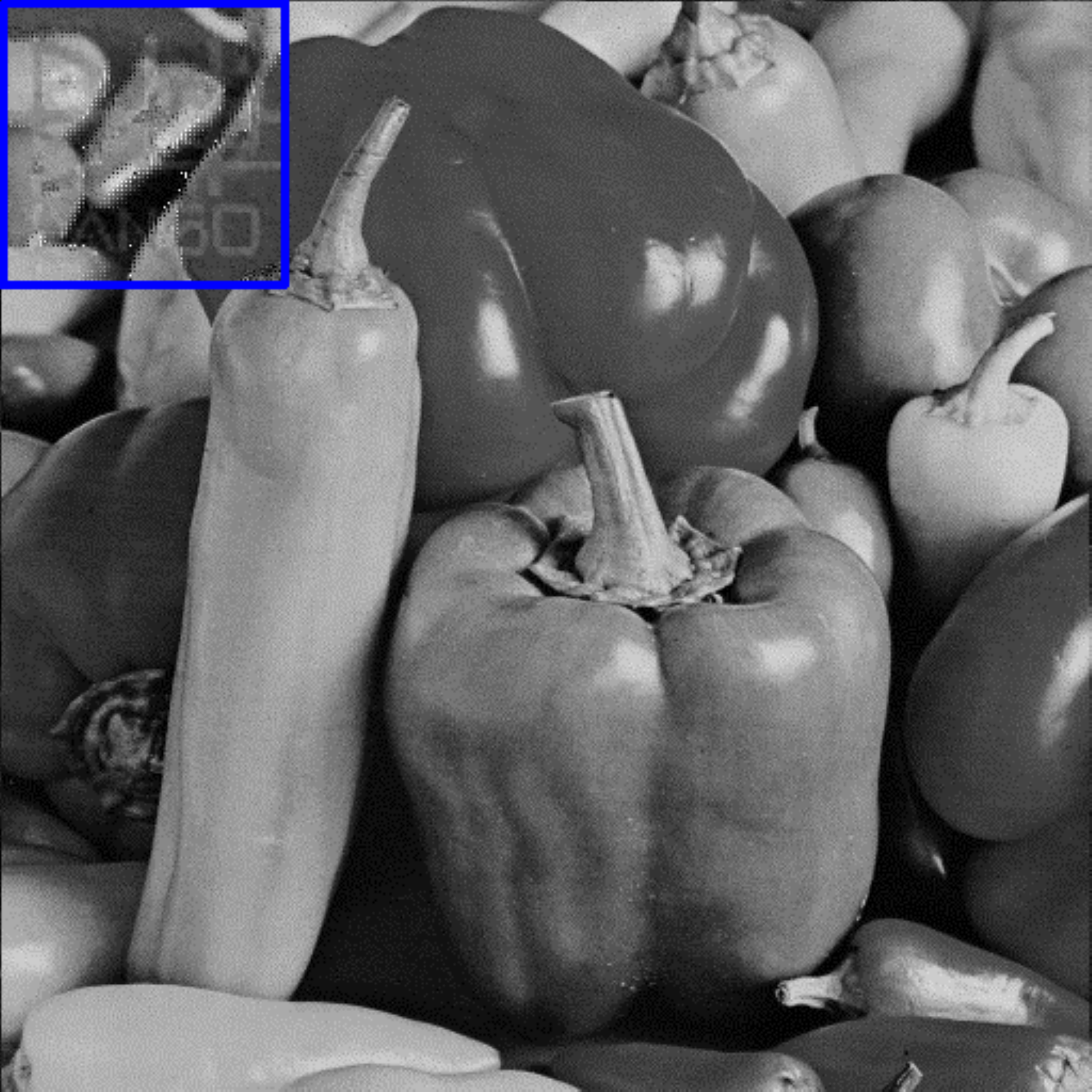}
\label{fig4.8c}
}
\subfigure[Initial watermarked images $\I^w$ and corresponding final watermarked image $\I^{w\prime\prime}$ obtained by \cite{b32}]{
\includegraphics[width=0.17\textwidth]{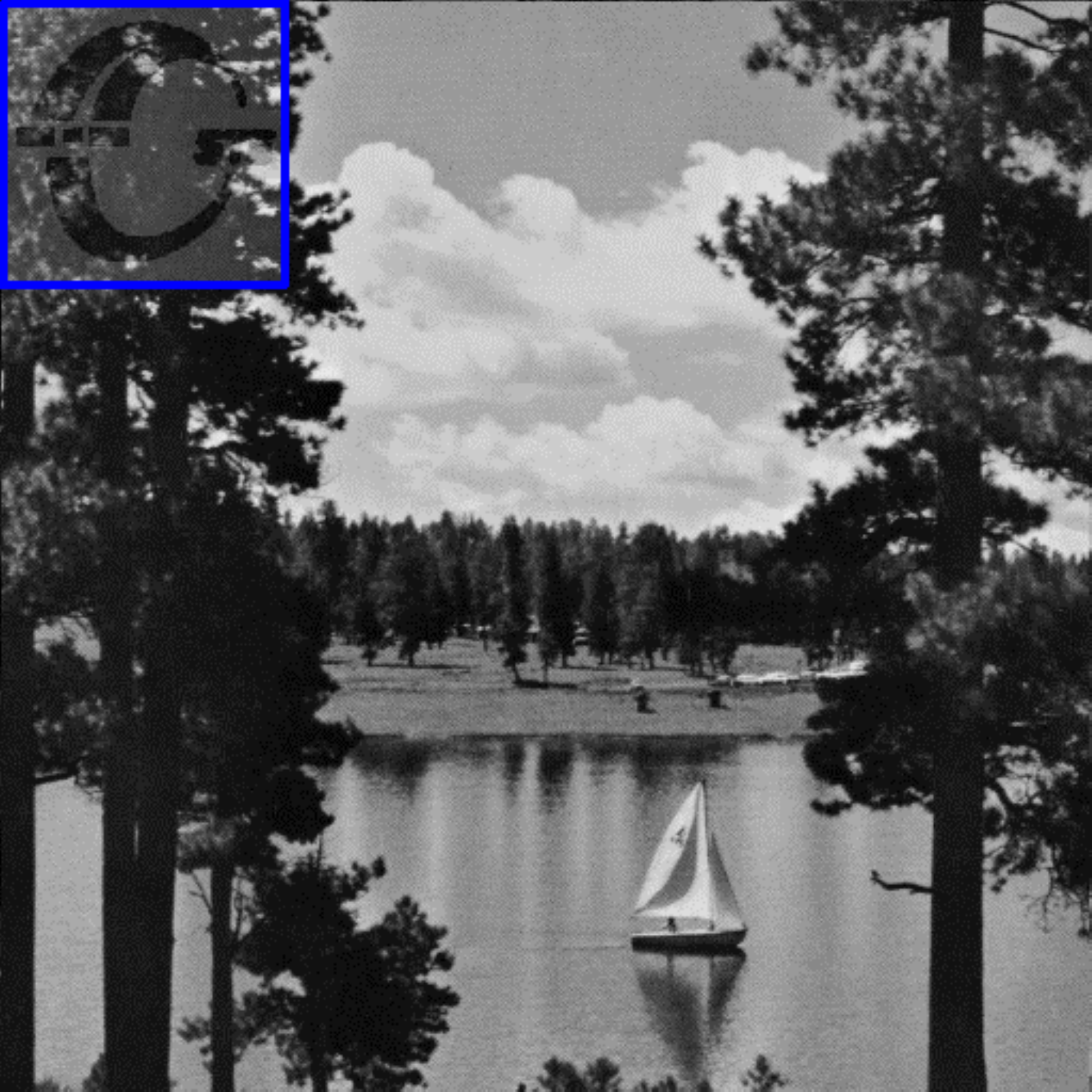}
\includegraphics[width=0.17\textwidth]{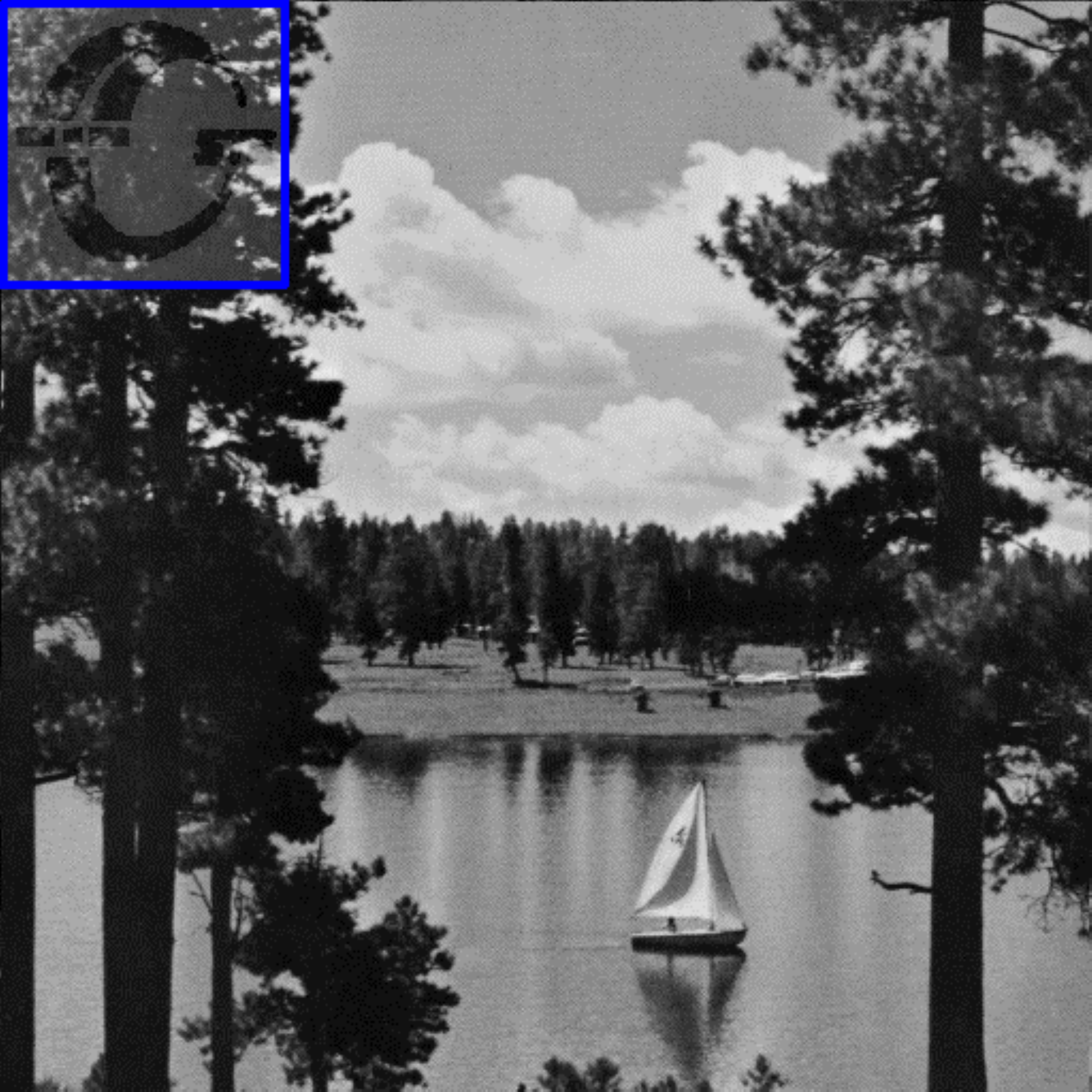}
\includegraphics[width=0.17\textwidth]{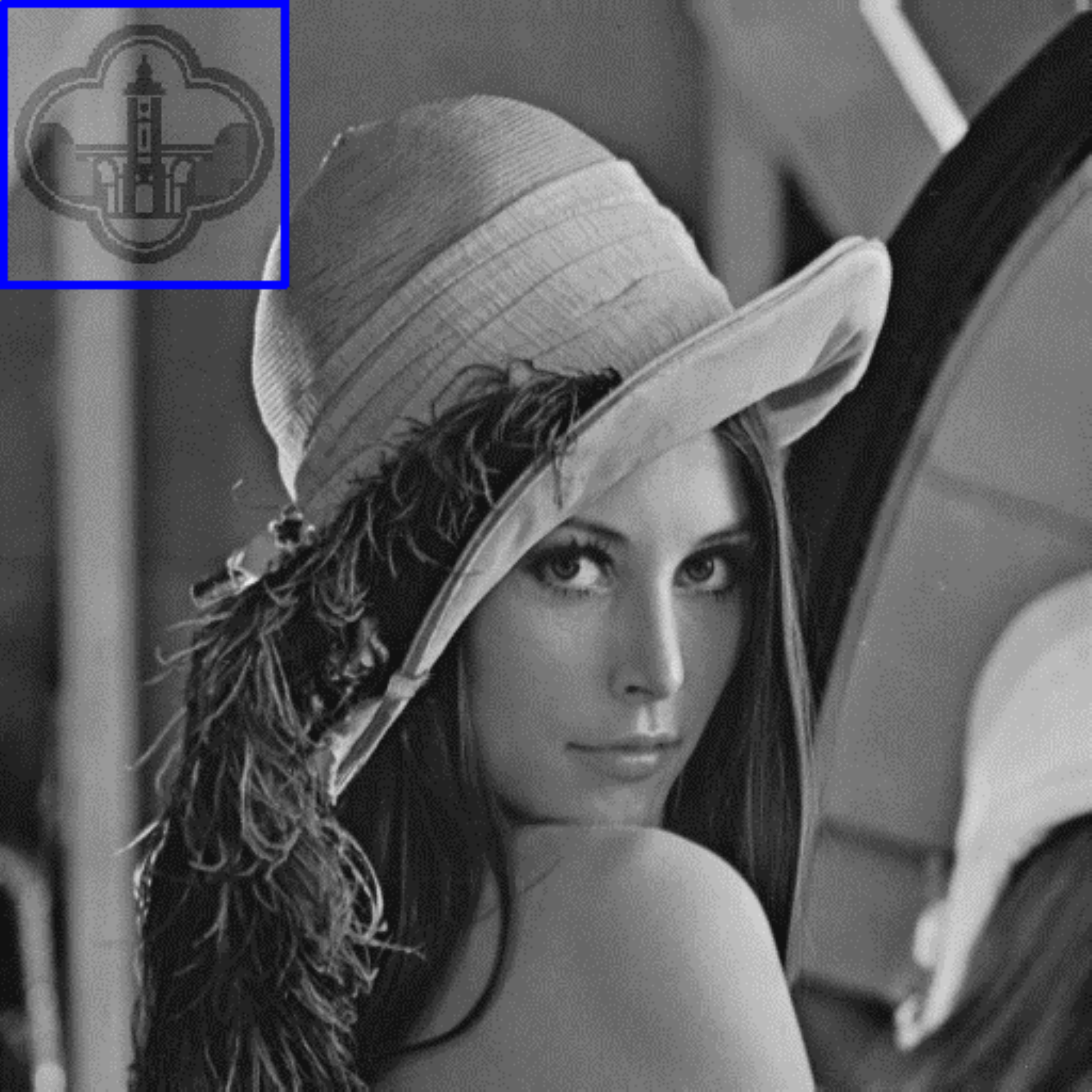}
\includegraphics[width=0.17\textwidth]{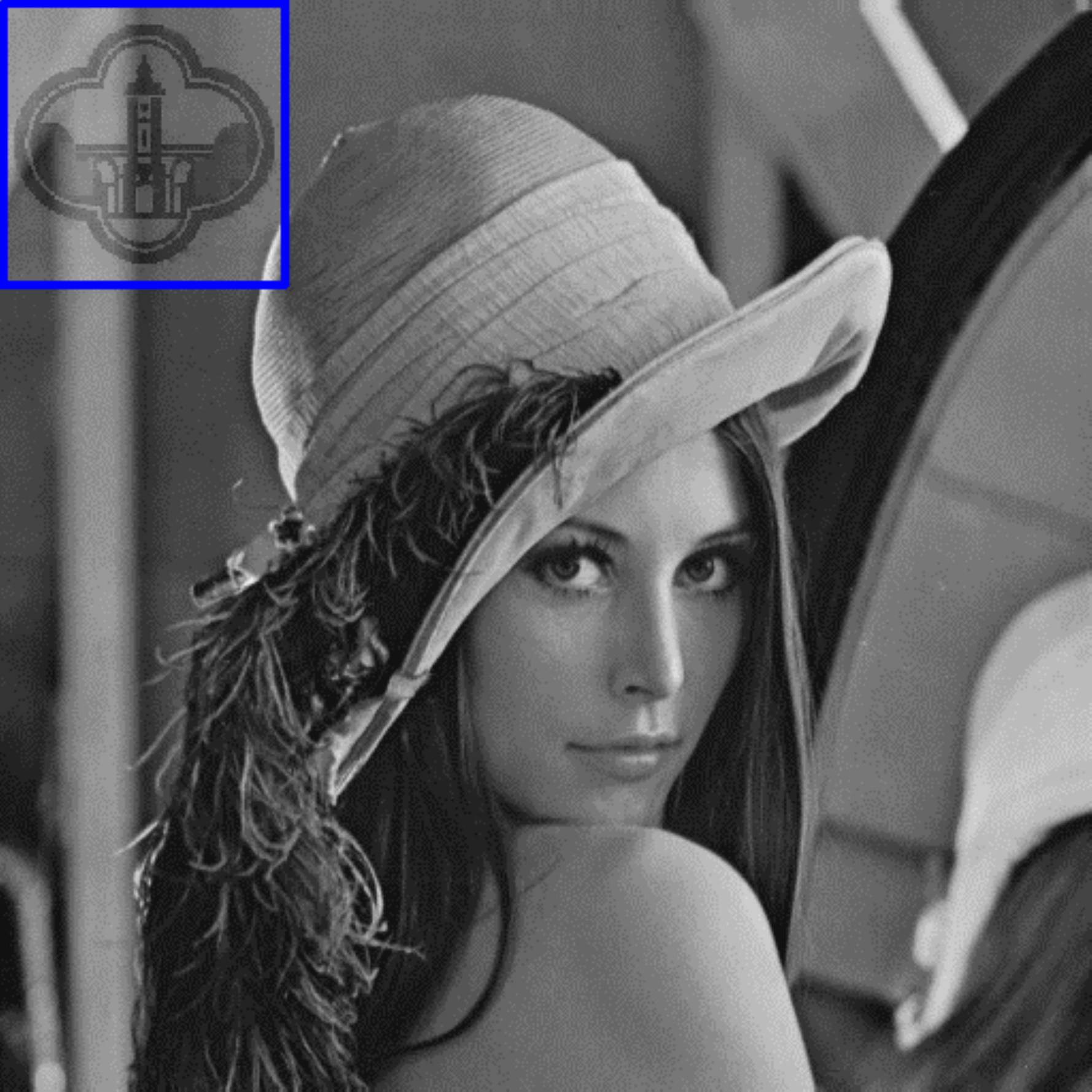}
\label{fig4.8d}
}
\subfigure[Initial watermarked images $\I^w$ and corresponding final watermarked image $\I^{w\prime\prime}$ obtained by \cite{b28}]{
\includegraphics[width=0.17\textwidth]{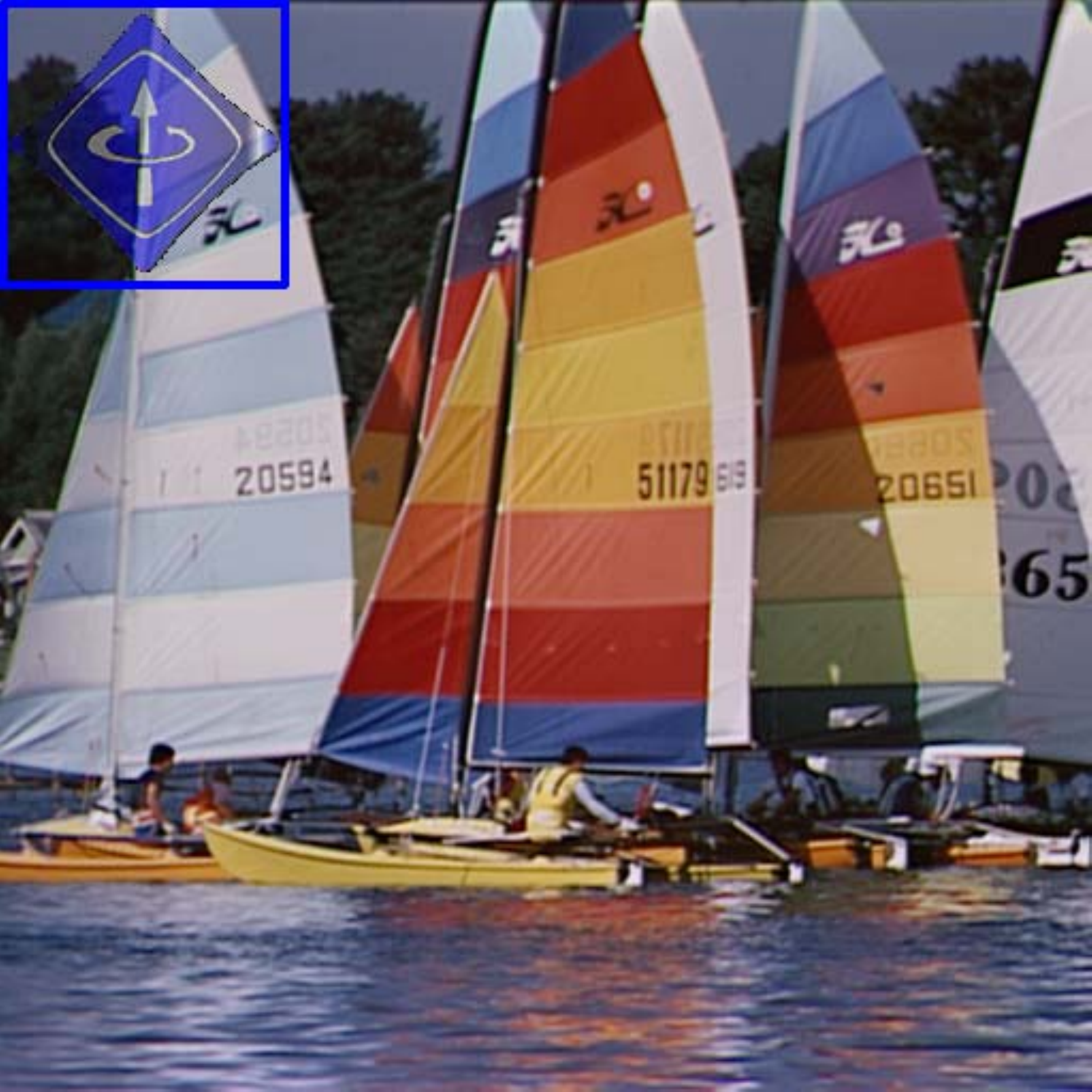}
\includegraphics[width=0.17\textwidth]{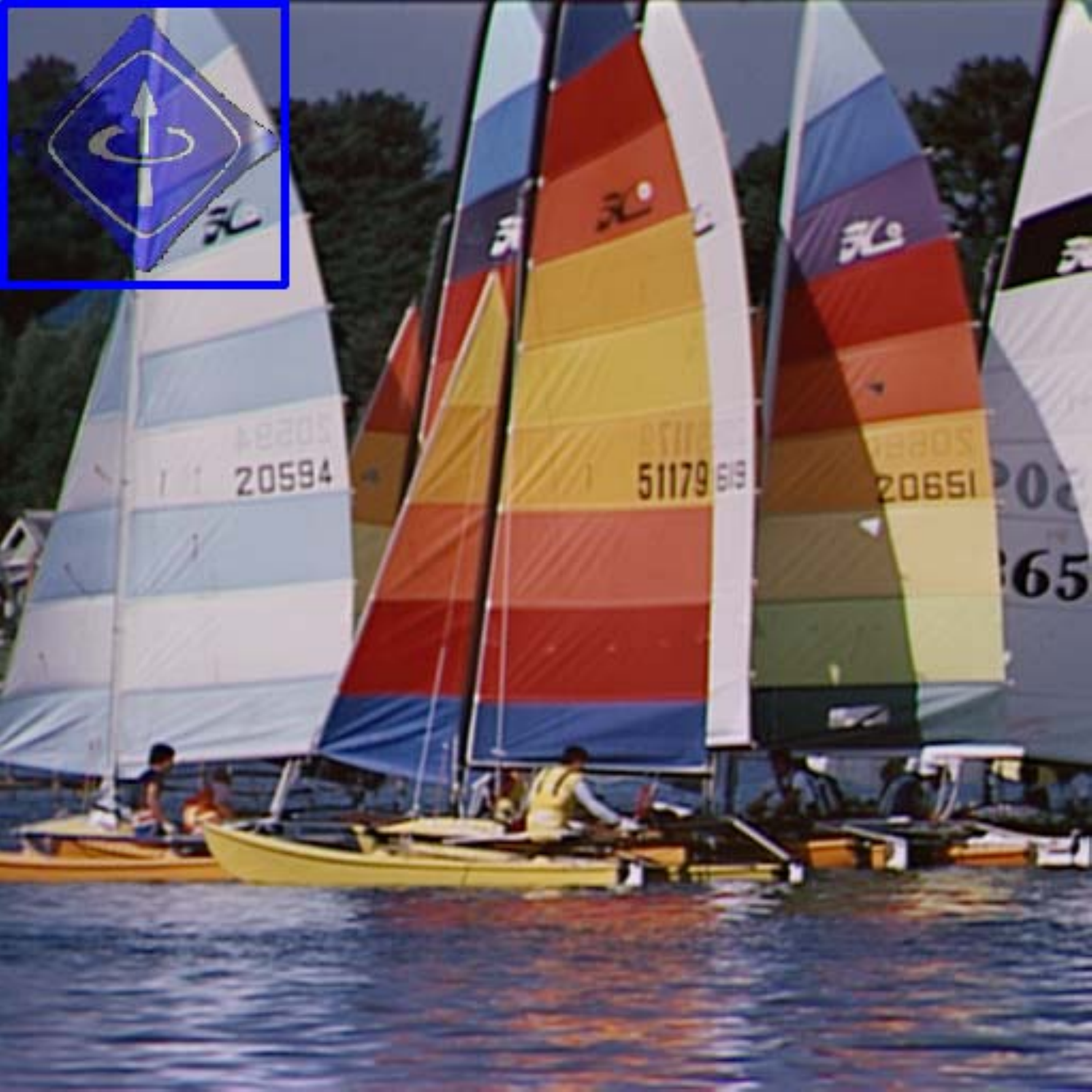}
\includegraphics[width=0.17\textwidth]{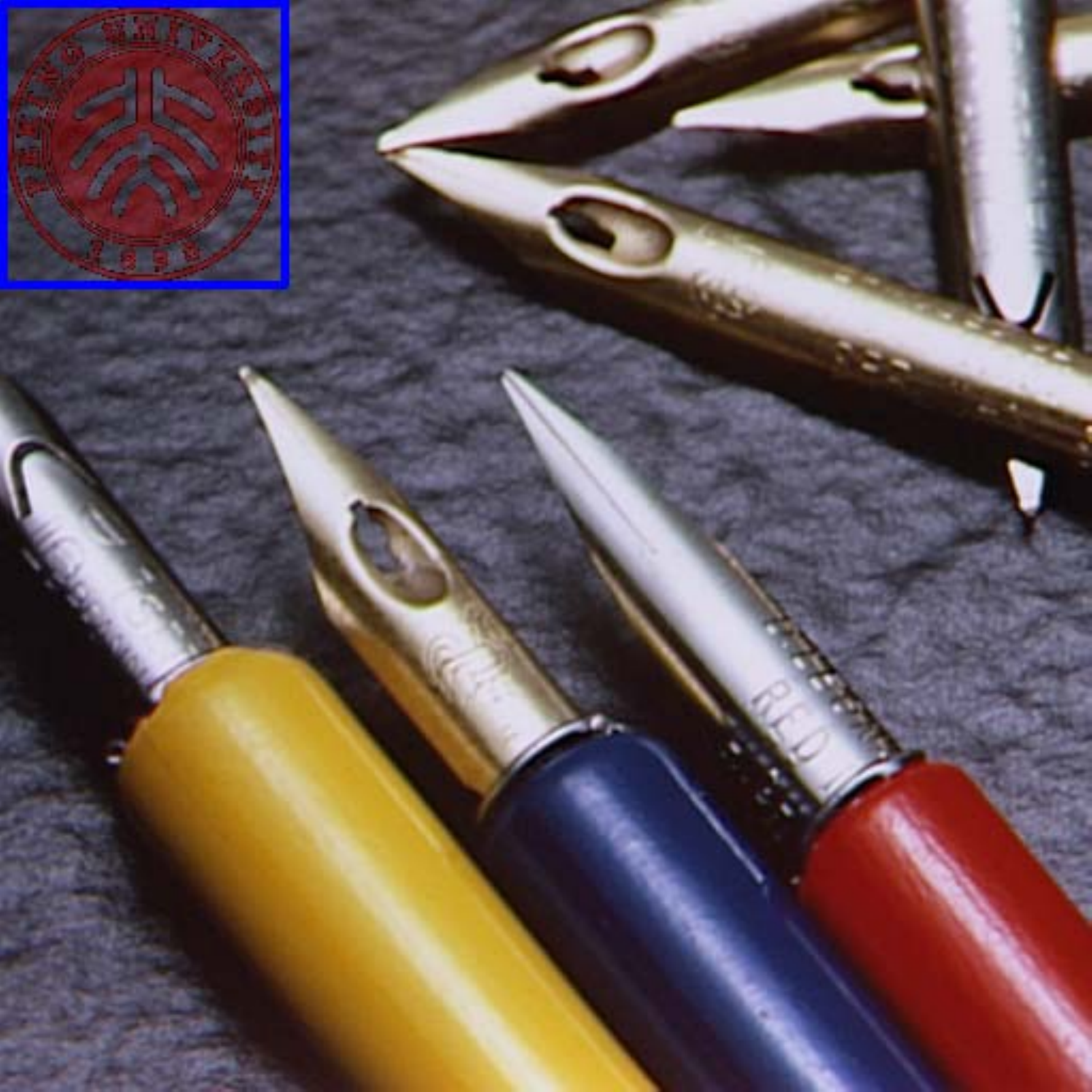}
\includegraphics[width=0.17\textwidth]{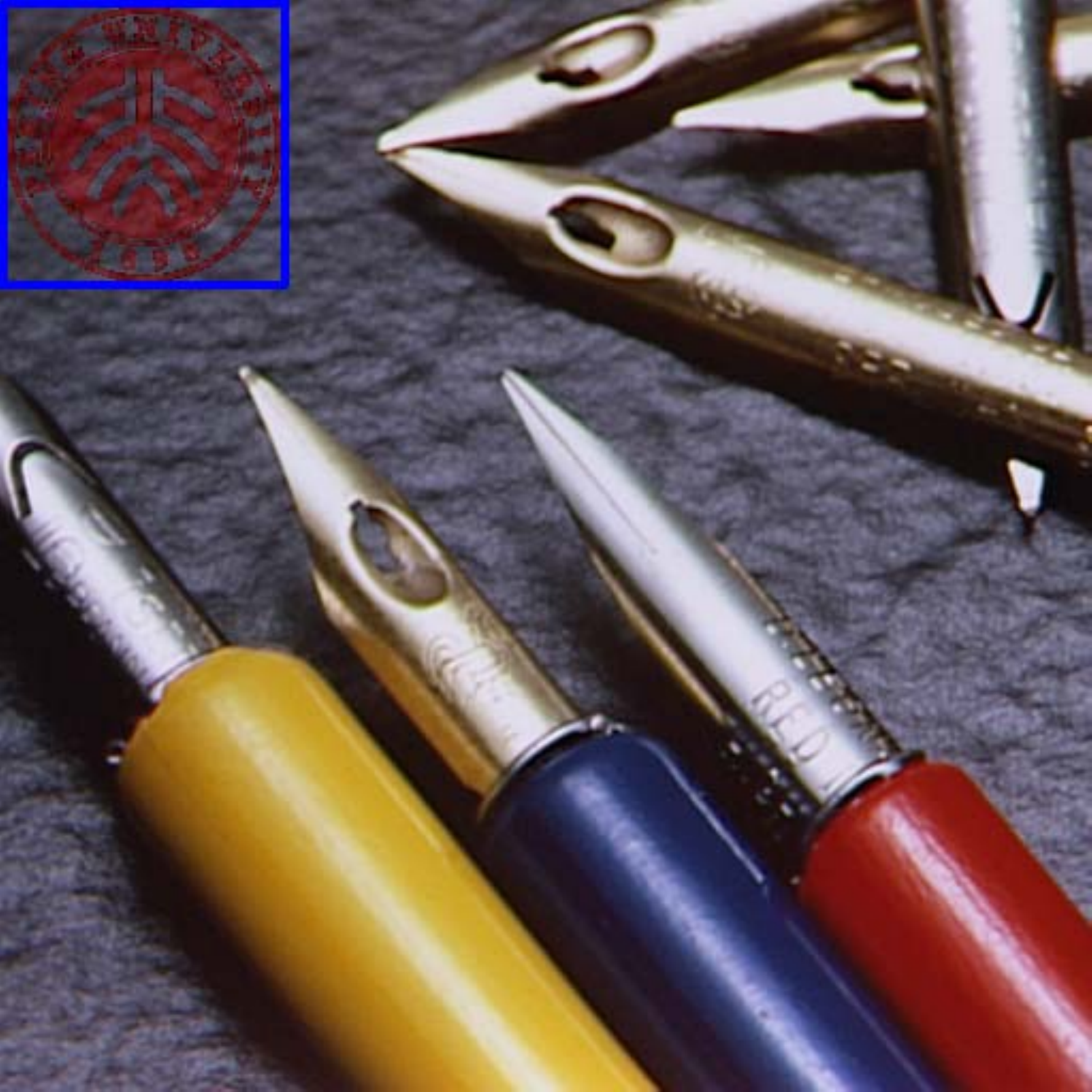}
\label{fig4.8e}
}
\subfigure[Initial watermarked images $\I^w$ and corresponding final watermarked image $\I^{w\prime\prime}$ obtained by \cite{alpha_fusion}]{
\includegraphics[width=0.17\textwidth]{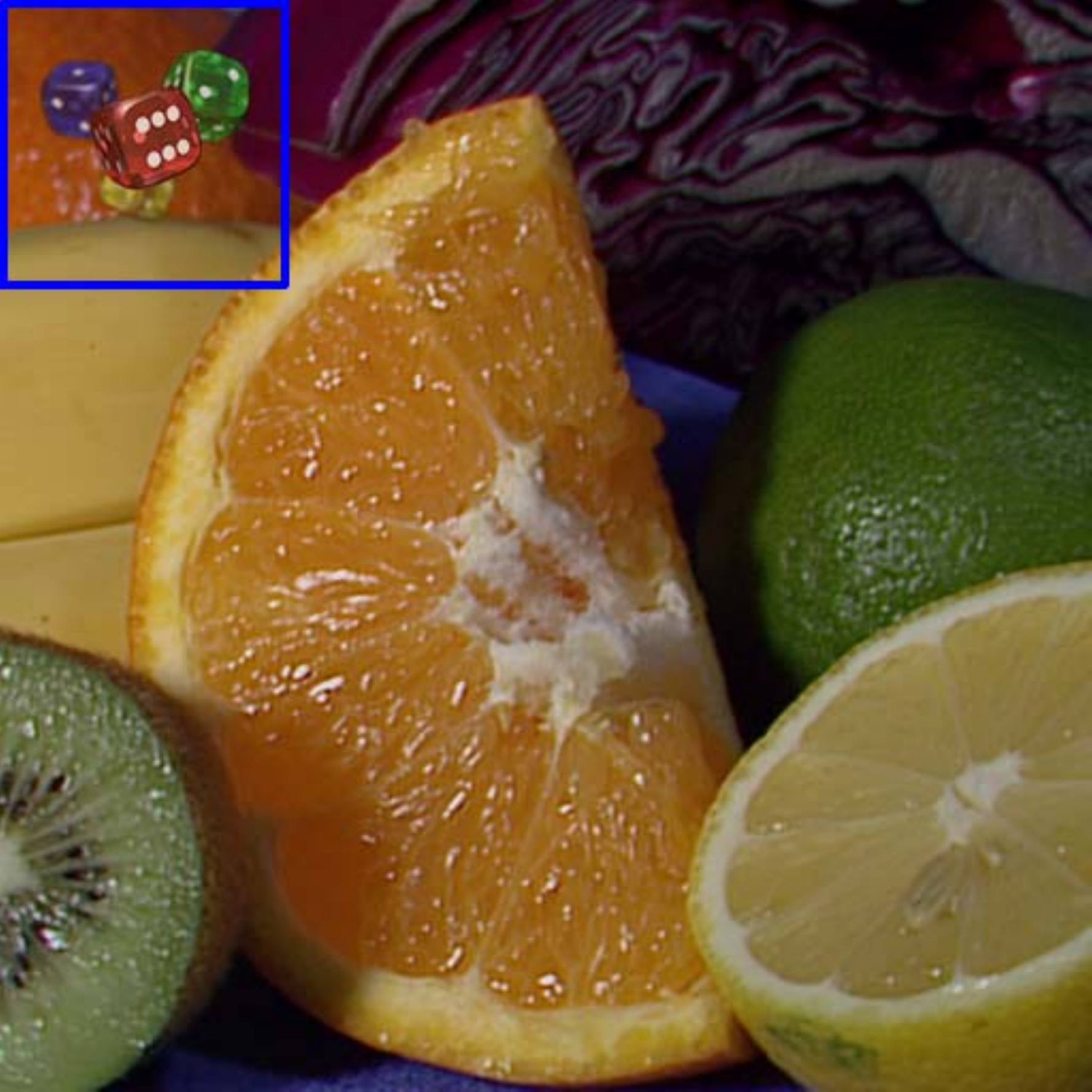}
\includegraphics[width=0.17\textwidth]{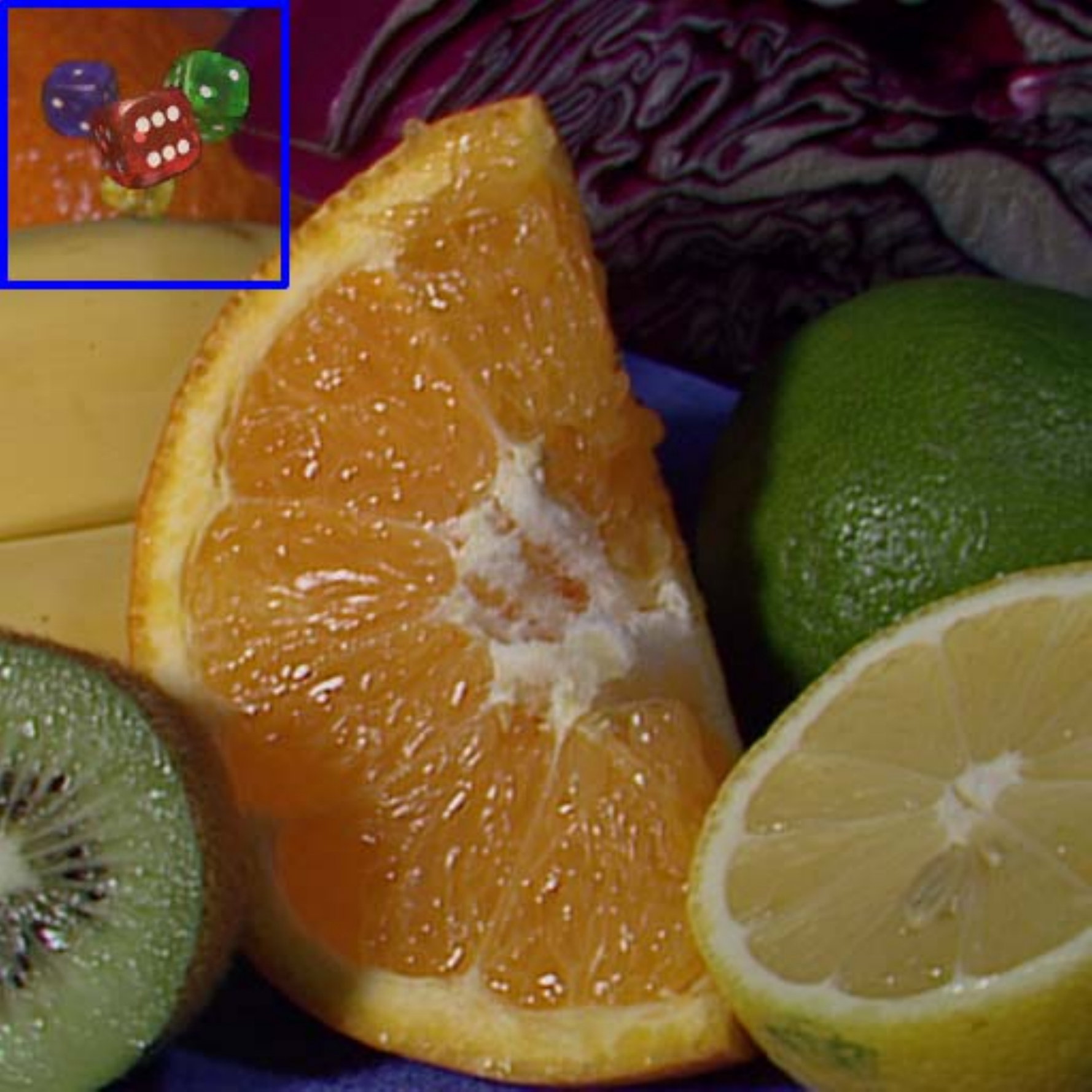}
\includegraphics[width=0.17\textwidth]{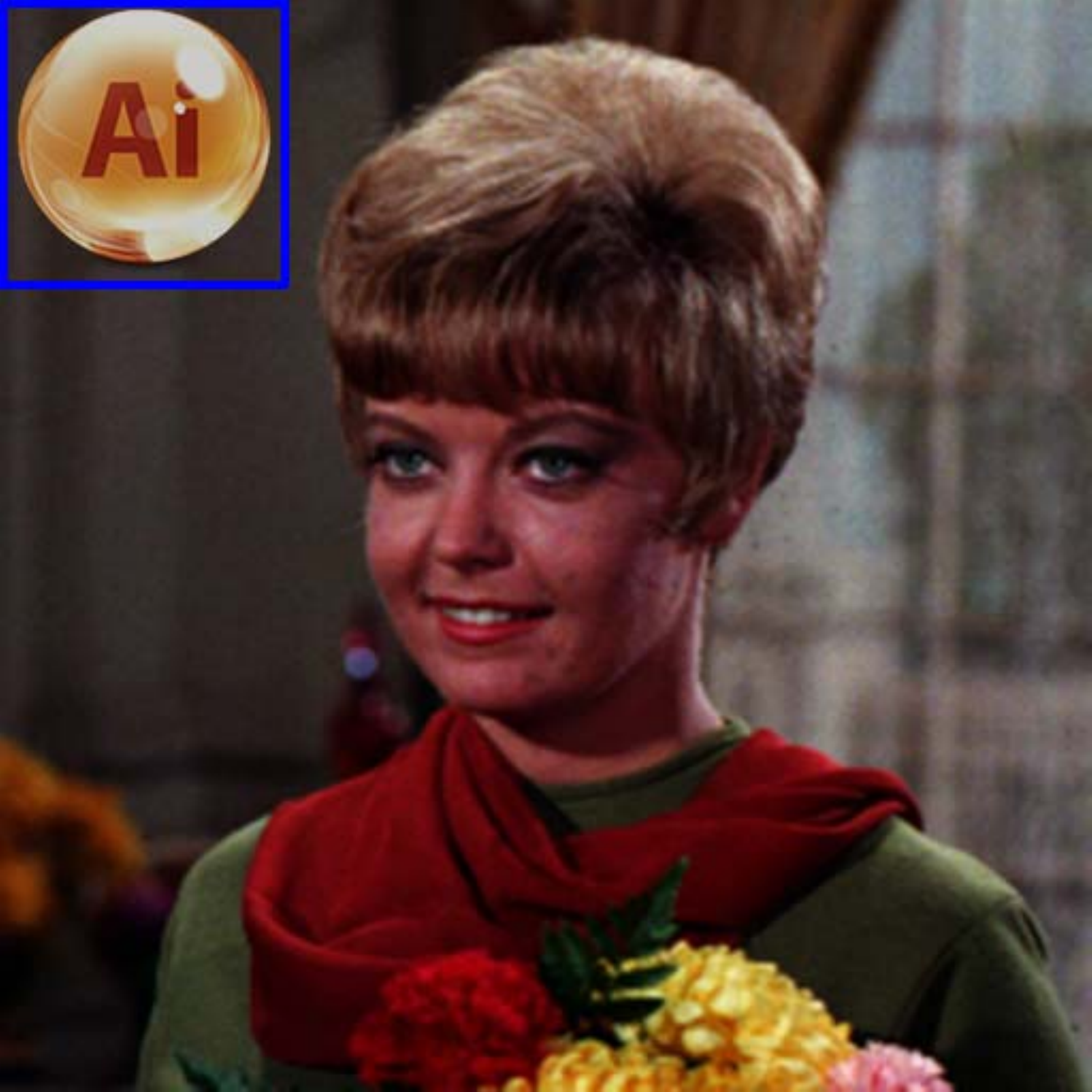}
\includegraphics[width=0.17\textwidth]{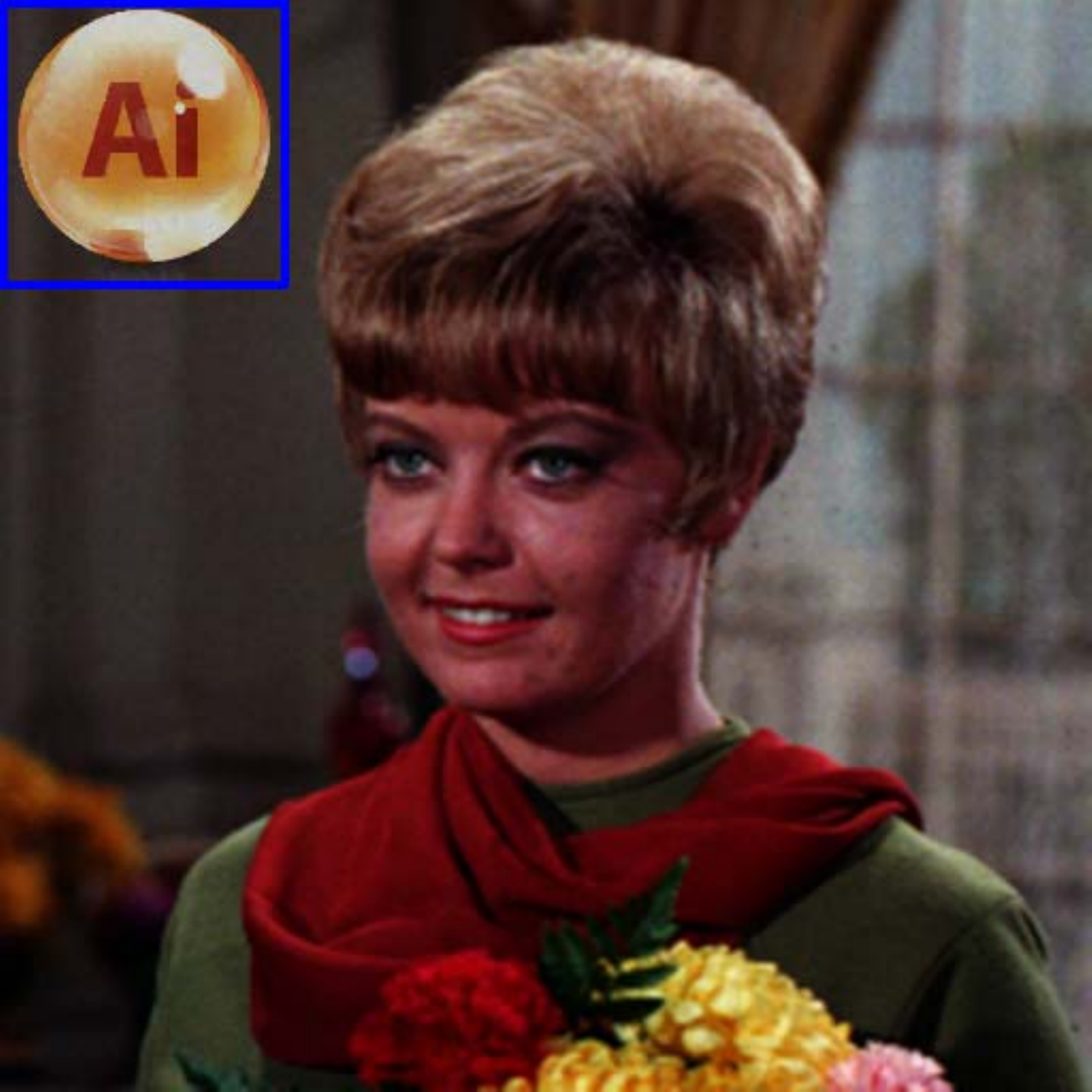}
\label{fig4.8f}
}
\caption{The embedding effects of watermarked images: from top to bottom, the first and third columns are the initial watermarked images $I^w$ obtained by the methods in \cite{b26}, \cite{b28}, \cite{b25}, \cite{b32}, \cite{b28} and \cite{alpha_fusion}, respectively; the second and fourth columns are the corresponding final watermarked images $\I^{w\prime\prime}$ obtained by our proposed method. Inside the blue box is ROI.}
\label{fig4.8}
\end{figure*}

In order to verify the versatility of the proposed framework, we employ different methods to perform visible watermark embedding, including the RVW methods in the spatial domain \cite{b25,b26,b28} and in the transform domain \cite{b32}, as well as the conventional method \cite{alpha_fusion}. The binary visible watermark images in Fig. \ref{fig4.1a}-\ref{fig4.1d} are embedded into the host images \textit{F-16}, \textit{Boat}, \textit{Barbara}, \textit{Cameraman}, \textit{Goldhill}, \textit{Pepper}, \textit{Sailboat} and \textit{Lena} respectively to obtain gray-scale watermarked images $\I^w$.
The RGB visible watermark images in Fig. \ref{fig4.1e}-\ref{fig4.1h} are embedded into the host images \textit{Yacht}, \textit{Fruits}, \textit{Pens}, \textit{Flower}, \textit{Couple}, \textit{Cable car}, \textit{Cornfield} and \textit{Girl} respectively to acquire RGB watermarked images $\I^w$.
The watermarked images $\I^w$ are processed with our proposed method to obtain the final watermarked image $\I^{w\prime\prime}$.
Due to the page limit, we present some representative initial watermarked images $\I^w$ in Fig. \ref{fig4.8}, where the watermarked images in Fig. \ref{fig4.8a}-\ref{fig4.8f} are obtained by the methods in \cite{b26},\cite{b28},\cite{b25},\cite{b32},\cite{b28} and \cite{alpha_fusion}, respectively. The corresponding final watermarked images $\I^{w\prime\prime}$ are presented in Fig. \ref{fig4.8} as well.
Fig. \ref{fig4.8} shows that modifications on the watermarked images $I^w$ are negligible, which does not affect the overall visual effect of the watermarked images, especially in the visible-watermarked area.

\begin{table*}[t]
  \caption{The versatility verification by embedding binary visible watermark images into gray-scale host images using different visible watermarking schemes.}
  \label{tab1}
  \centering
  \resizebox{0.9\textwidth}{!}{
  \begin{tabular}{c|c|c|c|c|c|c|c}
  \hline
  \hline
\textbf{Host Image}&\textbf{Methods}&$\bm{N_D}$ (bits)&$\bm{N_C}$ (bits) &$\bm{Ratio}$&$\bm{\mathrm{PSNR_{I}}}$ (dB)&$\bm{\mathrm{PSNR_{N}}}$ (dB)&$\bm{\mathrm{PSNR_{W}}}$ (dB)\\
  \hline
   \multirow{4}{*}{F-16}
   &Qi et al.\cite{b26}&131072&652&0.0050&56.76&73.89&44.80\\
    \cline{2-8}
   &Liu et al.\cite{b28}&131072&592&0.0045&69.90&74.34&59.65\\
   \cline{2-8}
   &Chen et al.\cite{b25}&131072&5913&0.0451&39.88&71.48&27.84\\
   \cline{2-8}
   &Yang et al.\cite{b32}&131072&1223&0.0093&50.17&77.19&38.13\\
     \hline
   \multirow{4}{*}{Boat}
   &Qi et al.\cite{b26}&131072&588&0.0045&58.48&62.41&48.51\\
   \cline{2-8}
   &Liu et al.\cite{b28}&131072&607&0.0046&60.53&66.61&49.64\\
   \cline{2-8}
   &Chen et al.\cite{b25}&131072&3337&0.0255&44.74&65.5&32.73\\
   \cline{2-8}
   &Yang et al.\cite{b32}&131072&1161&0.0089&49.03&66.45&37.06\\
    \hline
   \multirow{4}{*}{Barbara}
   &Qi et al.\cite{b26}&131072&1132&0.0086&53.31&76.41&41.29\\
   \cline{2-8}
   &Liu et al.\cite{b28}&131072&4703&0.0359&47.49&75.26&35.46\\
   \cline{2-8}
   &Chen et al.\cite{b25}&131072&22624&0.1726&37.89&60.31&25.87\\
   \cline{2-8}
   &Yang et al.\cite{b32}&131072&1927&0.0147&46.76&78.61&34.73\\
   \hline
   \multirow{4}{*}{Cameraman}
   &Qi et al.\cite{b26}&131072&542&0.0041&62.60&77.90&50.68\\
   \cline{2-8}
   &Liu et al.\cite{b28}&131072&550&0.0042&63.46&77.08&51.60\\
   \cline{2-8}
   &Chen et al.\cite{b25}&131072&737&0.0056&64.96&83.97&52.97\\
   \cline{2-8}
   &Yang et al.\cite{b32}&131072&1283&0.0098&48.15&84.25&36.11\\
 \hline
   \multirow{4}{*}{Goldhill}
   &Qi et al.\cite{b26}\cite{b26}&131072&720&0.0055&54.60&71.57&42.64\\
   \cline{2-8}
   &Liu et al.\cite{b28}&131072&4993&0.0381&45.84&70.99&33.81\\
   \cline{2-8}
   &Chen et al.\cite{b25}&131072&14361&0.1096&38.81&64.95&26.78\\
   \cline{2-8}
   &Yang et al.\cite{b32}&131072&1460&0.0111&49.44&69.79&37.44\\
 \hline
   \multirow{4}{*}{Pepper}
   &Qi et al.\cite{b26}&131072&955&0.0073&52.69&72.43&40.69\\
   \cline{2-8}
   &Liu et al.\cite{b28}&131072&2473&0.0189&47.27&67.91&35.26\\
   \cline{2-8}
   &Chen et al.\cite{b25}&131072&19235&0.1468&38.39&57.12&26.40\\
   \cline{2-8}
   &Yang et al.\cite{b32}&131072&1765&0.0135&47.57&69.48&35.56\\
 \hline
   \multirow{4}{*}{Sailboat}
   &Qi et al.\cite{b26}&131072&952&0.0073&53.37&70.79&41.41\\
   \cline{2-8}
   &Liu et al.\cite{b28}&131072&1185&0.0090&61.95&73.60&50.19\\
   \cline{2-8}
   &Chen et al.\cite{b25}&131072&24347&0.1858&37.85&57.08&25.86\\
   \cline{2-8}
   &Yang et al.\cite{b32}&131072&1910&0.0146&46.93&68.01&34.92\\
   \hline
    \multirow{4}{*}{Lena}
   &Qi et al.\cite{b26}&131072&696&0.0053&54.99&73.55&43.01\\
   \cline{2-8}
   &Liu et al. \cite{b28}&131072&550&0.0042&63.27&74.30&51.56\\
   \cline{2-8}
   &Chen et al.\cite{b25}&131072&4909&0.0375&41.90&70.05&29.87\\
   \cline{2-8}
   &Yang et al.\cite{b32}&131072&1778&0.0136&47.34&69.78&35.33\\
   \hline
   \hline
   \multicolumn{8}{c}{}\\
   \multicolumn{8}{c}{}\\
\end{tabular}
 }
\end{table*}

\begin{table*}[t]
  \caption{{The versatility verification by embedding color images using \cite{b28} and \cite{alpha_fusion}.}}
  \label{tab2}
  \centering
  \resizebox{0.9\textwidth}{!}{
  \begin{tabular}{c|c|c|c|c|c|c|c|c}
  \hline
  \hline
  \textbf{Host Image}&\textbf{Methods}&\textbf{Channel}&$\bm{N_D}$ (bits)&$\bm{N_C}$ (bits) &$\bm{Ratio}$&$\bm{\mathrm{PSNR_{I}}}$ (dB)&$\bm{\mathrm{PSNR_{N}}}$ (dB)&$\bm{\mathrm{PSNR_{W}}}$ (dB)\\
  \hline
   \multirow{6}{*}{Yacht}
    &\multirow{3}{*}{Liu et al.\cite{b28}}
    &R&131072&4252&0.0324&\multirow{3}{*}{42.53}&\multirow{3}{*}{68.29}&\multirow{3}{*}{30.50}\\
    \cline{3-6}
    &&G&131072&4357&0.0332&&&\\
    \cline{3-6}
    &&B&131072&7215&0.0550&&&\\
     \cline{2-9}
    &\multirow{3}{*}{Alpha fusion\cite{alpha_fusion}}
    &R&131072&4057&0.0310&\multirow{3}{*}{44.04}&\multirow{3}{*}{70.71}&\multirow{3}{*}{32.01}\\
    \cline{3-6}
    &&G&131072&3973&0.0303&&&\\
    \cline{3-6}
    &&B&131072&4023&0.0307&&&\\

\hline
\multirow{6}{*}{Fruits}
    &\multirow{3}{*}{Liu et al.\cite{b28}}
    &R&131072&4064&0.0310&\multirow{3}{*}{44.62}&\multirow{3}{*}{63.18}&\multirow{3}{*}{32.64}\\
    \cline{3-6}
    &&G&131072&3790&0.0289&&&\\
    \cline{3-6}
    &&B&131072&2930&0.0224&&&\\
     \cline{2-9}
    &\multirow{3}{*}{Alpha fusion\cite{alpha_fusion}}
    &R&131072&3022&0.0231&\multirow{3}{*}{47.09}&\multirow{3}{*}{63.28}&\multirow{3}{*}{35.14}\\
    \cline{3-6}
    &&G&131072&3020&0.0230&&&\\
    \cline{3-6}
    &&B&131072&3048&0.0233&&&\\
\hline
 \multirow{6}{*}{Pens}
    &\multirow{3}{*}{Liu et al.\cite{b28}}
    &R&131072&5511&0.0420&\multirow{3}{*}{41.56}&\multirow{3}{*}{66.69}&\multirow{3}{*}{29.53}\\
    \cline{3-6}
    &&G&131072&4788&0.0365&&&\\
    \cline{3-6}
    &&B&131072&5756&0.0439&&&\\
    \cline{2-9}
            &\multirow{3}{*}{Alpha fusion\cite{alpha_fusion}}
    &R&131072&3306&0.0252&\multirow{3}{*}{45.34}&\multirow{3}{*}{69.05}&\multirow{3}{*}{33.32}\\
    \cline{3-6}
    &&G&131072&2610&0.0199&&&\\
    \cline{3-6}
    &&B&131072&3071&0.0234&&&\\
\hline
 \multirow{6}{*}{Flower}
    &\multirow{3}{*}{Liu et al.\cite{b28}}
    &R&131072&2152&0.0164&\multirow{3}{*}{44.35}&\multirow{3}{*}{68.74}&\multirow{3}{*}{32.33}\\
    \cline{3-6}
    &&G&131072&3276&0.0250&&&\\
    \cline{3-6}
    &&B&131072&3800&0.0290&&&\\
     \cline{2-9}
    &\multirow{3}{*}{Alpha fusion\cite{alpha_fusion}}
    &R&131072&1661&0.0127&\multirow{3}{*}{45.21}&\multirow{3}{*}{76.20}&\multirow{3}{*}{33.17}\\
    \cline{3-6}
    &&G&131072&3347&0.0255&&&\\
    \cline{3-6}
    &&B&131072&3517&0.0268&&&\\

\hline
 \multirow{6}{*}{Couple}
    &\multirow{3}{*}{Liu et al.\cite{b28}}
    &R&131072&4163&0.0318&\multirow{3}{*}{44.15}&\multirow{3}{*}{66.91}&\multirow{3}{*}{32.13}\\
    \cline{3-6}
    &&G&131072&4547&0.0347&&&\\
    \cline{3-6}
    &&B&131072&4641&0.0354&&&\\
     \cline{2-9}
    &\multirow{3}{*}{Alpha fusion\cite{alpha_fusion}}
    &R&131072&4001&0.0305&\multirow{3}{*}{44.70}&\multirow{3}{*}{66.89}&\multirow{3}{*}{32.69}\\
    \cline{3-6}
    &&G&131072&3987&0.0304&&&\\
    \cline{3-6}
    &&B&131072&3711&0.0283&&&\\

\hline

 \multirow{6}{*}{Cable car}
    &\multirow{3}{*}{Liu et al.\cite{b28}}
    &R&131072&6666&0.0509&\multirow{3}{*}{41.73}&\multirow{3}{*}{64.88}&\multirow{3}{*}{29.71}\\
    \cline{3-6}
    &&G&131072&7008&0.0535&&&\\
    \cline{3-6}
    &&B&131072&5910&0.0451&&&\\
     \cline{2-9}
    &\multirow{3}{*}{Alpha fusion\cite{alpha_fusion}}
    &R&131072&3759&0.0287&\multirow{3}{*}{46.37}&\multirow{3}{*}{66.57}&\multirow{3}{*}{34.37}\\
    \cline{3-6}
    &&G&131072&3188&0.0243&&&\\
    \cline{3-6}
    &&B&131072&3482&0.0266&&&\\
\hline
 \multirow{6}{*}{Cornfield}
    &\multirow{3}{*}{Liu et al.\cite{b28}}
    &R&131072&12435&0.0949&\multirow{3}{*}{38.12}&\multirow{3}{*}{66.42}&\multirow{3}{*}{26.09}\\
    \cline{3-6}
    &&G&131072&8945&0.0682&&&\\
    \cline{3-6}
    &&B&131072&14715&0.1123&&&\\
     \cline{2-9}
    &\multirow{3}{*}{Alpha fusion\cite{alpha_fusion}}
    &R&131072&2001&0.0153&\multirow{3}{*}{42.53}&\multirow{3}{*}{64.40}&\multirow{3}{*}{30.52}\\
    \cline{3-6}
    &&G&131072&4557&0.0348&&&\\
    \cline{3-6}
    &&B&131072&7194&0.0549&&&\\

\hline
 \multirow{6}{*}{Girl}
    &\multirow{3}{*}{Liu et al.\cite{b28}}
    &R&131072&2551&0.0195&\multirow{3}{*}{43.85}&\multirow{3}{*}{71.57}&\multirow{3}{*}{31.82}\\
    \cline{3-6}
    &&G&131072&3511&0.0268&&&\\
    \cline{3-6}
    &&B&131072&4099&0.0313&&&\\
     \cline{2-9}
    &\multirow{3}{*}{Alpha fusion\cite{alpha_fusion}}
    &R&131072&1695&0.0129&\multirow{3}{*}{45.62}&\multirow{3}{*}{66.96}&\multirow{3}{*}{33.61}\\
    \cline{3-6}
    &&G&131072&3206&0.0245&&&\\
    \cline{3-6}
    &&B&131072&3582&0.0273&&&\\

    \hline
    \hline
\end{tabular}
}
\end{table*}

Table \ref{tab1} and Table \ref{tab2} list objective results including $N_D, N_C, Ratio, PSNR_{I}, PSNR_N, PSNR_W$, respectively, where $PSNR_{I}, PSNR_N, PSNR_W$ are PSNR values of the whole image, the non-visible-watermarked region, and the visible-watermarked region between the watermarked images $\I^w$ and $\I^{w\prime\prime}$, respectively.
Here, for color images in Table \ref{tab2}, the proposed method needs to be used separately for data compression and embedding in each channel. Since the data difference of each channel is usually relatively large, and the corresponding compression rate is also different, we investigate the $Ratio$ for each color channel separately. In contrast, PSNR generally measures the overall visual effect of the image. Therefore, we calculate the PSNR by averaging the PSNR values of R, G and B channels.
The results verify that the proposed method achieves excellent performance for any embedding methods, whether they are in the spatial domain or the transform domain, used for gray-scale or RGB images, reversible or irreversible, \etc.
On the one hand, the proposed method effectively reduces the data volume via compression.
The maximum amount of the compressed data ($N_C$) of different images compressed by our method is no more than 25000 bits. For most watermarked images, the amount of the compressed data is less than 10000 bits, and even less than 1000 bits for some embedding methods. Such results indicate that our method has high compression efficiency for the difference image, thus leading to efficient embedding.
For example, for the Lena image with size of $512\times512$ pixels, while more than 50000 bits of information can be embedded by using conventional RDH methods, such as histogram-modification \cite{b5}, pixel-value-ordering \cite{NEW_PVO} or difference-expansion \cite{NEW_DE}, the proposed compression method only need no more than 25000 bits. Therefore, the proposed framework can be used for almost all available reversible data hiding methods due to the high compression efficiency for effective embedding.

On the other hand, by comparing the values of $PSNR_{I}$, $PSNR_N$, and $PSNR_W$ in Tables \ref{tab1} and \ref{tab2}, we observe that the modification of the watermarked images using our method mainly focuses on the visible-watermarked regions, while the PSNR of the non-visible-watermarked area is very high ($\geq$ 50 dB), with that of nearly half of the images higher than 70 dB.

Further, to evaluate the performance for images taken with digital cameras, we use the images in the {\it Kodak24} dataset \cite{kodakdb} as the host images, and adopt methods \cite{b28} and \cite{alpha_fusion} to embed visible watermark images in Fig. \ref{fig4.1e}-\ref{fig4.1h} into the host images, respectively.

\begin{figure*}[]
\centering
  \includegraphics[width=0.45\textwidth]{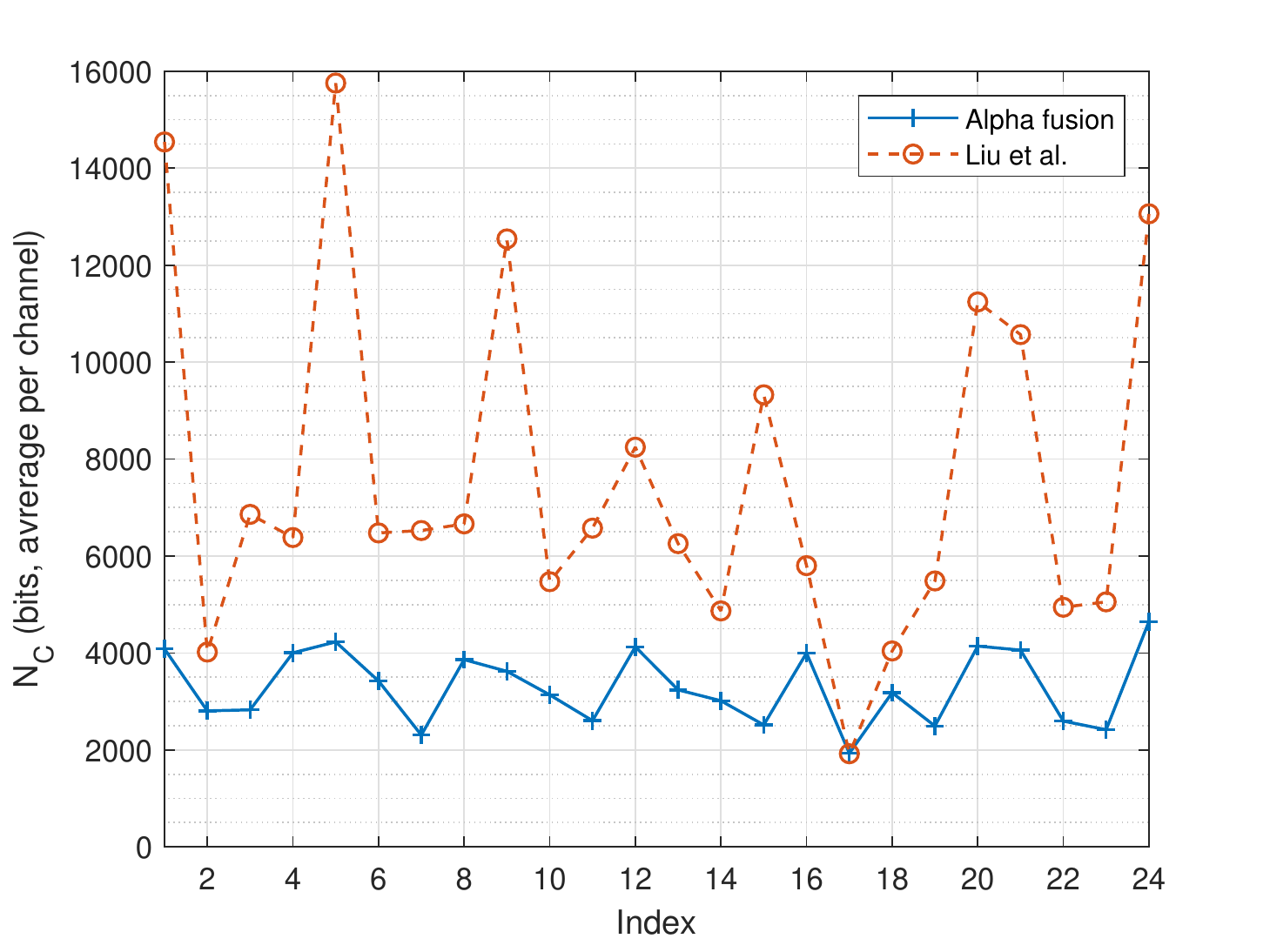}
 \caption{Data compression performance ($N_c$) of our proposed method adopting RVW methods \cite{b28} and \cite{alpha_fusion}, on the {\it Kodak24} dataset.}
   \label{fig:kodak_N_C}
\end{figure*}

\begin{figure*}[]
\centering
\subfigure[]{
  \includegraphics[width=0.45\textwidth]{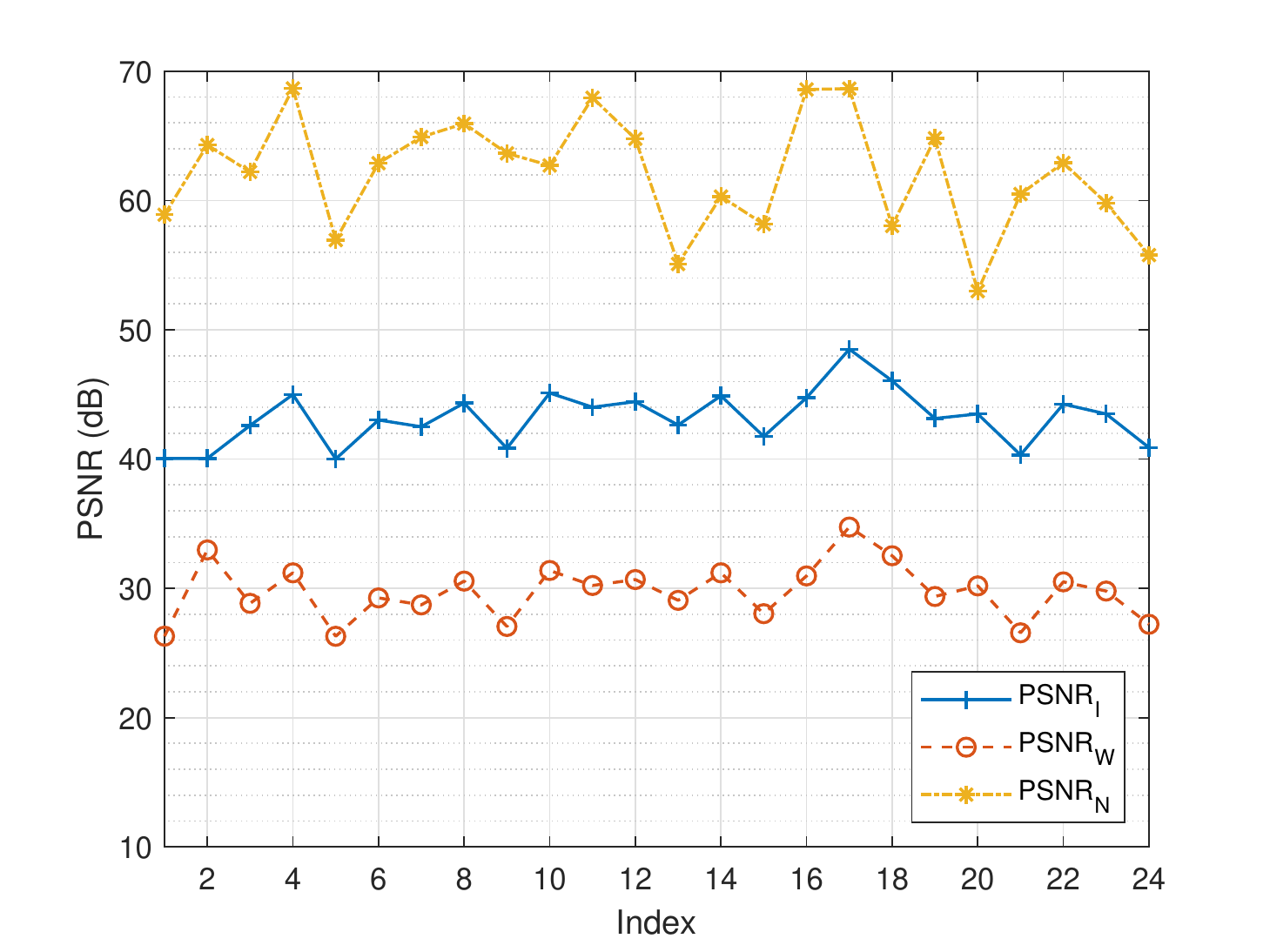}
  \label{fig:kodak_liu_PSNR}
}
\subfigure[]{
  \includegraphics[width=0.45\textwidth]{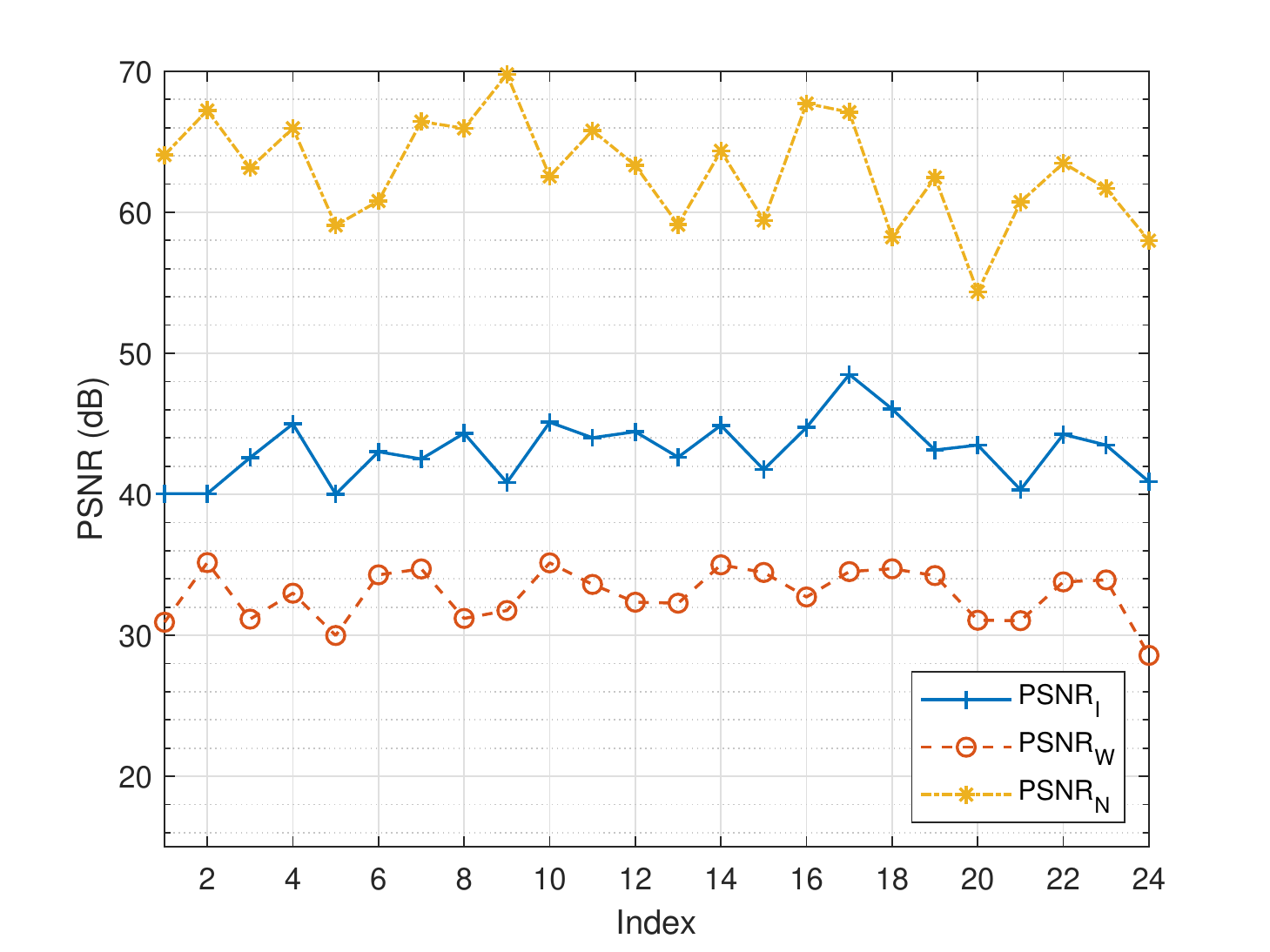}
  \label{fig:kodak_alpha_PSNR}
}
 \caption{PSNR performance of our proposed method adopting RVW methods \cite{b28} (a) and \cite{alpha_fusion} (b), on the {\it Kodak24} dataset.}
   \label{fig:kodak_PSNR}
\end{figure*}

The experimental results are shown in Fig.~\ref{fig:kodak_N_C} and Fig.~\ref{fig:kodak_PSNR}, respectively. The results in Fig.~\ref{fig:kodak_N_C} show that the amount of the compressed data $N_C$ is less than 20000 bits for all watermarked images, and even less than 10000 bits for most images.
In addition, by comparing the values of $PSNR_{I}$, $PSNR_N$, and $PSNR_W$ in Fig.~\ref{fig:kodak_PSNR}, the PSNR of the non-visible-watermarked area is still very high ($\geq$ 55 dB), which demonstrates that our method leads to significant compression performance while the quality of the watermarked images is well preserved.


As mentioned above, the image quality of the visible-watermarked areas is degraded a bit due to the additional modifications, which however is negligible to human perception.
In all, our method leads to significant compression performance while the quality of the watermarked images is well preserved.

To summarize, the proposed framework enables a variety of existing visible watermarking schemes to be reversible. If necessary, the visible watermark image can be completely removed, and the original host image can be restored without any distortion, which is applicable to both RGB images and gray-scale images.
This validates the strong versatility of the proposed RVW framework.

\subsection{Performance Comparison with Competitive RVW Methods}
\label{s4.3}

\begin{figure*}[htbp]
\centering
\subfigure[]{
\begin{minipage}{0.2\textwidth}
    \centering
    \centering{\includegraphics[width=\textwidth]{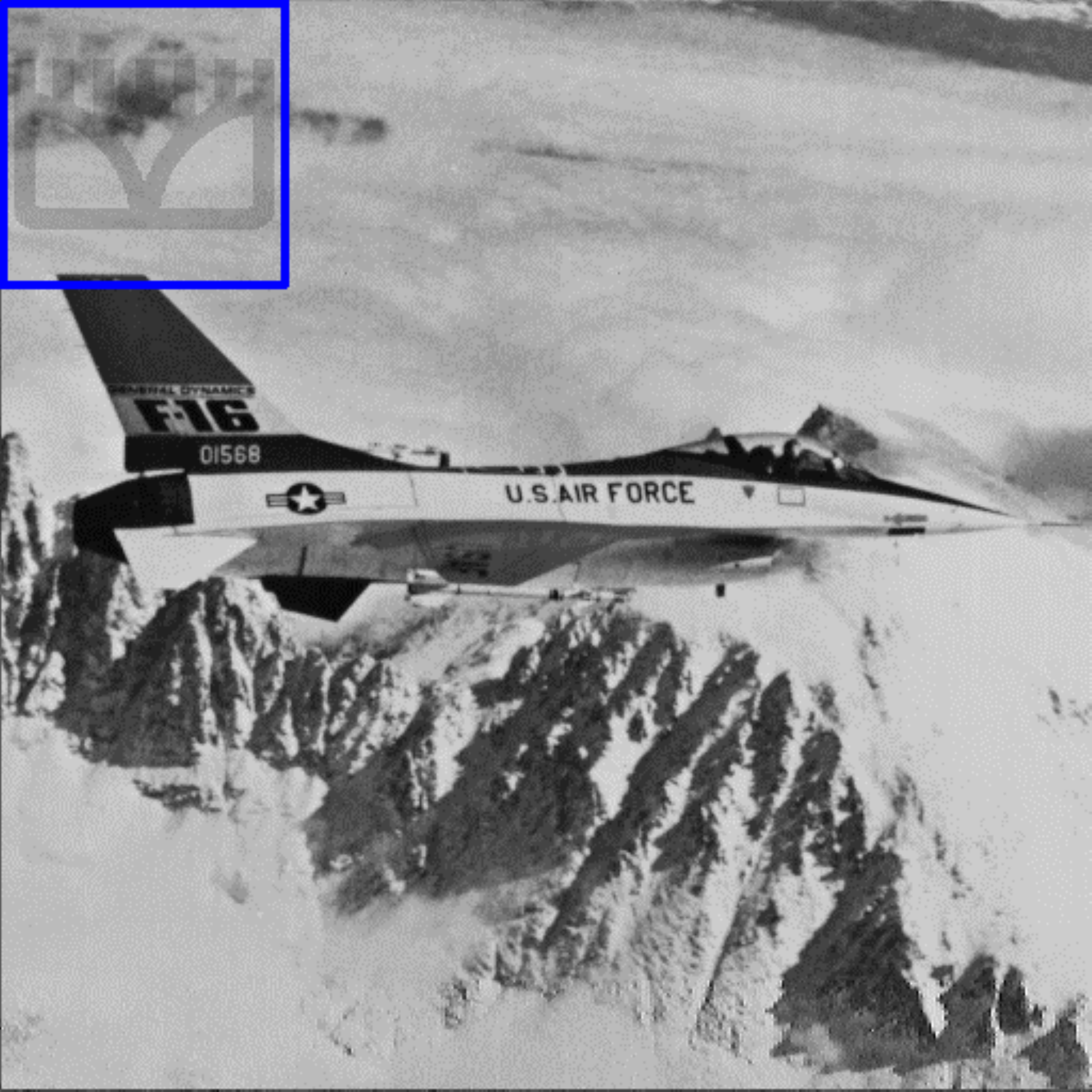}}
\\[5pt]
    \centering{\includegraphics[width=\textwidth]{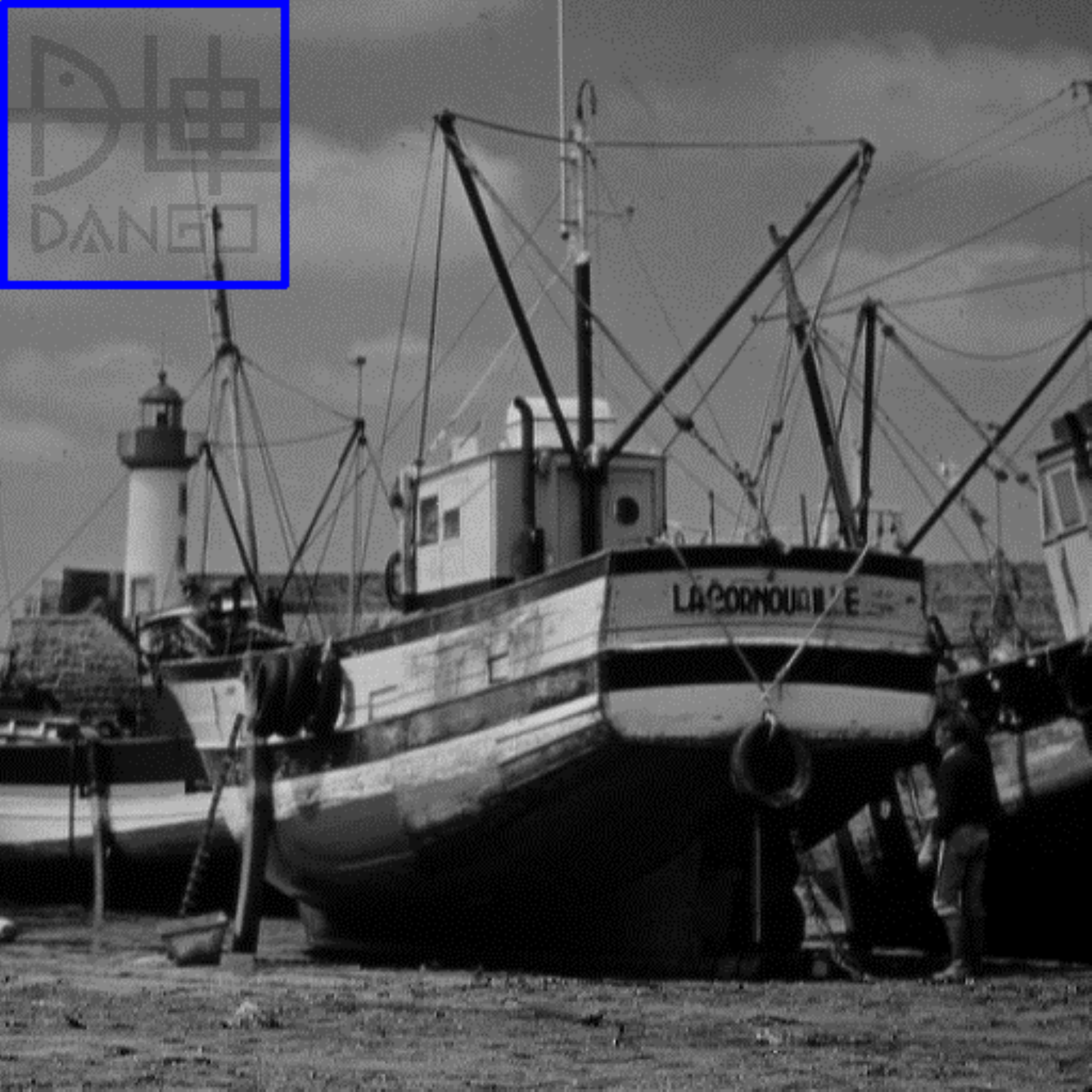}}
    \hspace{1pt}
  \end{minipage}%
  \label{fig4.9.1}
}
\subfigure[]{
\begin{minipage}{0.2\textwidth}
    \centering
\centering{\includegraphics[width=\textwidth]{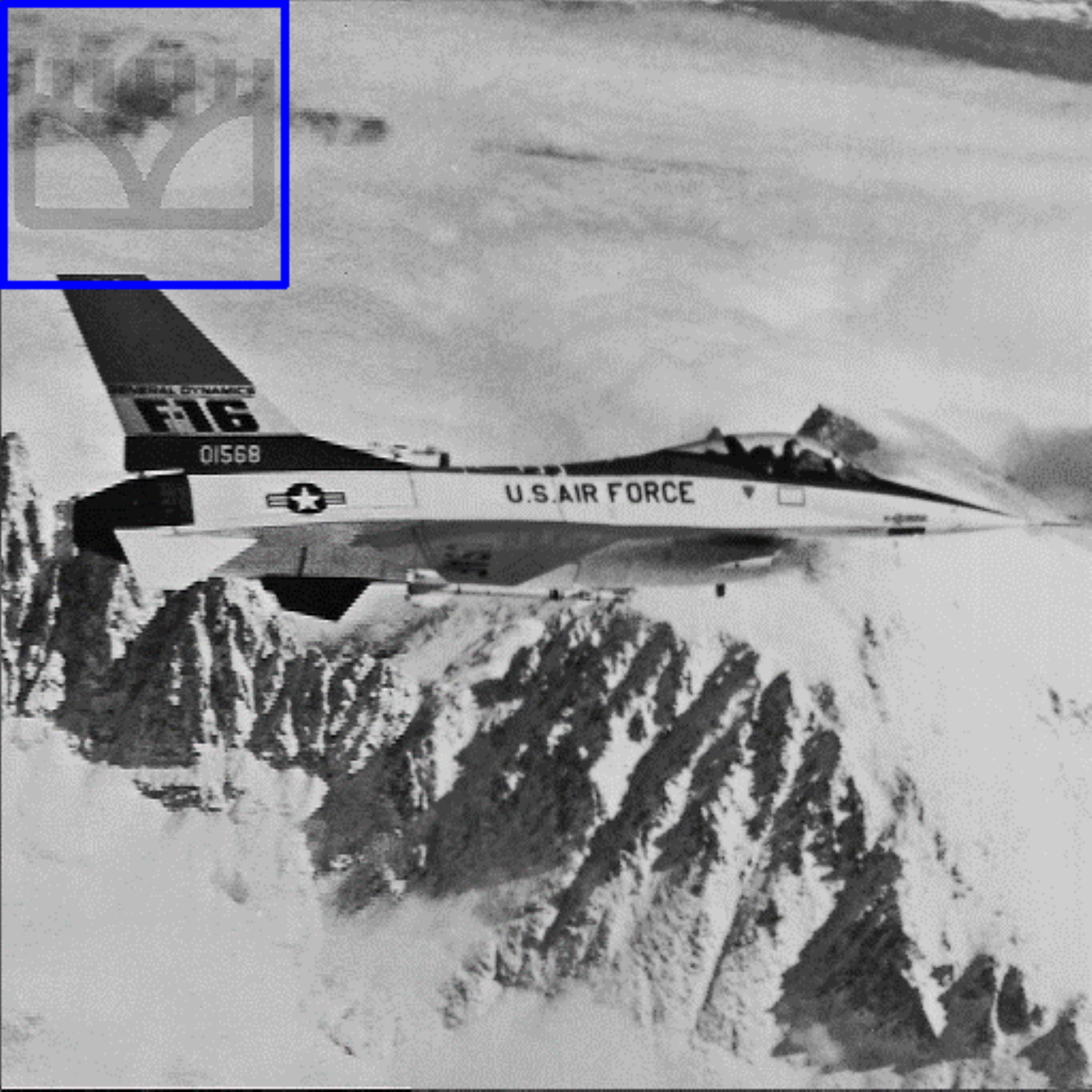}}
\\[5pt]
\centering{\includegraphics[width=\textwidth]{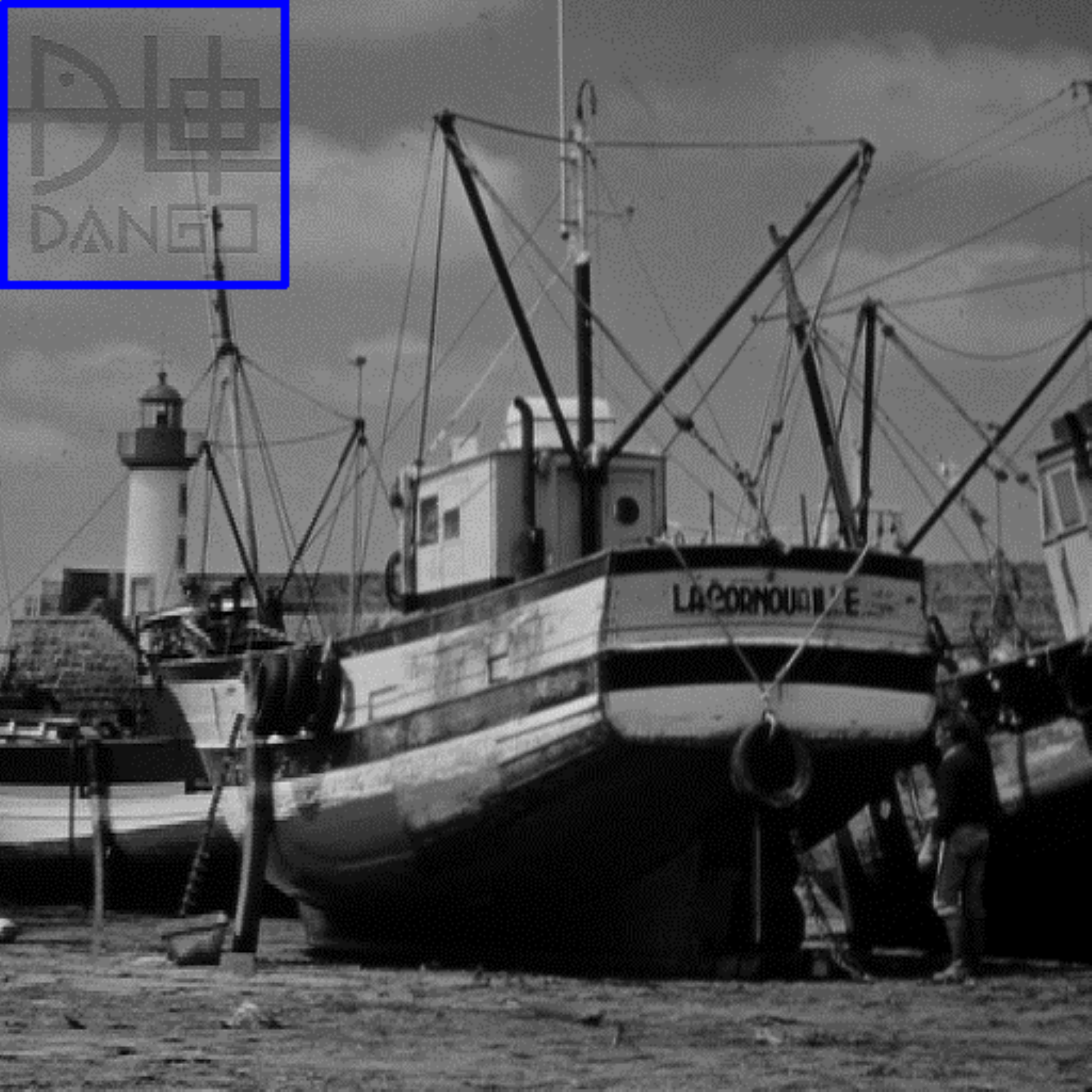}}
\hspace{1pt}
  \end{minipage}%
  \label{fig4.9.2}
  }
 \subfigure[]{
  \begin{minipage}{0.2\textwidth}
    \centering{\includegraphics[width=\textwidth]{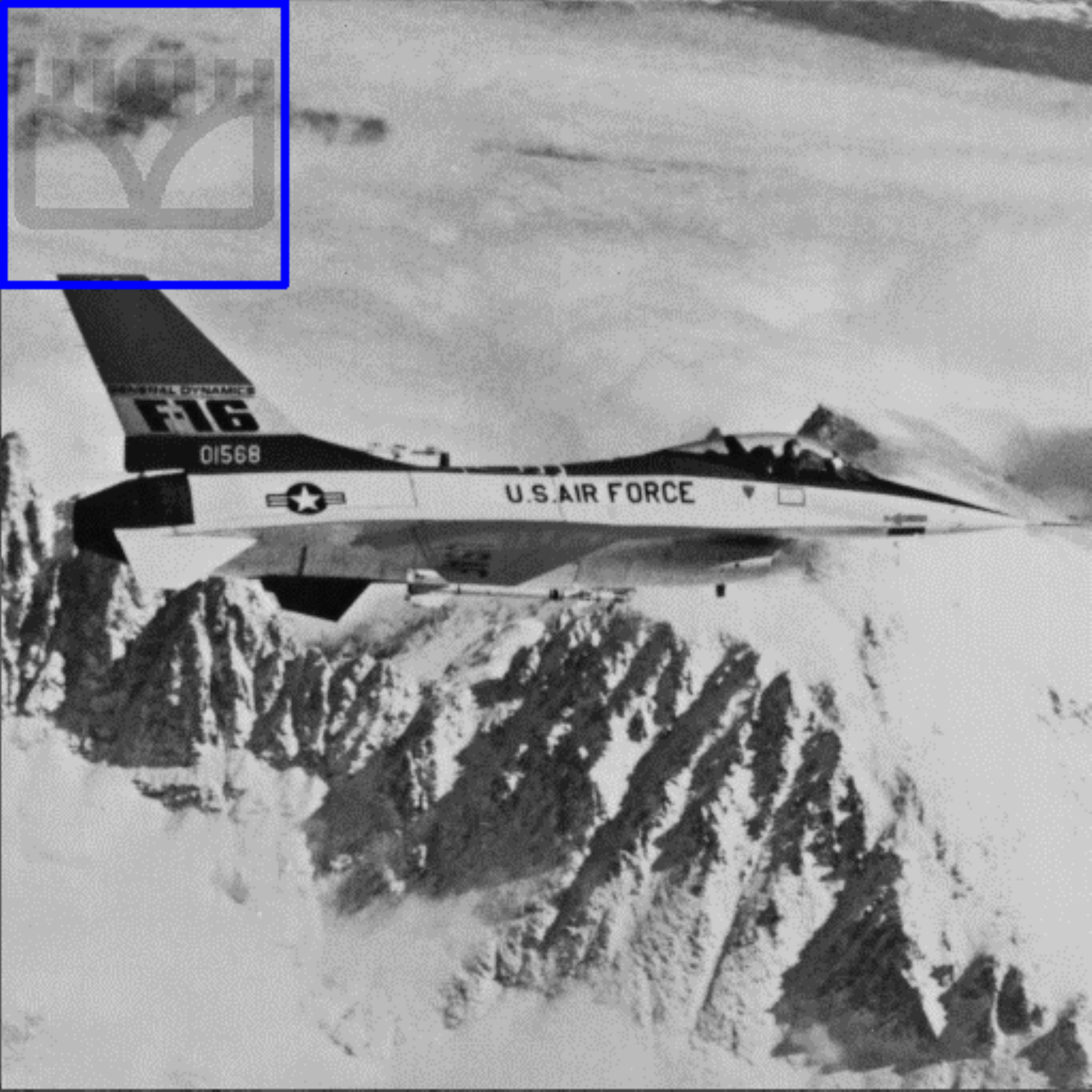}}
\\[5pt]
\centering{\includegraphics[width=\textwidth]{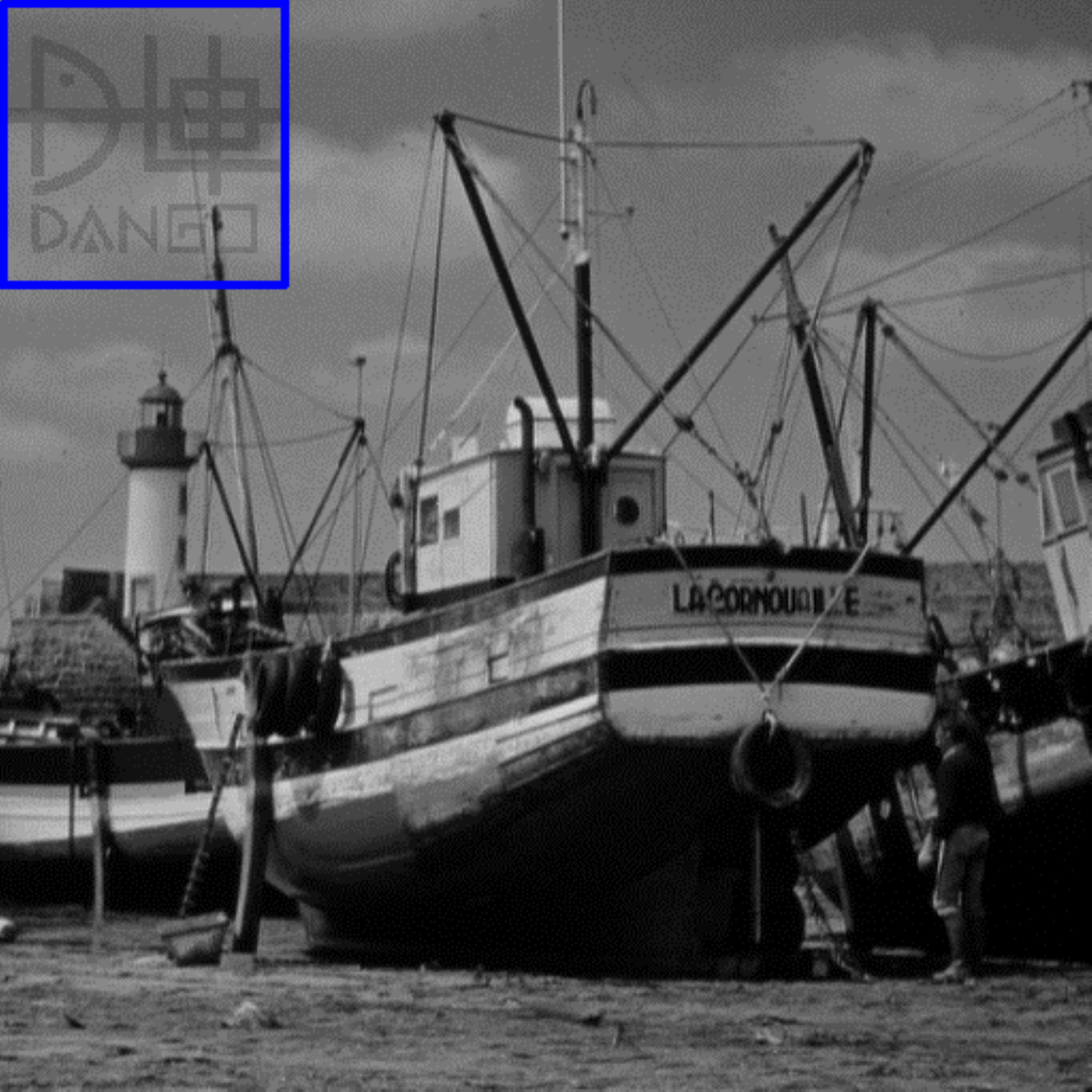}}
\hspace{1pt}
  \end{minipage}%
\label{fig4.9.3}
  }
\caption{Comparison of the final watermarked images from (a) \cite{b39}, (b) \cite{b27}, and (c) our proposed method. Inside the blue box is ROI. }
\label{fig4.9}
\end{figure*}

Finally, we compare the performance of the proposed method with competitive RVW approaches \cite{b27,b39} introduced in Section \ref{sec:related}.
We first embed the visible watermark images in Fig. \ref{fig4.1a}-\ref{fig4.1d} into the host images \textit{Pens}, \textit{Fruits}, \textit{Flower}, \textit{Couple}, \textit{Cornfield}, \textit{Cable car}, \textit{Girl}, \textit{Yacht} using the method in \cite{b39} to obtain the watermarked images $I^w$. Then, the RMR is constructed by deploying the methods in \cite{b39}, \cite{b27}, and our proposed method, respectively.
We show some visual results in Fig. \ref{fig4.9}, and present the compressed data volume $N_C, PSNR_{I}, PSNR_N, PSNR_W$ obtained by these methods in Table \ref{tab3}.

\begin{table*}[t]
  \caption{Performance Comparison among our proposed method, Yang et al. \cite{b39} and Qin et al. \cite{b27}.}
  \label{tab3}
  \centering
  \resizebox{0.9\textwidth}{!}{
  \begin{tabular}{c|c|c|c|c|c|c|c}
  \hline
  \hline
\textbf{Host Image}&\textbf{Methods}&$\bm{N_D}$ (bits)&$\bm{N_C}$ (bits) &$\bm{Ratio}$&$\bm{\mathrm{PSNR_{I}}}$ (dB)&$\bm{\mathrm{PSNR_{N}}}$ (dB)&$\bm{\mathrm{PSNR_{W}}}$ (dB)\\
  \hline
   \multirow{3}{*}{F-16}
   &Yang et al.\cite{b39}&131072&13192&0.1006&40.12&39.84&$-$\\
   \cline{2-8}
   &Qin  et al.\cite{b27}&131072&49505&0.3777&41.68&34.91&35.61\\
   \cline{2-8}
   &Ours&\textbf{131072}&\textbf{651}&\textbf{0.0050}&\textbf{59.73}&\textbf{79.23}&\textbf{47.73}\\
   \hline

    \multirow{3}{*}{Boat}
   &Yang  et al.\cite{b39}&131072&13584&0.1036&41.91&41.63&$-$\\
   \cline{2-8}
   &Qin  et al.\cite{b27}&131072&39020&0.2977&42.98&39.35&38.40\\
   \cline{2-8}
   &Ours&\textbf{131072}&\textbf{637}&\textbf{0.0049}&\textbf{58.72}&\textbf{66.57}&\textbf{47.40}\\
   \hline

   \multirow{3}{*}{Barbara}
   &Yang  et al.\cite{b39}&131072&14776&0.1127&34.35&34.07&$-$\\
   \cline{2-8}
   &Qin  et al.\cite{b27}&131072&50094&0.3822&40.64&37.19&29.77\\
   \cline{2-8}
   &Ours&\textbf{131072}&\textbf{742}&\textbf{0.0057}&\textbf{55.60}&\textbf{79.32}&\textbf{43.58}\\
   \hline

   \multirow{3}{*}{Cameraman}
   &Yang  et al.\cite{b39}&131072&12568&0.0959&59.59&59.31&$-$\\
   \cline{2-8}
   &Qin  et al.\cite{b27}&131072&36636&0.2795&51.17&59.61&36.89\\
   \cline{2-8}
   &Ours&\textbf{131072}&\textbf{623}&\textbf{0.0048}&\textbf{65.55}&\textbf{84.40}&\textbf{53.56}\\
   \hline

   \multirow{3}{*}{Goldhill}
   &Yang  et al.\cite{b39}&131072&14008&0.1069&46.56&46.28&$-$\\
   \cline{2-8}
   &Qin  et al.\cite{b27}&131072&55226&0.4213&39.55&35.47&33.52\\
   \cline{2-8}
   &Ours&\textbf{131072}&\textbf{688}&\textbf{0.0052}&\textbf{55.52}&\textbf{71.46}&\textbf{43.59}\\
   \hline

   \multirow{3}{*}{Pepper}
   &Yang  et al.\cite{b39}&131072&13752&0.1049&38.64&38.36&$-$\\
   \cline{2-8}
   &Qin  et al.\cite{b27}&131072&52272&0.3988&40.86&37.17&32.33\\
   \cline{2-8}
   &Ours&\textbf{131072}&\textbf{804}&\textbf{0.0061}&\textbf{55.14}&\textbf{69.06}&\textbf{43.27}\\
   \hline

   \multirow{3}{*}{Sailboat}
   &Yang  et al.\cite{b39}&131072&14512&0.1107&36.69&36.41&$-$\\
   \cline{2-8}
   &Qin  et al.\cite{b27}&131072&47241&0.3604&40.24&35.08&29.74\\
   \cline{2-8}
   &Ours&\textbf{131072}&\textbf{933}&\textbf{0.0071}&\textbf{54.96}&\textbf{74.61}&\textbf{42.96}\\
   \hline

   \multirow{3}{*}{Lena}
   &Yang  et al.\cite{b39}&131072&17880&0.1364&39.44&39.16&$-$\\
   \cline{2-8}
   &Qin  et al.\cite{b27}&131072&43940&0.3352&44.31&40.73&39.48\\
   \cline{2-8}
   &Ours&\textbf{131072}&\textbf{589}&\textbf{0.0045}&\textbf{58.71}&\textbf{78.33}&\textbf{46.72}\\
   \hline
   \hline

\end{tabular}
}
\end{table*}

We observe from Fig. \ref{fig4.9} that, the overall visual qualities of the watermarked images are almost the same.
However, as discussed in Section~\ref{sec:intro} and Section~\ref{sec:related}, \cite{b39} is restricted to a specific visible watermarking scheme, while \cite{b27} is generic but requires the prior knowledge of the locations of visible watermark embedding.
Further, Table \ref{tab3} shows that the compression performance of our proposed method significantly outperforms the methods in \cite{b27} and \cite{b39}.
Specifically, our results achieve compression gain by 66 times on average when compared with \cite{b27}, and gain by 21 times when compared with \cite{b39}.
Meanwhile, both the $PSNR_{I}$ and $PSNR_N$ obtained by the proposed method are the highest.
In particular, for the non-visible-watermarked area, $PSNR_ N$'s are almost up to $70$ dB or even higher, which gives credits to the smaller volume of the reconstruction data packets compressed by our proposed method.

In addition, it is worthwhile noting that in the method \cite{b39}, the visible-watermarked area has not been modified. However, in the method \cite{b27}, the visible-watermarked area has been modified during RDH, so the quality of the visible-watermarked area has decreased. Similarly, in our proposed method, to compensate for the lossy compression error, the error matrix $\e$ is explicitly superimposed on the visible-watermarked area. Even so, the image quality of our proposed method in the visible-watermarked area is still significantly better than the method \cite{b27}.
Moreover, the image quality in the non-visible-watermarked area and the overall image quality of the proposed method are better than those of both methods \cite{b27,b39}.

In summary, compared with the competitive methods in \cite{b27} and \cite{b39}, our proposed method not only effectively compresses the difference between the original host image and the watermarked image and thus significantly reduces the capacity of the hidden information, but also improves the quality of the watermarked images.
Further, our method is compatible with any visible watermarking schemes, thus providing strong versatility.



\section{Conclusion}
\label{sec:conclude}
In this paper, we propose a versatile framework for reversible visible watermarking, which is generic to various visible watermarking algorithms in a blind manner and meanwhile leads to satisfactory visual quality of the watermarked image. 
We develop a versatile framework for the complete removal of the visible watermark image and lossless restoration of the original host image, with the assistance of a reconstruction data packet---a compressed version of the difference image between the watermarked image and the host image. 
The key is to achieve compact compression of the difference image for efficient embedding of the reconstruction data packet. 
In particular, we propose regularized Graph Fourier Transform (GFT) coding, where the difference image is smoothed via the graph Laplacian regularizer and then efficiently encoded by multi-resolution GFTs in an approximately optimal manner.
Experimental results demonstrate the versatility and compression efficiency of the proposed method, which promotes the application of visible watermarking in the field of multimedia data security, copyright protection and so on.



\ifCLASSOPTIONcaptionsoff
  \newpage
\fi

\bibliographystyle{IEEEtran}
\bibliography{7_ref}

\end{document}